\documentclass[aps,prd,amsmath,floats,floatfix, twocolumn, superscriptaddress,nofootinbib,showpacs]{revtex4-1}
\usepackage{soul}
\usepackage{graphicx}
\usepackage{dcolumn}
\usepackage{bm}
\usepackage{xcolor}
\usepackage{hyperref}
\usepackage{placeins}
\usepackage[deletedmarkup=xout,commentmarkup=uwave,commandnameprefix=ifneeded]{changes}
\usepackage{orcidlink}

\definechangesauthor[name=Harrison, color=teal]{HS}

\begin{document}

\preprint{APS/123-QED}

\title{The impact of waveform systematics and Gaussian noise on the interpretation of GW231123}


\author{
S.~Bini}
\email{bini@caltech.edu}
\affiliation{LIGO Laboratory, California Institute of Technology, Pasadena, CA 91125, USA}

\author{K.~Kr\'ol}
\affiliation{Astronomical Observatory, University of Warsaw, Al. Ujazdowskie 4, 00-478 Warszawa, Poland}

\author{K.~Chatziioannou}
\affiliation{LIGO Laboratory, California Institute of Technology, Pasadena, CA 91125, USA}
\affiliation{TAPIR, California Institute of Technology, Pasadena, CA 91125, USA}

\author{M.~Isi}
\affiliation{Columbia University, Department of Astronomy, 550 West 120th Street, New York, NY, 10027, USA}
\affiliation{Center for Computational Astrophysics, Flatiron Institute, New York, NY 10010, USA}

\date{\today}

\begin{abstract}
GW231123 is an exceptional gravitational-wave event consistent with the merger of two massive, highly-spinning black holes. 
Reliable inference of the source properties is crucial for accurate interpretation of its astrophysical implications. 
However, characterization of GW231123 is challenging: only few signal cycles are observed and different signal models result in systematically different parameters.
We investigate whether the interpretation of GW231123 is robust against model systematics and Gaussian detector noise.
We show that the model systematics observed in GW231123 can be reproduced for a simulated signal based on the numerical-relativity surrogate model \textsc{NRSur7dq4}. 
Simulating data using the maximum-likelihood \textsc{NRSur7dq4} waveform for GW231123 and no noise realization, we closely recover the systematics observed for the real signal. 
We then explore how the headline properties of GW231123 are impacted by Gaussian detector noise. 
Using the \textsc{NRSur7dq4} maximum-likelihood waveform and different noise realizations, we consistently find support for large masses, high spin magnitudes (median $\chi_1\geq 0.7$), and high spin precession (median $\chi_\mathrm{p}\geq 0.68$).
The spin in the direction of the angular momentum ($\chi_\mathrm{eff}$) fluctuates more.
Finally, again comparing to simulated signals, we show that any differences in the GW231123 inference based on each separate detector are not statistically significant.
These results show that the properties of GW231123, and most importantly the high mass and high spin magnitudes inferred by \textsc{NRSur7dq4}, are robust.
\end{abstract}

\maketitle

\section{Introduction}

The exceptional gravitational-wave (GW) event, GW231123\_135430 shortened
as GW231123, is consistent with the merger of two massive black holes (BHs) with total mass $190{-}265\,M_\odot$ and high component spins~\cite{GW231123}. 
The event was observed by the LIGO Livingston and LIGO Hanford detectors~\cite{ LIGOScientific:2014pky} during the first part of the fourth observing run (O4a) and it is part of the subsequent GWTC-4 catalog~\cite{gwtc4-intro, GWTC4method, gwtc4results}. 
The event's mass is of particular astrophysical interest: pair instability is expected to preclude BHs with masses ${\sim} 60{-}130M_\odot$ \cite{2019ApJ...887...53F, Woosley:2021xba, Hendriks:2023yrw}. 
The primary BH of GW231123 lies within or beyond the mass gap, while the likely values for the secondary BH span the entire gap.
In addition, the high component spins (${\sim} 0.9$ and $0.8$) further challenge the astrophysical interpretation: 
models based on isolated binary formation predict small natal spins, and at most one spinning BH~\cite{Belczynski:2017gds, Qin:2018vaa, Fuller:2019sxi, vanSon:2020zbk}.  
Hierarchical mergers, i.e., mergers of BHs that were formed from previous mergers~\cite{Gerosa:2021mno}, may populate the mass gap. 
In this scenario, the spins are expected to be ${\sim} 0.7$ \cite{Gerosa:2021mno}, while higher spins, such as GW231123's, constrain the parent binaries to unequal masses
and dense astrophysical environments to overcome the merger recoils~\cite{GW231123}. 
For GW231123, though, the primaries of both parent binaries may lie in the mass gap as well, suggesting an origin in multiple previous mergers \cite{GW231123}.

Since its announcement, numerous potential formation scenarios for GW231123 have been investigated.  
Within the hierarchical merger scenario, Ref.~\cite{delfavero2025prospects} proposed formation via fourth and third generation BHs in active galactic nuclei (AGN).
Ref.~\cite{paiella2025assembling} discussed hierarchical mergers in low metallicity clusters, while Ref.~\cite{passenger2025gw231123} explored the scenario of two second-generation cluster BHs and argued that it is unlikely given that both component spins are high. 
Other scenarios include the merger of two BHs originating from massive, low metallicity stars with strong magnetic fields \cite{gottlieb2025spinning}; BHs formed from the direct collapse of massive, highly spinning stars \cite{croon2025can} and chemically homogeneous evolution \cite{popa2025very}; 
the merger of primordial BHs \cite{de2025gw231123, yuan2025gw231123}; population III stars \cite{liu2025formation,tanikawa2025gw231123}; a binary black hole (BBH) system with sustained accretion, which may occur either in AGN disks or in metal-poor or metal-free Population
III stars \cite{bartos2025accretion} or in dense star clusters \cite{kirouglu2025beyond}; and cosmic strings \cite{Cuceu:2025fzi}.

Accurate inference of the source properties of GW231123 is crucial for discerning the various formation channels. 
However, a number of factors complicate the analysis and interpretation of GW231123.
Among a population of events, the event identified as the ``highest-mass'' one is likely the one with the most extreme statistical noise fluctuation~\cite{Fishbach:2019ckx,Mandel:2025qnh}.
Moreover, the observed data from such high-mass signals are merger-dominated, with only few waveform cycles in the sensitive frequency band.
In such cases, spin inference may be driven by as little data as an individual waveform cycle~\cite{Miller:2023ncs, Miller:2025eak}. 
Finally, the inferred parameters differ substantially among five waveform models \cite{GW231123}: a direct surrogate to numerical relativity simulations \textsc{NRSur7dq4} (NRSur \cite{Varma:2019csw}), two frequency-domain phenomenological model whose primary difference is the treatment of spin dynamics \textsc{IMRPhenomXPHM} (XPHM \cite{Colleoni:2024knd}) and  \textsc{IMRPhenomXO4a} (XO4a \cite{Thompson:2023ase}), one time-domain phenomenological model \textsc{IMRPhenomTPHM} (TPHM \cite{Estelles:2021gvs}), and one time-domain model based on the effective-one-body framework \textsc{SEOBNRv5PHM} (SEOBNR \cite{Ramos-Buades:2023ehm}). Using simulated data, Ref.~\cite{Ray:2025rtt} explored the possibility that unmodeled non-Gaussian noise coincident with the event in the two detectors could impact its source inference. 
To simplify matters to some extent, Ref.~\cite{Jan:2025zcm} re-analyzed GW231123 using a waveform model that includes both orbital eccentricity and precession (TEOBResumS-Dali \cite{Nagar:2024dzj}), and found that the exclusion of eccentricity has minimal impact on the inference of GW231123.

While the three time-domain models (NRSur, TPHM, SEOBNR) yield broadly consistent results, those differ to results obtained with the two frequency-domain models (XPHM and XO4a).
The median mass ratio between the binary BHs is ${\sim}0.8$ for the time-domain models, ${\sim}0.6$ for XPHM, and ${\sim}0.4$ for XO4a. 
The spin magnitude of the secondary BH is high for the time-domain models, but uninformative for XPHM and XO4a.
The time-domain models recover a small inclination angle and a median distance of ${\sim}1.9\,$Gpc, ${\sim}2.7\,$Gpc, and ${\sim}2.3\,$Gpc for NRSur, TPHM, SEOBNR respectively.
XPHM supports an edge-on system at ${\sim}0.9\,$Gpc, while XO4a finds a nearly face-on system at distance ${\sim}3.5\,$Gpc. 
To complicate things further, an analysis based solely on the signal ringdown finds some support for even higher remnant mass and spin that are not consistent with any model~\cite{siegel2025gw231123}.
These competing results motivate this study.

Starting with the waveform systematics, Ref.~\cite{GW231123} presented model mismatches against numerical waveforms as well as inference on simulated signals.
As expected, NRSur is more faithful to existing simulations, as it is directly interpolated from them.
Analyzing simulated data based on two high-mass and high-spin simulations yielded in one case less prominent systematics than GW231123 and in the other case uniform biases towards lower secondary mass. However, even though the mass and spin magnitude of these simulations are in the one-dimensional GW231123 posterior credible region, the waveforms are different from GW231123, and thus their simulated data are not representative of the real signal.
The inability to reproduce the waveform systematics using simulated signals is cited as motivation to explore alternative interpretations \cite{Hu:2025lhv, Goyal:2025eqo}.
We instead simulate a signal much more faithful to GW231123 using the maximum-likelihood waveform obtained from the NRSur analysis.

Contrary to Ref.~\cite{GW231123}, we \emph{can} reproduce the waveform systematics observed in GW231123 in the absence of detector noise.  
We obtain similar results when simulating data based on the XO4a maximum-likelihood waveform.
Even though the inferred parameters differ significantly between NRSur and XO4a for GW231123, the maximum-likelihood waveforms are similar since they are both directly fitted to the data, see Fig.~\ref{fig:maxL_3models}.
Even with systematics, therefore, they are a more faithful representation of the observation than any existing numerical simulation.
Overall, the GW231123 systematics are consistent with a signal described by existing waveforms.
Our results are consistent with Ref.~\cite{Jan:2025zcm} based on the TEOBResumS-Dali maximum-likelihood waveform, and the NRSur, SEOBNRv5PHM, and TEOBResumS-Dali waveforms.

Next, we investigate whether the astrophysically-relevant conclusions drawn from GW231123, namely the high mass and high spins, are qualitatively affected by the exact realization of Gaussian noise. 
The impact of Gaussian noise on posterior distributions is well understood~\cite{cutler1994gravitational,Cutler:2007mi,Vallisneri:2007ev}: noise shifts the maximum-likelihood point and the posterior width encodes the expected shifts. 
Though this happens for all parameters and all posteriors in an expected way, noise shifts might affect restricted ``figures of merit" and their corresponding astrophysical interpretation.
For example, if the relevance of an exceptional observation hinges on the probability that its spin is larger than (say) 0.5, then the answer to \emph{that} very restricted question might be highly sensitive to noise.
We explore the degree of such sensitivity with simulated GW231123-like signals in various Gaussian noise realizations and inference with NRSur, XPHM, and XO4a. 
We show that for NRSur, the component masses and spin magnitudes are consistently inferred to be high, while the effective aligned spin varies more significantly.
Combining the studies of noise realization and systematics, we also find that the degree of manifested modeling systematics depends on the exact noise realization.

Finally, we consider the possibility of non-Gaussian noise.
Searches for non-Gaussian excess power (glitches) in the time-frequency window around the event determined that nearby glitches have no measurable effect \cite{GW231123}. 
However, subthreshold glitches ($\mathrm{SNR}<5$) overlapping the signal could still impact inference while remaining undetected~\cite{udall2025inferring}.
Moreover, the data of each LIGO detector separately yield posteriors that are not identical, raising concerns about data quality. 
Given that the two detectors experience different noise realizations, we explore the degree of difference between Livingston-only and Hanford-only posteriors expected in purely Gaussian noise.
Comparing Gaussian-noise simulations to results for GW231123 using each LIGO detector independently, we recover larger differences than the ones observed in GW231123 in ${\sim}30\%$ of the realizations for the total mass and mass ratio, and ${\sim} 38\%$ and ${\sim} 16\%$ for the spins. 
Thus, the differences observed in GW231123 are consistent with Gaussian noise.
We conclude by evaluating the prospect of detecting an event like GW231123 with the mid-2030s LIGO sensitivity (LIGO A$\#$).

The paper is structured as follows: Section \ref{sec:method} recalls the expected uncertainty on the measured source parameters induced by Gaussian noise and model inaccuracies, describes how we simulate a GW231123-like signal and the parameter estimation configuration, and introduces the Jensen–Shannon divergence as a measure of similarity between posterior distributions. 
Section \ref{sec:wf_syst} focuses on waveform systematics for GW231123. 
Next, we present the impact of Gaussian noise on the measurement of masses and spins in Sec.~\ref{sec:impact}, and on waveform systematics in Sec.~\ref{sec:Gaussian_noise_wf_system}. 
Section \ref{sec:HvsL} focuses on the single-detector inference on GW231123, and finally Sec.~\ref{sec:future} shows prospects with future sensitivity. We conclude in Sec.~\ref{sec:conclusions}.

\section{Method}
\label{sec:method}

We perform a set of simulations of GW231123-like signals into various realizations of Gaussian noise. 
This section describes the motivation for such approach (Sec.~\ref{subsec:motivation}), the choice of the simulated signal (Sec.~\ref{subsec:inj}), and the inference (Sec.~\ref{subsec:PE}). 
Finally, we define a metric employed to compare posterior distributions (Sec.~\ref{sec:JS}).

\subsection{Motivation for a simulation campaign}\label{subsec:motivation}

The impact of Gaussian noise \cite{cutler1994gravitational, branchesi2023science} and modeling systematics \cite{Lindblom:2008cm, miller2005accuracy,Cutler:2007mi} on GW inference can be analytically estimated in the Fisher framework under a high SNR approximation.
Although the approximation breaks down for GW231123, the main results are still qualitatively informative and we recap them here for motivation.
The framework relies on the Fisher Information Matrix $\Gamma$, defined by
\begin{equation}
    \Gamma_{jk} = \langle \partial_j \mathbf{h}|\partial_k  \mathbf{h}\rangle\,,
\end{equation}
where $\mathbf{h}(\theta)$ is the GW waveform characterized by parameters $\theta$. 
Boldface refers to quantities in multiple GW detectors. 
The term $\langle \cdot|\cdot\rangle$ is the noise weighted inner product defined for a specific detector as:
\begin{equation}
    \langle a | b \rangle = 4~\mathrm{Re}\int_{0}^\infty \frac{\tilde a^*(f) \tilde b(f)}{S_n(f)}\,,
\end{equation} 
where $S_n(f)$ is the noise spectral density (PSD), with multiple detectors added in quadrature.
An overview of the basic principle of the Fisher matrix formalism and a discussion of the strengths and limitations of this approach can be found in Ref.~\cite{Vallisneri:2007ev}. 

For a given GW signal, different realizations of the detector Gaussian noise will result in different maximum-likelihood parameters. 
For large SNR, or equivalently when the GW waveform is a linear function of the parameters, the maximum-likelihood parameters will follow a normal distribution centered on the true value.
The difference between the maximum-likelihood parameters $\theta^i_{\mathrm{ML}}$ and the true parameters $\theta_{\mathrm{TR}}^i$ due to detector noise $\mathbf{n}$ is given by:
\begin{equation}\label{eq:error_mL}
    \Delta \theta^i \equiv \theta_{\mathrm{ML}}^i - \theta_{\mathrm{TR}}^i \approx (\Gamma^{-1}({\theta_{\mathrm{ML}}}))^{ij}\langle \partial_j \mathbf{h}({\theta_{\mathrm{ML}}}) | \mathbf{n} \rangle\,.
\end{equation}
The meaning of this equation is that the waveform model fits the noise by adjusting its parameters, expressed by the derivative. 
The inverse of the Fisher matrix provides the variance-covariance of the statistical uncertainty $\Delta \theta^i$: 
\begin{equation}
    \bigl \langle \Delta \theta^i \Delta \theta^j \bigr \rangle_n = ( \Gamma^{-1})^{ij}+\mathcal{O}(\mathrm{SNR}^{-1})\,,
    \label{eq:varianceF}
\end{equation}
where $\langle \cdot \rangle_n$ indicates averaging  over realizations of the noise. 
In other words, the Fisher matrix provides a practical first-order estimate of how the maximum-likelihood estimate is distributed and how well the parameters of a given GW signal can be measured.

In addition to statistical uncertainty due to the noise, there might be additional systematic errors due to inaccuracies in the GW model.
To describe the impact of systematics in the high SNR limit, we follow Ref.~\cite{Cutler:2007mi} and introduce two manifolds in the parameter space: one for the true waveform $\mathbf{h}_{\mathrm{TR}}(\theta^i)$, and one for the approximate waveform  $\mathbf{h}_{\mathrm{AP}}(\theta^i)$. 
These manifolds are parametrized by the same coordinates (the masses, spins of the binary system, etc). The GW data contain the true waveform and the detector noise $\mathbf{n}$:
\begin{equation}
    \mathbf{s} = \mathbf{h}_{\mathrm{TR}}(\theta_{\mathrm{TR}}) +\mathbf{n}\,.
\end{equation}
Our best (i.e., highest likelihood) estimate of the source properties is given when the approximate waveform is the closest to the data $\mathbf{s}$. 
Hence, to find the maximum-likelihood parameters $\theta^i_{\mathrm{ML}}$, we draw the normal from $\mathbf{s}$ to the manifold of the approximate waveform $\mathbf{h}_{\mathrm{AP}}(\theta^i)$:
\begin{equation}
    \langle  \partial_j \mathbf{h}_{\mathrm{AP}}(\theta^i_{\mathrm{ML}}) | \mathbf{s} - \mathbf{h}_{\mathrm{AP}}(\theta^i_{\mathrm{ML}})  \rangle = 0\,.
\end{equation}
From the equation above, it is possible to obtain the difference between the maximum-likelihood and the true parameters $\Delta \theta^i$~ \cite{Cutler:2007mi}. 
At leading order, it
is the sum of a statistical contribution from the Gaussian noise $\Delta_n \theta^i$ and a contribution from the waveform systematics $\Delta_{\mathrm{th}} \theta^i$:
\begin{equation}
\begin{split}
& \Delta \theta^i  = \Delta_n \theta^i + \Delta_{\mathrm{th}} \theta^i \,, \\ 
& \Delta_n \theta^i = (\Gamma^{-1} ( \theta_{\mathrm{ML}}))^{ij}  \langle \partial_j \mathbf{h}_{\mathrm{AP}}( \theta_{\mathrm{ML}})|\mathbf{n} \rangle \,,\\
& \Delta_{\mathrm{th}} \theta^i = (\Gamma^{-1} ( \theta_{\mathrm{ML}}))^{ij} \langle \partial_j \mathbf{h}_{\mathrm{AP}}(\theta_{\mathrm{ML}})| \mathbf{h}_{\mathrm{TR}}(\theta_{\mathrm{\mathrm{ML}}})- \mathbf{h}_{\mathrm{AP}}(\theta_{\mathrm{\mathrm{ML}}}) \rangle\,.
\end{split}
\label{eq:error_sum}
\end{equation}
The statistical uncertainty $\Delta_n \theta^i $ decreases for louder signals as SNR$^{-1}$, and is independent from the systematic error which does not depend on the SNR.

These results describe the impact of Gaussian noise and waveform systematics on inference in the limit of high SNR. 
When the waveform has several parameters and relatively low SNR, as GW231123, instead simulations are required. 

\subsection{Simulating a GW231123-like signal}\label{subsec:inj}

\begin{figure*}
        \centering 
        \includegraphics[width=0.44\textwidth]{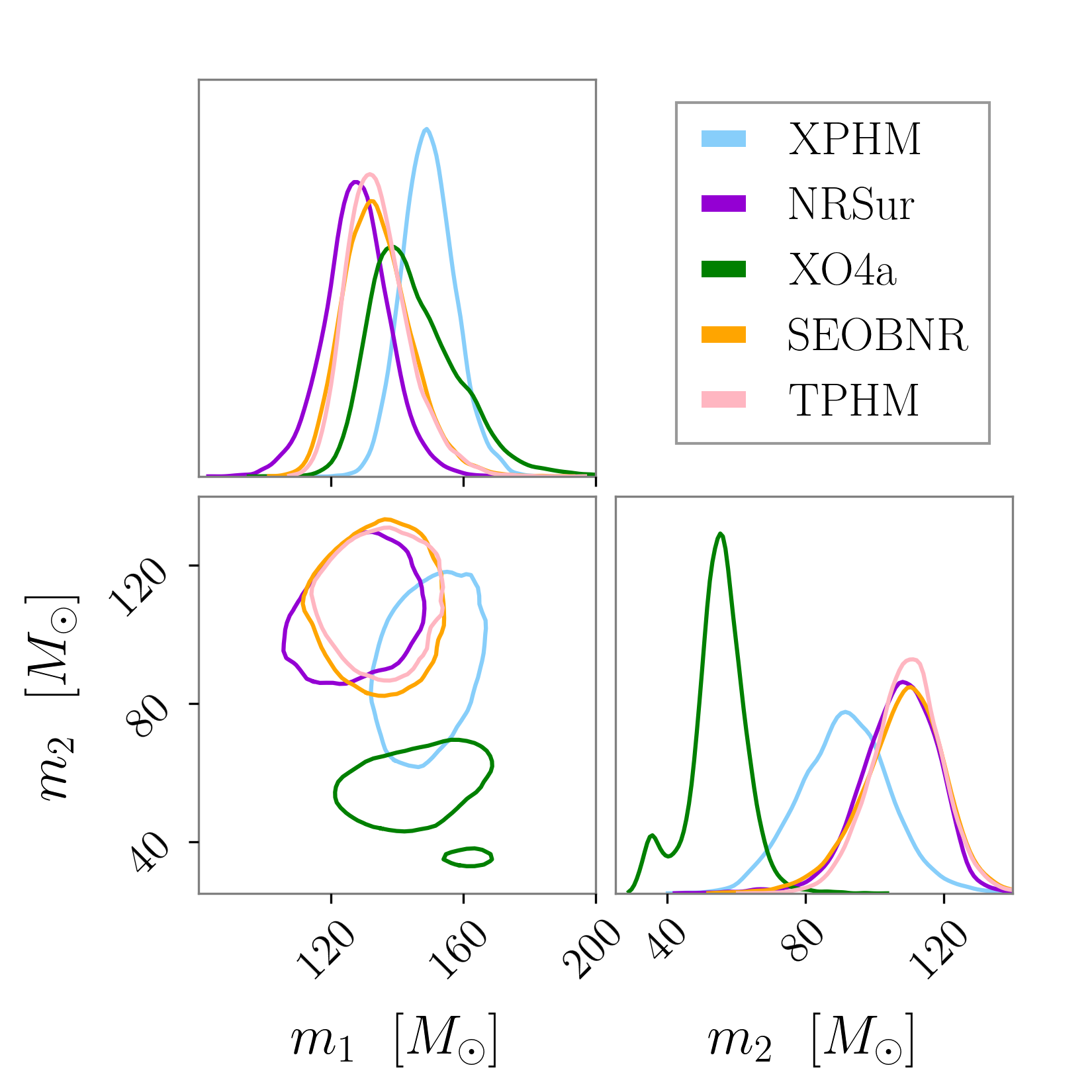}
        \includegraphics[width=0.54\textwidth]{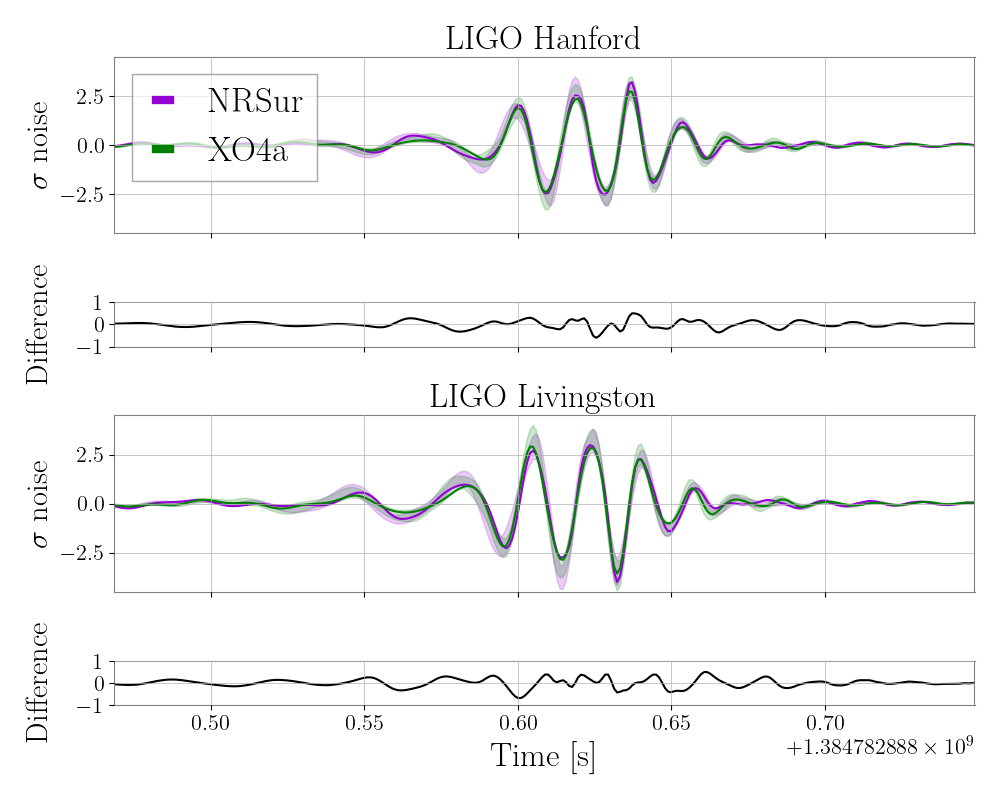}
        
        \caption{(\textit{Left})  Marginalized posteriors for the source-frame component masses of GW231123 inferred with five waveform models \cite{GW231123,zenodo_v2}.
        (\textit{Right}) Whitened waveforms for GW231123 according to NRSur (violet), and XO4a (green) in LIGO Hanford (first panel) and LIGO Livingston (third panel). The solid lines indicate the maximum-likelihood waveforms, while the bands shows the 90\% credible intervals. The difference between the maximum-likelihood waveforms are reported in the second and fourth panels. There are strong systematics between NRSur and XO4a in the component masses, however the maximum-likelihood waveforms are similar; their difference remains less than half a standard deviation with respect to the noise. 
        The waveform 90\% credible intervals also fully overlap.
        We use the NRSur maximum-likelihood waveform to simulate GW231123-like signals.
        } 
        \label{fig:maxL_3models} 
\end{figure*}

To investigate waveform systematics and the impact of Gaussian noise, we need to simulate a signal that approximates GW231123. 
Ref.~\cite{GW231123} uses high-mass and high-spin BBH numerical waveforms.  
However, the number of such simulations is limited and it is unlikely that the parameters of these simulations correspond closely to GW231123.

Here, we instead use maximum-likelihood waveforms from the actual GW231123 analyses. 
The maximum-likelihood parameters, under the assumed waveform model, by definition generate the signal that maximizes the likelihood, i.e., minimizes the residual and the noise. 
Their advantage is that (by definition) they allow for simulations with signal morphology the closest to the data possible, and closer than any of the NR waveforms available. 
We consider the maximum-likelihood waveforms from the NRSur and XO4a analyses: the former is the model with the lowest mismatch against high-mass and high-spin simulations~\cite{GW231123}, while the latter is the model that yields posteriors that are the most discrepant among the five models used.

Despite the large differences in the maximum-likelihood parameters, the maximum-likelihood \emph{waveforms} are very similar.
The left panel of Fig.~\ref{fig:maxL_3models} shows the GW231123 posterior distributions for the source-frame component masses with all models~\cite{GW231123}. 
The systematics are evident; in particular, the 2-dimensional 90\% credible levels for NRSur and XO4a do not overlap at all. 
The right panel of Fig.~\ref{fig:maxL_3models} shows that the NRSur and XO4a maximum-likelihood waveforms are still very similar in both detectors. 
The difference, obtained by subtracting the XO4a waveform from NRSur, is distributed evenly across the entire signal.
The two models also yield similar maximum-likelihood values (218.1 and 218.6 for NRSur and XO4a, respectively, with arbitrary normalization), meaning that they fit the data comparably well (the goodness-of-fit will be further discussed in Sec.~\ref{sec:Gaussian_noise_wf_system}).
Thus, even if the source properties inferred with NRSur and XO4a strongly differ, there is little disagreement on the morphology of the signal.

For the analyses presented in the main body, we simulate GW231123-like signals using the NRSur maximum-likelihood waveform.  
In App.~\ref{app:wf_syst}, we repeat the analysis using the XO4a maximum-likelihood waveform instead. 
The maximum-likelihood NRSur parameters are listed in Table~\ref{tab:maxL}: the simulated system is composed of two massive BHs with similar masses ($129\,M_\odot$ and 117\,$M_\odot$) with extremely high component spin magnitudes (0.96 and 0.98), misaligned with respect to the orbital angular momenta ($\chi_{\mathrm{eff}}= -0.02$, $\chi_{\mathrm{p}}= 0.96$). 
See Ref.~\cite{gwtc4-intro} for a complete review of the parameters that describe a compact binary coalescence.

\begin{table}
\begin{tabular}{l l}
\hline
Parameter & Value \\
\hline
Source-frame primary mass,  $m_1$/$M_\odot$ & $129$ \\
Source-frame secondary mass, $m_2$/$M_\odot$ & $117$ \\
Source-frame total mass,  M/$M_\odot$ & $246$ \\
Mass ratio, $q=m_1/m_2$& $0.91$ \\
Primary spin magnitude, $\chi_1$ & $0.96$ \\
Secondary spin magnitude, $\chi_2$ & $0.98$ \\
Effective inspiral spin, $\chi_{\rm{eff}}$ & $-0.02$ \\ 
Effective precessing spin, $\chi_{\rm{p}}$ & $0.96$ \\
Inclination angle, $\theta_{\rm{JN}}$/rad & $1.4$ \\
Luminosity distance, $D_{\rm{L}}$/Mpc & $1121$ \\
\hline
\end{tabular}
\caption{GW231123 maximum-likelihood parameters inferred with the waveform model NRSur and used to simulate a GW231123-like signal.}\label{tab:maxL}
\end{table}

\subsection{Estimation of source properties}
\label{subsec:PE}

A number of waveform models are employed to assess the impact of systematics on GW inference.
The main GWTC-4 results were obtained with XPHM and SEOBNR, while further models were used for high-mass systems (NRSur) and for systems with asymmetric masses or evidence of precession (XO4a)~\cite{GWTC4method}. 
A fifth model was used exclusively for GW231123 (TPHM). 
For most events, the source properties inferred with different models are consistent~\cite{gwtc4results}, with exceptions discussed in Sec.~3.6 of Ref.~\cite{gwtc4results}. 
Among them, the largest systematics are observed for GW231123.
 
We employ three waveform models to infer the properties of the simulated signals: NRSur, XPHM, XO4a. 
All models describe precessing, quasi-circular binaries including higher multipole content. 
NRSur interpolates between numerical-relativity simulations in the time-domain, while XPHM and XO4a use a combination of analytical and numerical information in the frequency-domain.  
Further details are reported in Ref.~\cite{GW231123} and references therein. 
We downselect to these three models because NRSur has the lowest mismatches against numerical waveforms~\cite{GW231123}, and XPHM and XO4a result in posteriors with the largest difference compared to NRSur.
We omit SEOBNR and TPHM from now on, as they both give similar results to NRSur for GW231123.

We simulate data using both different noise relations and a zero-noise realization, i.e., assuming that the noise is exactly zero 
at all frequencies, while the PSD is finite.
Other than Sec.~\ref{sec:future} which focuses on future prospects, we employ the median PSD estimated by BayesWave \cite{Littenberg:2014oda} from data around GW231123 (6\,s before and 2\,s after the merger time)~\cite{GW231123}. 
Zero-noise simulations are often used when comparing to the real-noise analysis~\cite{Pankow:2018qpo}, as they correspond to an average over noise realizations~\cite{Nissanke:2009kt}. 
To investigate the impact of Gaussian noise, we simulate the same signal in 20 random Gaussian noise realizations.

All analyses are listed in Table \ref{tab:list_pe_runs}.
Inference is performed with BILBY \cite{Ashton:2018jfp, Romero-Shaw:2020owr} using the same settings and prior distributions as GWTC-4~\cite{GWTC4method}. 
Since the initial announcement and first data release of GW231123~\cite{zenodo_v2}, results have been updated following a change to the likelihood normalization in BILBY \cite{Talbot:2025vth}, making them consistent with GWTC-4 \cite{gwtc4results}. 
As a result, the reported SNR is lowered from 22.6 (median value) to 20.7, but all parameter posteriors are minimally affected and the key results and conclusions are unchanged. 
Below, all results referring to GW231123 or zero-noise simulations are obtained with the updated BILBY likelihood.
However, since the Gaussian noise simulations were performed before the BILBY update release and the differences are minimal, we present their
pre-update version.  
We use a uniform prior over the detector-frame component masses, uniform on the spin magnitudes and isotropic in spin orientations, isotropic priors on the binary orientation, and uniform in comoving volume and comoving time so that the priors on the sky location are isotropic. 
The NRSur model further restricts the mass ratio to $m_1/m_2 > 1/6$, while the other models allow for $m_1/m_2>1/10$ \cite{GW231123}. 

\begin{table}
\centering
\begin{tabular}{l p{3cm} p{2.2cm} l}
\hline
 Detector(s) & Noise Realization & Model &  Section \\ \hline
    H, L          & zero & NRSur, XPHM, XO4a    &  \ref{sec:wf_syst} \\ \hline
    H, L        & 20 Gaussian & NRSur, XPHM, XO4a    &  \ref{sec:impact}, \ref{sec:Gaussian_noise_wf_system} \\ \hline

    H     & 20 Gaussian &  NRSur    & \ref{sec:HvsL}  \\ \hline
    L    & 20 Gaussian &   NRSur   & \ref{sec:HvsL} \\ \hline

   LIGO A\#  & zero &    NRSur     & \ref{sec:future} \\ \hline
  
\end{tabular}
\caption{List of all analyses presented below. We report the detector(s) used, whether the simulation is performed in zero-noise or in multiple Gaussian noise realizations, the waveform approximant used for the model, and the section where results are discussed.}\label{tab:list_pe_runs}
\end{table}

\subsection{Comparing posterior distributions}\label{sec:JS}

We quantitatively compare posteriors with the Jensen--Shannon (JS) divergence \cite{Lin:1991zzm}. 
The JS divergence between distributions $Q_1$ and $Q_2$, is defined as
\begin{equation*}
    {\rm JS}(Q_1 || Q_2) = \frac{1}{2} {\rm D_{KL}}(Q_1 || M) + \frac{1}{2} {\rm D_{KL}}(Q_2 || M)\,,
\end{equation*}
where $M = \frac{1}{2}(Q_1+Q_2)$, and ${\rm D_{KL}}(Q_1 || Q_2)$ is the Kullback--Leibler divergence defined as
\begin{equation*}
    {\rm D_{KL}}(Q_1||Q_2) = \sum_{x\in X} Q_1(x) \log_2 \frac{Q_1(x)}{Q_2(x)}\,.
\end{equation*}
The JS divergence ranges within $[0,1]$, with ${\rm JS}=0$ indicating identical distributions. 
There is no unique JS threshold below which two distributions would be considered to be in good agreement, and the interpretation of JS values varies from application to application. 
For this reason, in Sec.~\ref{sec:Gaussian_noise_wf_system} and~\ref{sec:HvsL}, we only compare JSs between different analyses rather than adopt a threshold.

\section{Waveform systematics in GW231123}
\label{sec:wf_syst}

The inferred source properties of GW231123 are systematically different across the various waveform models, with posteriors for multiple parameters not overlapping within the 90\% credible level \cite{GW231123}.
While the most reliable way to assess the accuracy of a model is by comparing to numerical relativity, a complementary way is to try to reproduce the systematics observed in the data with simulated signals. 
If similar levels of systematics as for the event of interest are found in simulated signals (typically in the absence of noise), then in that region of the parameter space systematics can be attributed to internal modeling differences. 
Otherwise, when the waveform systematics cannot be reproduced, i.e., the various models are in agreement or the systematics are qualitatively different, then this raises the possibility of additional analysis systematics, for example in the noise model or calibration. 

Tens of simulations were carried out in preparation of Ref.~\cite{GW231123} to explore systematics by simulating signals via numerical waveforms. 
For the majority of the high-mass, highly-precessing simulations, Ref.~\cite{GW231123} reported that it was not possible to reproduce the degree of systematics observed in GW231123.
Two examples are presented in detail. 
In the first case, there are no systematics and the five models correctly infer the primary mass and slightly underestimate the secondary mass. 
In the second case, the secondary mass is significantly underestimated for all models leading to a measured mass ratio $\lesssim 0.3$ instead of the correct value of $1$.
But there is still no sign of systematics in mass inference among the five models as the 90\% credible levels essentially overlap.

\begin{figure*}
        \centering 
        \includegraphics[width=0.49\textwidth]{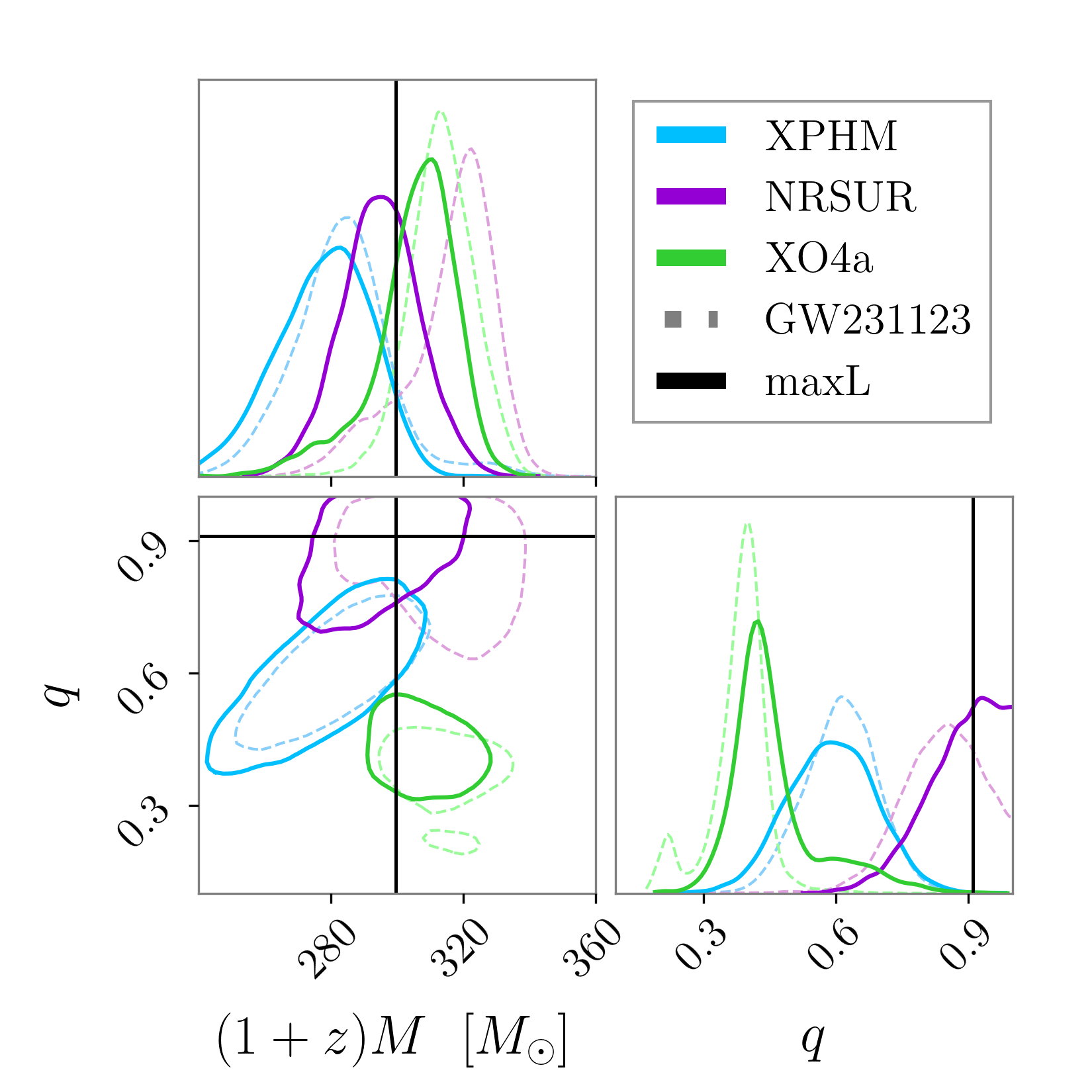}
        \includegraphics[width=0.49\textwidth]{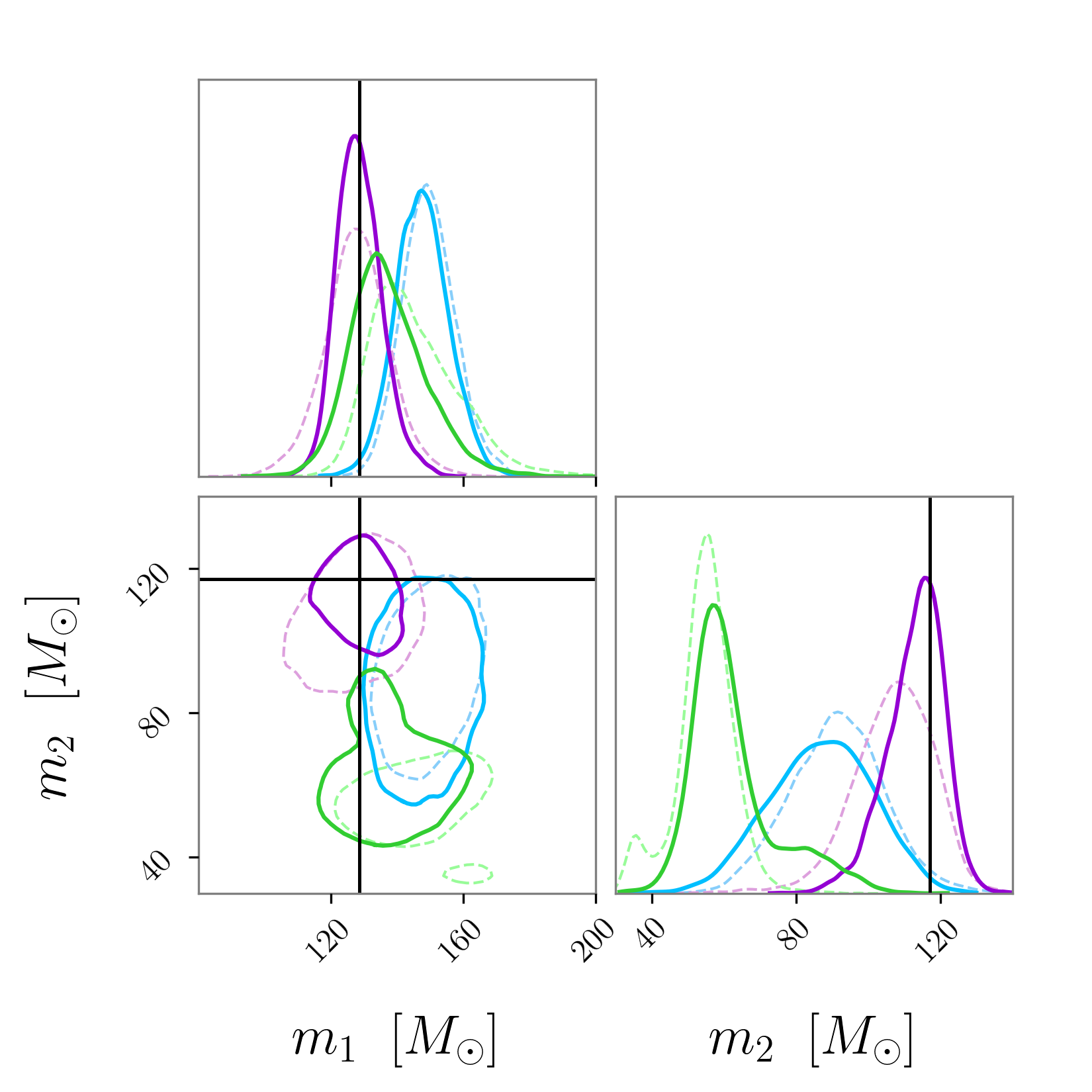}
        \includegraphics[width=0.32\textwidth]{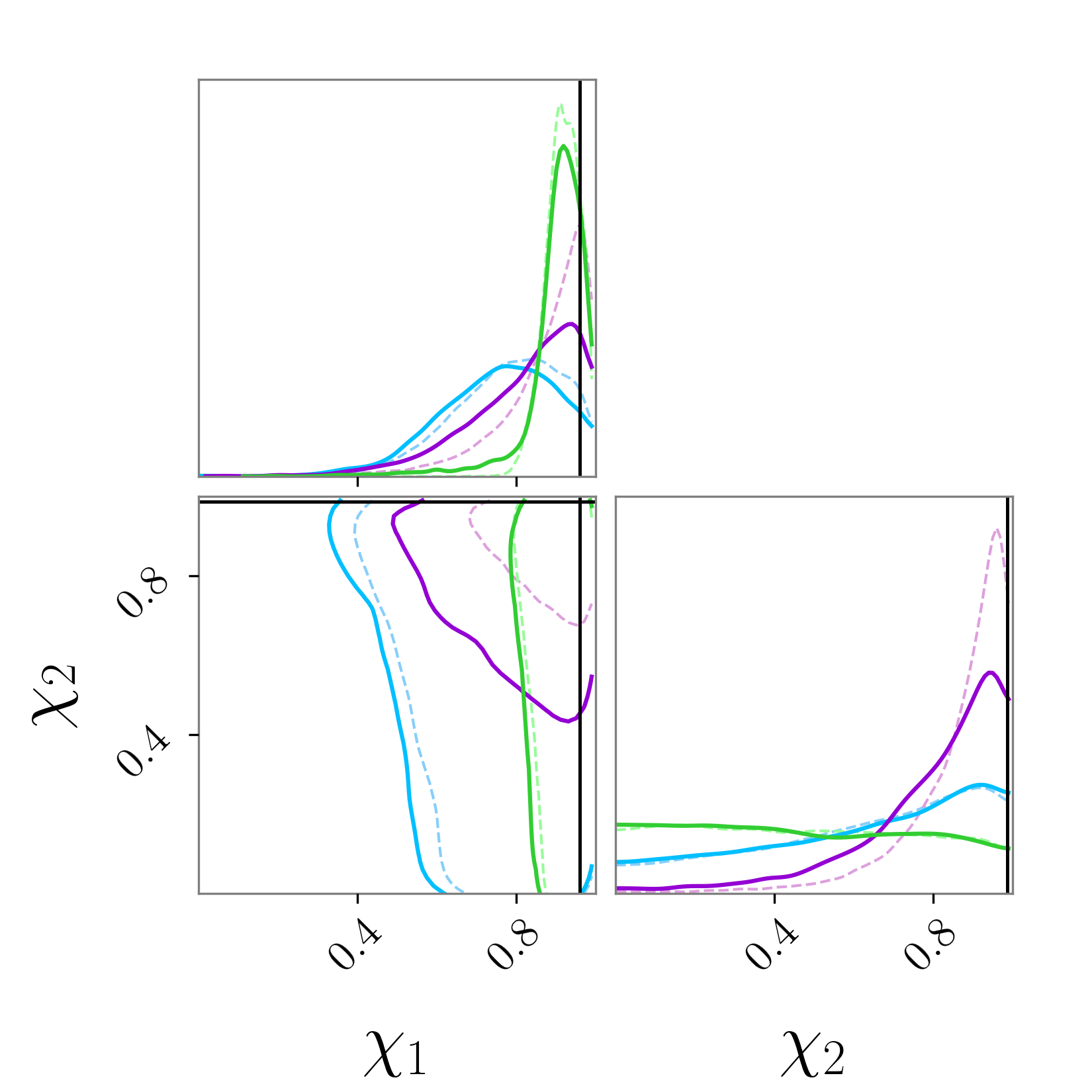}
        \includegraphics[width=0.32\textwidth]{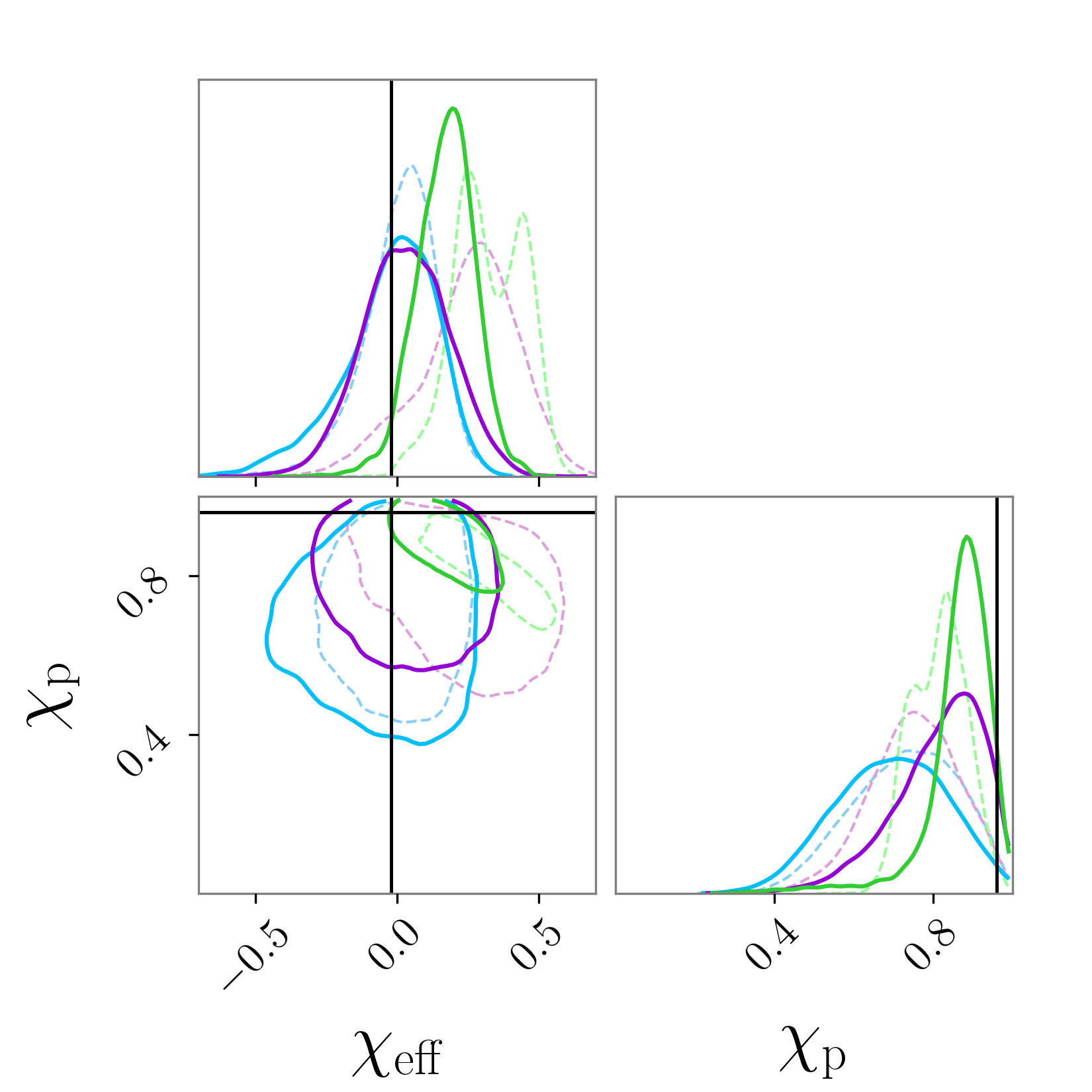}
        \includegraphics[width=0.32\textwidth]{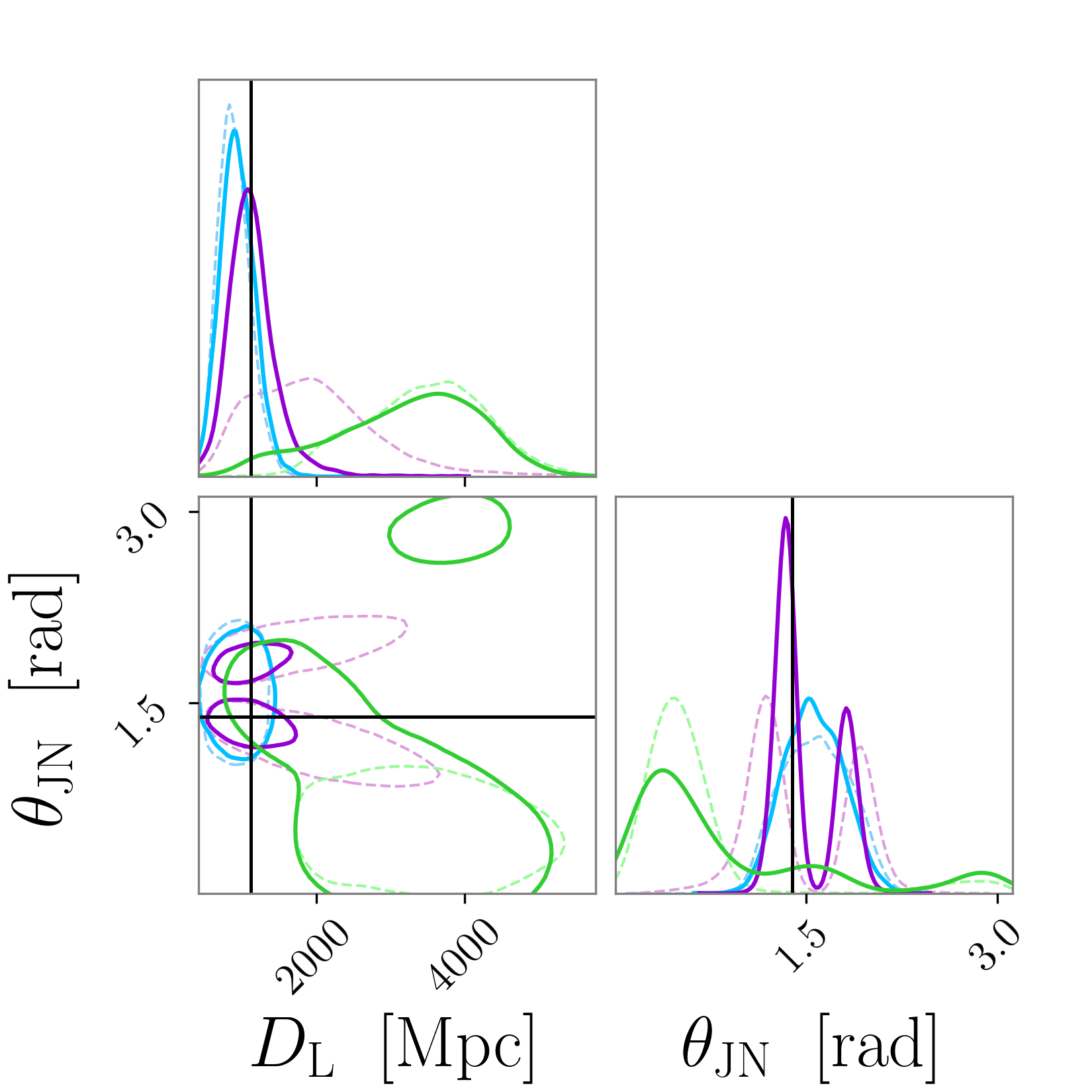}
        \caption{ 
        Marginalized posteriors for selected parameters for GW231123 (dashed lines, from Ref.~\cite{GW231123,zenodo_v2}) and from the NRSur maximum-likelihood (maxL) simulations in zero-noise (solid lines) for three waveform models. The black lines indicate the true value for the simulation. Contours show the 90\% credible levels. From top left: detector-frame total binary mass and mass ratio, primary and secondary source-frame masses, spin magnitudes, $\chi_{\mathrm{eff}}$ and $\chi_{\mathrm{p}}$, luminosity distance and inclination.
        We are able to reproduce the degree of waveform systematics observed in GW231123 with a simulated signal in absence of noise.
        } 
        \label{fig:wf_syst} 
\end{figure*}

Here, we repeat this analysis but instead of numerical simulations, we use a waveform that is more similar to the GW231123 data: the maximum-likelihood waveform from the NRSur analysis.
We then compare the waveform systematics observed in GW231123, with the systematics observed in this zero-noise simulation with NRSur, XPHM and XO4a. 
Figure \ref{fig:wf_syst} compares inference for selected parameters for GW231123 (dotted lines) and the simulation (solid lines). 
Starting with the masses, for GW231123, there are evident systematics in both total mass and mass ratio, with almost no overlap in the joint 90\% contour~\cite{GW231123}. 
The systematics in the maximum-likelihood simulation are qualitatively similar to the ones observed for GW231123, again with limited overlap in the two-dimensional 90\% contours. 
The secondary mass posterior with XO4a exhibits a long tail towards higher masses that relaxes systematics with respect to the other models, but the main peaks support a lower value, similarly to GW231123. 
NRSur posteriors peak at the true value, while XPHM and especially XO4a underestimate the secondary component mass.

The degree of systematics is lower for spin parameters, but can still be qualitatively reproduced by the simulation.
All models find a high primary spin magnitude ($\chi_1 \sim 0.9$ for NRSur and XO4a, $\chi_1 \sim 0.8$ for XPHM). 
NRSur finds support for high spin magnitude for the secondary BH as well ($\chi_2 \sim 0.9$), while the XPHM and especially XO4a results are not informative, i.e., broadly consistent with the prior. 
Concerning $\chi_{\mathrm{eff}}$ and $\chi_{\mathrm{p}}$, NRSur infers GW231123 to have an aligned spin consistent with zero ($\chi_{\mathrm{eff}}=0.27_{-0.35}^{+0.24}$) and high precession ($\chi_{\mathrm{p}} = 0.76^{+0.19}_{-0.17}$), XPHM also infers null aligned spin and precession ($\chi_{\mathrm{p}} = 0.74^{+0.21}_{-0.21}$), while XO4a results in a bimodal posterior peaking at ${\sim}0.25$ and ${\sim}0.45$ for $\chi_{\mathrm{eff}}$ and bimodal posterior for $\chi_{\mathrm{p}}$ peaking at ${\sim} 0.75$ and ${\sim} 0.85$.
Results on the simulation are similar with NRSur and XPHM accurately finding a $\chi_{\mathrm{eff}}$ consistent with null, while XO4a is biased towards positive values. 
Similarly, in the simulation NRSur and XO4a find strong spin precession  ($\chi_{\mathrm{p}}=0.83_{-0.14}^{+0.16}$, and $\chi_{\mathrm{p}}=0.88_{-0.35}^{+0.06}$, respectively), while XPHM obtains a lower value ($\chi_{\mathrm{p}}=0.70_{-0.20}^{+0.25}$).
The quoted ranges correspond to 90\% high posterior density (HPD) intervals. 

Finally, similar results are obtained for the luminosity distance and the inclination.
For GW231123 there are significant systematics: NRSur supports a BBH at a closer distance ($1.9^{+1.7}_{-1.0}\,$Gpc) with high inclination, but has a long tail towards larger distances. 
XPHM infers a closer distance ($0.9^{+0.4}_{-0.3}\,$Gpc) for an edge-on system, while XO4a supports a larger distance ($3.5^{+1.2}_{-1.4}\,$Gpc) with face-off orientation. 
The differences in the localization of the source reflect differences in the measurement of the higher-order modes: XPHM retrieves high SNR in modes higher than the fundamental (SNR $\sim5.0$ in the (3,3) mode and ${\sim}$3.3 in the (4,4) mode), while NRSur and XO4a find lower SNR in the higher order modes (SNR $\sim2.3$ in (3,3) and ${\sim}2.9$ in (4,4) for NRSur, and ${\sim}4.5$ in (3,3) and ${\sim}1.9$ in (4,4) for XO4a). 
The systematics are similar for the simulation, though with respect to GW231123 NRSur finds a better constrained distance and XO4a has a long tail at lower distances. 
The true value is in the 90\% credible interval for all the three models. 
As for GW231123, XPHM finds high SNR in the high order modes (SNR $\sim4.9$ in (3,3) and SNR $\sim3.1$ in (4,4)), while NRSur and XO4a find lower SNR in the higher order modes. 

We repeat the same analysis simulating a signal with the XO4a maximum-likelihood waveform and show the results in Appendix \ref{app:wf_syst}. 
As expected from Fig.~\ref{fig:maxL_3models}, we again find large systematics between NRSur, XO4a, and XPHM similarly to the systematics observed in GW231123.

This comparison does not inform on which model is more accurate: this question is best answered with mismatches between each model and numerical simulations \cite{cutler1994gravitational}. 
Indeed, the mismatch study on high-mass and high-spin simulations reported in Ref.~\cite{GW231123} finds that NRSur is on average the most accurate model. 
Here, we perform a consistency test: we are able to reproduce the systematics observed in GW231123 assuming the signal is given by the best-fit NRSur waveform to the data. 
Combining the results of the mismatch and the systematics studies suggests that GW231123 is consistent with a signal described most accurately by NRSur and its inferred source properties are the most robust.
  

\section{Impact of Gaussian noise on mass and spin measurements}
\label{sec:impact}

In this section, we investigate the impact of different Gaussian noise realizations on inference by simulating the same signal based on the NRSur maximum-likelihood waveform, and adding it to 20 different Gaussian noise realizations generated from the GW231123 PSD. 
For each realization, we analyze the data with NRSur, XPHM and XO4a. 
Our goal is to assess how much the posterior distributions fluctuate, and especially the impact on ``astrophysically significant"  
results, such as the probability that the masses are in the mass gap and whether the posteriors rule out zero spin.
While we do \emph{not} expect the true value to always lie within some arbitrary significance level (typically 90\% is quoted), this study allows us to assess whether quoting specific results from a single event is meaningful~\cite{Biscoveanu:2021nvg,Xu:2022zza}.

  \begin{figure*}
        \includegraphics[width=0.95\textwidth]{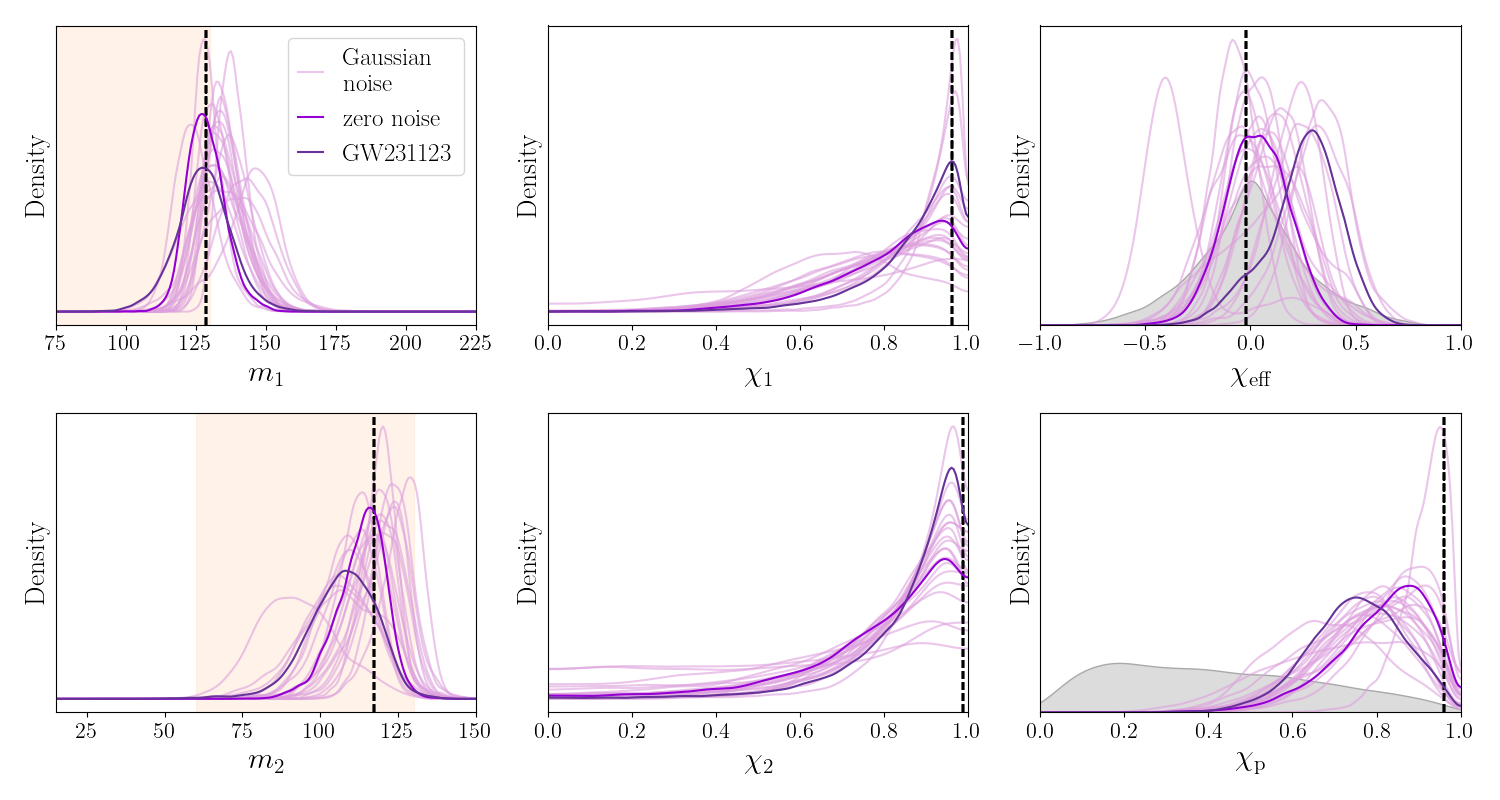} 
        \caption{Effect of Gaussian noise on the primary and secondary masses (left column), on the spin magnitudes (central column) and on the effective inspiral and effective precessing spin (right column). The light purple lines shows the posteriors when simulating data with the NRSur maximum-likelihood waveform in 20 Gaussian noise realizations, the purple line indicates the case of zero-noise, and the dark purple line corresponds to the GW231123 result.
        The black dotted line indicates the true value, the orange shaded area the pair-instability mass gap ($\sim$ 60 --
        130$\,M_\odot$), and the gray areas are the prior distributions. The waveform model employed to simulate and recover the signal is NRSur.
        }
        \label{fig:Gaussian} 
    \end{figure*}

We start with the NRSur recovery, shown in Fig. \ref{fig:Gaussian}.
The component masses are well recovered: for the primary mass the true value is always in the 90\% credible interval; the same is true for the secondary mass in all cases except for one, for which the mass is underestimated (92$_{-19}^{+24}\,M_\odot$ instead of 117$\,M_\odot$).  
The primary mass always lies at the superior edge of the mass gap (60--130$M_\odot$), while the secondary mass is mostly within the gap, or slightly above. 
The true primary spin magnitude is extremely high ($\chi_1 = 0.96$) and it is always included in the 90\% posterior HPD interval.
In most cases (16 over 20), the $\chi_1$ median is above 0.8, and never below 0.7. 
The true value of $\chi_2$ is always in the 90\% interval, with most of the posteriors (17 over 20) supporting spin magnitudes $\geq0.8$. 
Concerning the effective inspiral spin, the true value is contained in the 90\% credible interval for 16 over 20 noise realizations, but the posteriors vary significantly: from $\chi_{\mathrm{eff}}=-0.38_{-0.18}^{+0.60}$ to $\chi_{\mathrm{eff}}=+0.34_{-22}^{+22}$. The effective precession spin posteriors 14 times over 20 include the extreme true value in the 90\% HPD interval, and in half of the cases the $\chi_\mathrm{p}$ median value is $\geq0.8$ and never below 0.68.
The above variation shows what is expected under random noise fluctuations.

\begin{figure*}
        \includegraphics[width=0.95\textwidth]{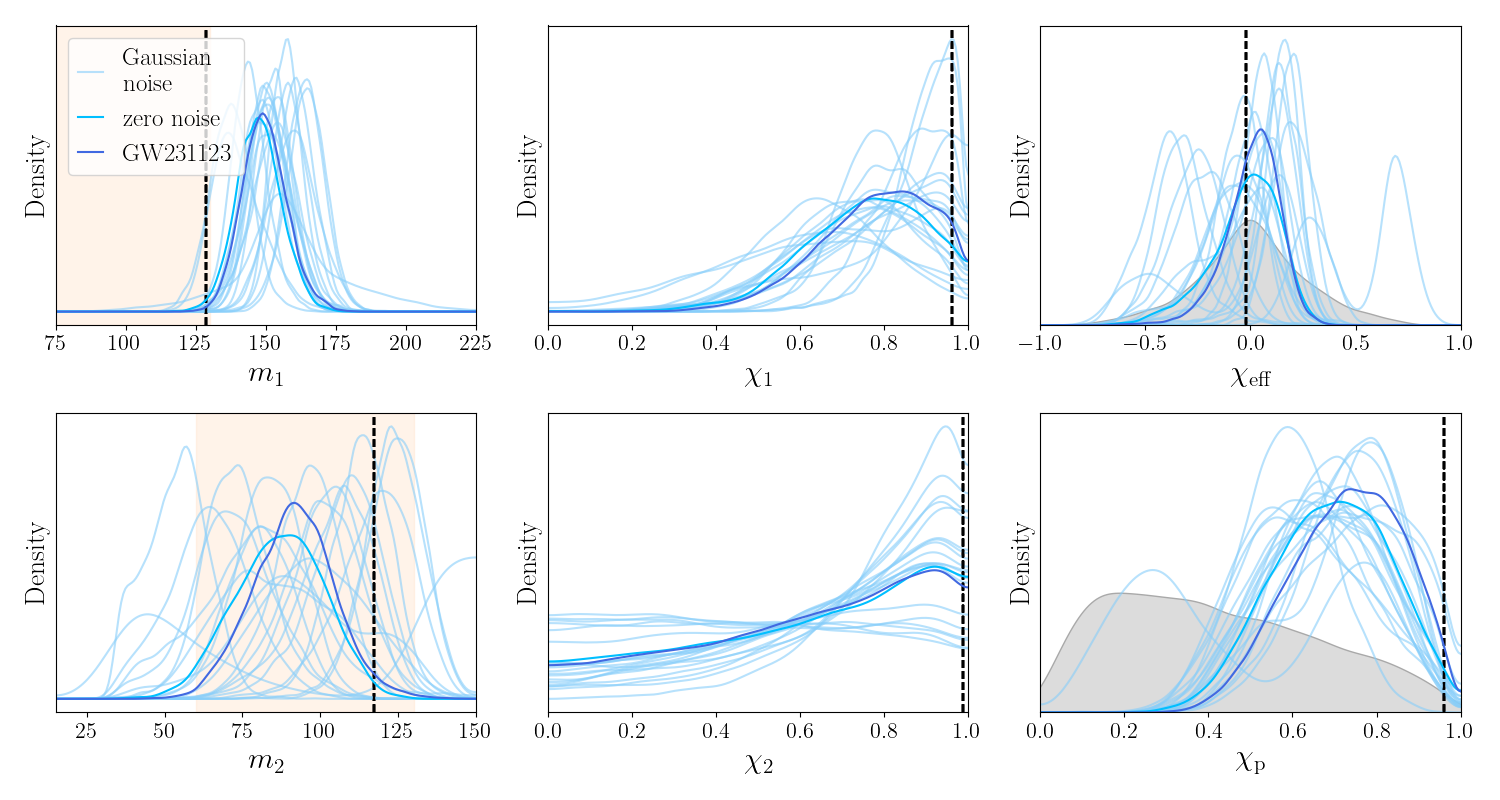} 
        \caption{Same as Fig. \ref{fig:Gaussian} using the XPHM waveform model to infer the source properties.} 
        \label{fig:Gaussian_XPHM} 
\end{figure*}

We repeat the same analysis using the XPHM model to recover the source properties of the same NRSur maximum-likelihood simulation in Gaussian noise.  
Figure \ref{fig:Gaussian_XPHM} shows that the primary component mass is often overestimated, and for 17 noise realizations over 20 the true value is not included in the 90\% credible interval. 
The posteriors of the secondary mass shift remarkably, oscillating from $54_{-17}^{+15}\,M_\odot$ to $124_{-17}^{+14}\,M_\odot$ and in 9 cases the true value is not included in the 90\%  interval. 
The primary mass is mostly above the mass gap, while the secondary mass fluctuates below, within, and above the gap.
The $\chi_1$ posteriors include in 17 cases over 20 the extreme true value, and the median values are always above 0.6. 
Concerning $\chi_2$, the true value is always in the 90\% HPD interval, in 14 cases there is support for high spins magnitude (median $\geq0.7$), while in 6 cases the posteriors are close to the priors, so they are not informative.
The measurements of $\chi_{\mathrm{eff}}$ oscillate significantly between $\chi_{\mathrm{eff}}=-0.39_{-22}^{+21}$ and $\chi_{\mathrm{eff}}=0.60_{-0.42}^{+0.19}$, and in 7 cases over 20, the true value is not included in the 90\% interval. 
The posteriors for $\chi_\mathrm{p}$ range from $0.46_{-0.33}^{+0.40}$ to $0.76_{-0.26}^{+0.19}$ and include the true value in the 90\% HPD interval only in two cases. 
These large variations are now the combination of \emph{both} random noise and waveform systematics.

\begin{figure*}
        \centering 
        \includegraphics[width=0.95\textwidth]{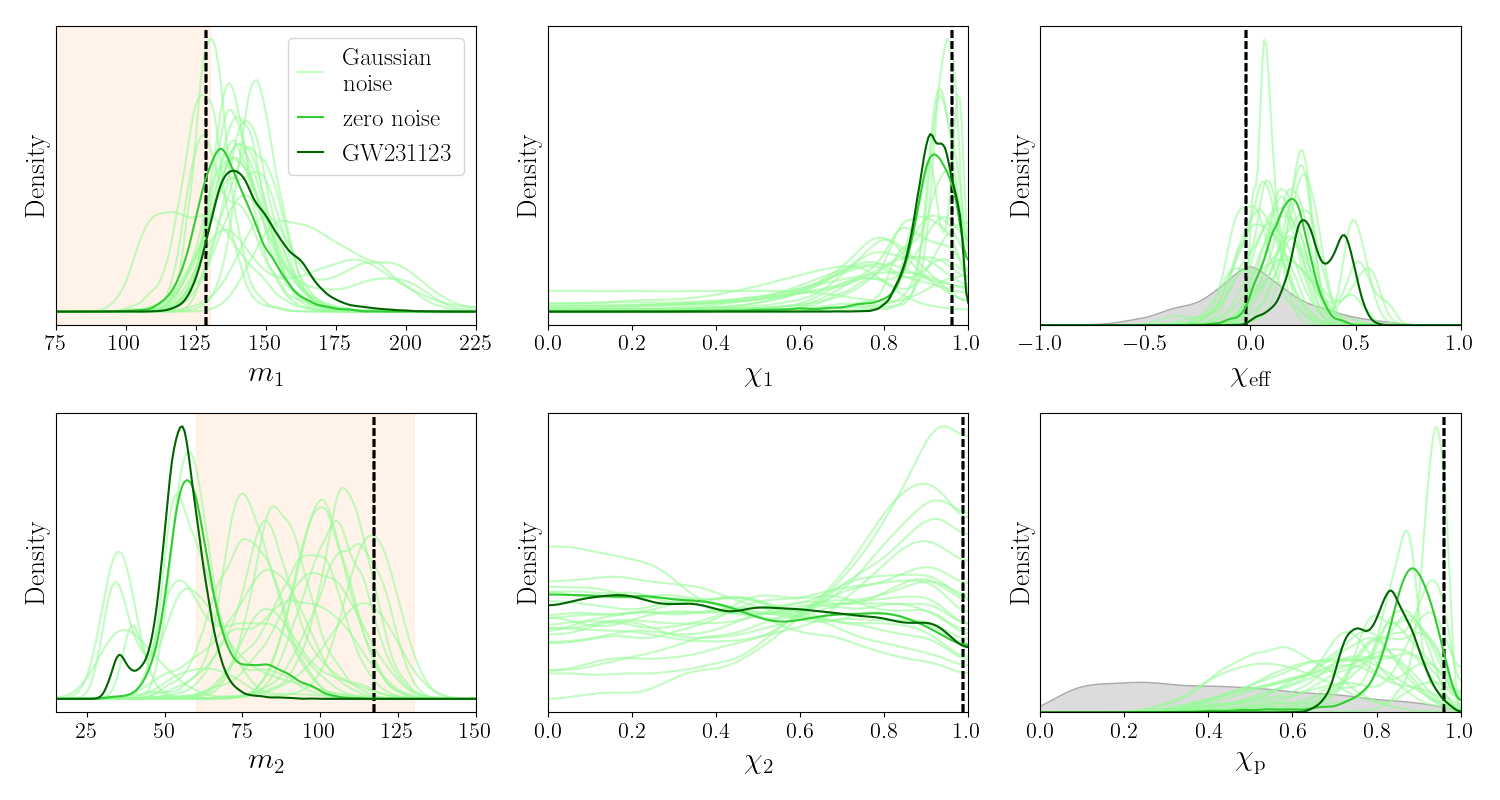} 
        \caption{Same as Fig. \ref{fig:Gaussian} using the XO4a waveform model to infer the source properties.} 
        \label{fig:Gaussian_XO4a} 
\end{figure*}

Finally, we use the XO4a model to recover the same simulation in Fig. \ref{fig:Gaussian_XO4a}. 
The primary component mass is retrieved in most of the cases, except for three instances for which the true value is not included in the 90\% interval. 
The posterior distributions of the secondary component mass shift considerably towards lower values (ranging from $56_{-24}^{+34}\,M_\odot$ to $114_{-58}^{+14}\,M_\odot$) and in 14 cases out of 20 the true value is not included in the 90\% interval. 
The primary mass is mostly above the mass gap, while the secondary mass fluctuates below and within the gap.
The spin magnitude vary between $0.48_{-0.44}^{+0.49}$ to $0.94_{-0.07}^{+0.04}$ for $\chi_1$.
In 19 cases over 20, the true $\chi_1$ value is in the 90\% HPD interval, while for $\chi_2$ the same is true in 5 over 20 cases.
Again, we often recover an uninformative posterior and, in some cases, support for high values. 
For $\chi_{\mathrm{eff}}$ the posteriors fluctuate between $0.00_{-0.16}^{+0.20}$ and $0.30_{-0.22}^{+0.26}$ and the true value is not included in the 90\% interval in 11 out of 20 cases.  
For $\chi_{\mathrm{p}}$ posteriors range from $0.56_{-0.20}^{+0.20}$ to $0.94_{-0.07}^{+0.04}$, always supporting spin precession, and the true value is included in the HPD interval only in 12 cases over 20.
Again, these variations combine the effects of random noise and systematics.

Overall, the impact of Gaussian noise on the key mass and spin GW231123 results with NRSur is minor: the major features (component masses within or above the mass gap and high spin magnitudes), are consistently inferred.
Due to a combination of waveform systematics and Gaussian noise, XO4a and XPHM tend to underestimate the secondary component mass across noise realizations. 
The total mass is known to be biased towards lower values for edge-on configurations \cite{GW231123}:
GW signals from face-on and face-off binary systems are louder, which leads to larger prior preference for smaller inclination angles, and so for larger distances and lower masses. 
Both XPHM and XO4a consistently find support for high spin magnitudes for the primary spin, while the measurement is often not informative for the secondary.
The impact of Gaussian noise is more pronounced for $\chi_\mathrm{eff}$ and $\chi_\mathrm{p}$.
NRSur and XO4a accurately infer the extreme $\chi_\mathrm{p}$ true value in 4 cases over 20 while XPHM never does so. 
Nevertheless, they always find support for strong precession, and for all noise realizations the $\chi_\mathrm{p}$ posteriors are significantly different from the prior. 
The posteriors of $\chi_{\mathrm{eff}}$ fluctuate significantly according to all models. 

Similar investigations on spin inference for massive events have been presented in Refs.~\cite{Biscoveanu:2021nvg, Xu:2022zza, Miller:2025eak}, as the spin uncertainty increases with the total mass.
In particular, Ref.~\cite{Xu:2022zza} examined the impact of Gaussian noise for systems similar to the high-mass event GW190521 \cite{LIGOScientific:2020iuh} using NRSur. 
In agreement with our results, they found that masses are typically well measured, while the impact is more pronounced for spins, with the true value of $\chi_{\mathrm{eff}}$ and $\chi_{\mathrm{p}}$ sometimes outside the 90\% credible interval; they suggested that spin precession is not measurable for SNR $\leq$ 45. 
Instead, Ref.~\cite{Miller:2025eak} employed a time-domain framework to investigate the spin precession measurability, and found it to be measurable in weaker signals, starting at an SNR of 14. 
This is because it is not just the SNR of a signal that matters, but also the specific signal morphology and the extrinsic angular configuration. 
Our results show that for a strongly precessing signal with SNR $\sim20$, all waveform models infer consistently high precession ($\chi_{\mathrm{p}}\geq 0.7$), while the measurement of $\chi_{\mathrm{eff}}$ is significantly impacted by Gaussian noise.

\section{Interplay between Gaussian noise and waveform systematics}
\label{sec:Gaussian_noise_wf_system}

\begin{figure}
        \centering 
        \includegraphics[width=0.49\textwidth]{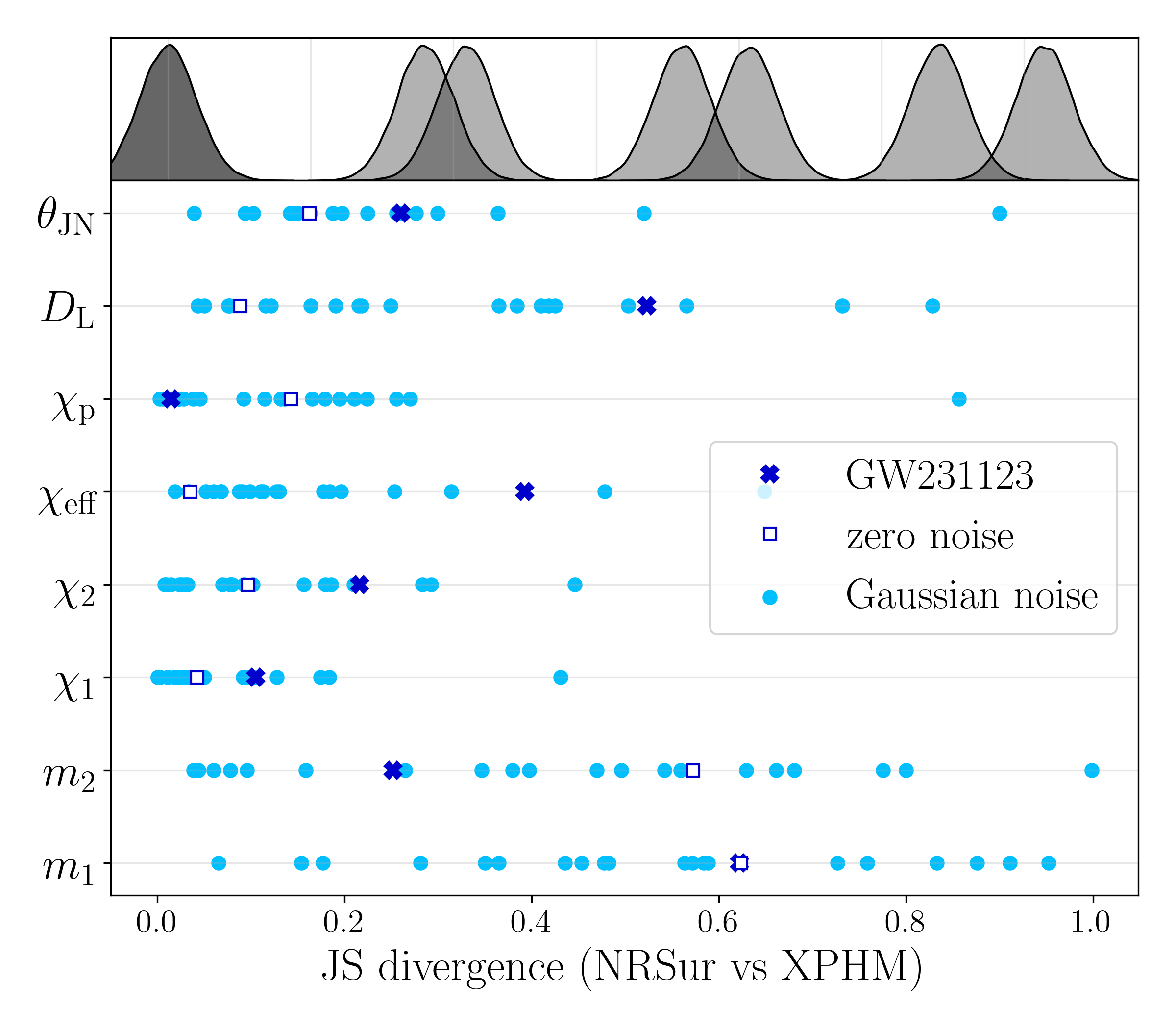}
        \includegraphics[width=0.49\textwidth]      {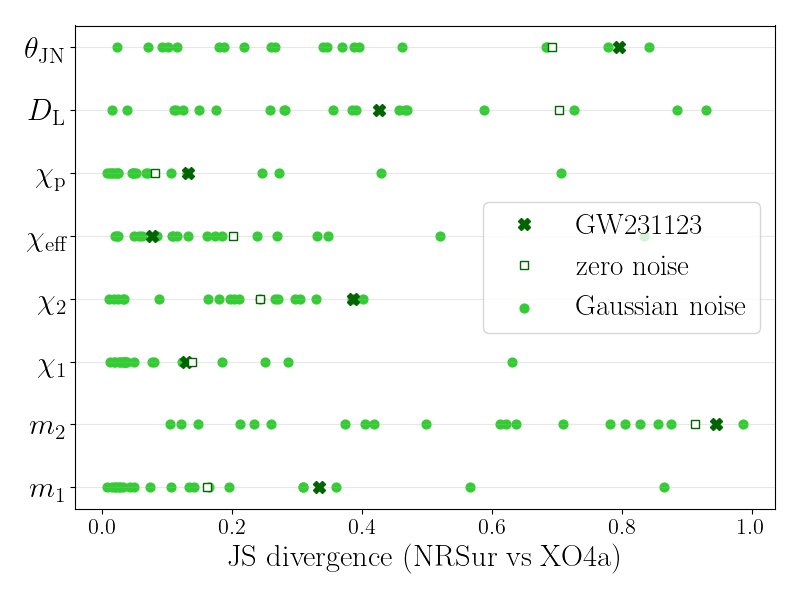}
        \caption{One-dimensional JS divergence between the posteriors computed using NRSur and XPHM (\textit{top}) and NRSur and XO4a  (\textit{bottom}) for various parameters. The  crosses indicate the JS divergence for GW231123~\cite{GW231123}, the empty squares indicate the divergence for the zero-noise maximum-likelihood simulation, and the dots refer to the  simulations with 20 Gaussian noise realizations. The curves at the top illustrate the JS divergences values in a simple case of two Gaussian distribution with same standard deviation. Gaussian noise can both increase or decrease the waveform systematics with respect to zero noise. Larger systematics than the ones observed in GW231123 can be reproduced with simulations in Gaussian noise.
        } 
        \label{fig:1D_JS_NRSUR_XPHM} 
\end{figure}

Visual inspection of Figs.~\ref{fig:Gaussian}--\ref{fig:Gaussian_XO4a} suggests that the degree of systematics manifested depends on the exact noise realization. 
To first order, the statistical uncertainty and the systematic error on the maximum-likelihood parameters, Eq.~\eqref{eq:error_sum}, are however independent. 
Our simulations show that this first order calculation is therefore not sufficient and statistical and systematic errors do correlate. 
Indeed, Gaussian noise shifts might increase or reduce the observed waveform systematics with respect to the zero-noise case shown in Fig.~\ref{fig:wf_syst} as they affect each waveform differently. 
We quantify how the manifest systematics vary in the presence of Gaussian noise by computing the one-dimensional JS divergence, defined in Sec.~\ref{sec:JS}, between the posterior distributions measured by different waveform models in the same noise realization.
Fig. \ref{fig:1D_JS_NRSUR_XPHM} shows the JS divergence between NRSur and XPHM (top), and between NRSur and XO4a (bottom) posteriors for the component source-frame masses, the spins, the luminosity distance and the inclination. For comparison, we also show the JS divergence for GW231123.

Starting with NRSur vs XPHM, we find that for 6 noise realizations over 20, the differences between the primary and secondary source-frame mass are larger than in the zero-noise case. 
For $\chi_{\mathrm{eff}}$ ($\chi_\mathrm{p}$) the divergences are larger than the zero-noise case in 18 (8) cases. 
The waveform systematics observed in the simulations sometimes are larger than those of GW231123: 6 times over 20 for $m_1$, 14 times for $m_2$, 2 times for $\chi_{\mathrm{eff}}$ and 17 times for  $\chi_\mathrm{p}$.

Switching, to NRSur vs XO4a, we find similar results.
For 7 noise realizations over 20, the divergence in $m_1$ is larger than in the zero-noise case. 
The JS of the secondary mass spans the entire range of values, and only in one noise realization is it higher than in zero-noise. 
In 6 noise realizations, $\chi_{\mathrm{eff}}$ and $\chi_{\mathrm{p}}$ have larger divergences than in zero-noise. 
Only in 3 and 2 cases over 20, the JS divergences for the luminosity distance and the inclination are larger than in the zero-noise simulation.
The simulations in Gaussian noise exhibit larger systematics than in GW231123 3 times over 20 for $m_1$, once for $m_2$,  14 and 4 times for $\chi_{\mathrm{eff}}$ and $\chi_{\mathrm{p}}$, 7 and 1 for the distance and the inclination.

These results suggest that Gaussian noise might often manifest worse the waveform systematics than the no-noise case, and that the degree of systematics observed in GW231123 is not exceptional and can be reproduced with simulations in Gaussian noise. 
For few parameters ($m_2$ and $\theta_{\mathrm{JN}}$) the JS between NRSur and XO4a in GW231123 is close to the maximum value and larger than the JS obtained in the majority of the simulations.
Therefore a potential inability to reproduce the systematics in a single noise realization is not significant given the limited statistics considered.\\

We conclude this section discussing the Bayes factor for different waveform models for GW231123, reported in Ref.~\cite{GW231123}. 
For intuition, the Bayes factor $B_{A,B}$ of model $A$ with respect to model $B$ can be approximated as \cite{Romano:2016dpx}:
\begin{equation}
    \label{eq:BF}
    \ln B_{A,B} \simeq \ln (\Lambda_{\mathrm{ML}}) + \ln \bigg ( \frac{\Delta V_A / V_A}{\Delta V_B / V_B} \bigg)\,,
\end{equation}
where the first term is the ratio of the maxima of the likelihood functions $\Lambda_{\mathrm{ML}}$ for the two models $A$ and $B$, and the second term encodes the characteristic spread $\Delta V$ of the posterior around its maximum and the total parameter space prior volume $V$. 
This latter term penalizes models that have a larger prior volume than required to fit the data. 
In GW231123, all models except XPHM obtain a larger evidence than NRSur, and in particular XO4a obtained a Bayes factor of at least 140:1 over NRSur \cite{GW231123}. 
The fact that such differences are not indicative of one model being more accurate than another is known \cite{Hoy:2022tst, hoy2025incorporation}, and here we expand on this point.

For GW231123, the maximum logarithmic likelihoods are 218.1, 210.7, 218.6 for NRSur, XPHM and XO4a, respectively. 
The difference between NRSur and XO4a is minimal, meaning that the two waveform models fit the data equally well, see also Fig.~\ref{fig:maxL_3models}. 
Hence, the difference in the Bayes factor in favor of XO4a ($\ln B_{\mathrm{NRSur, XO4a}}$ = -4.94, or $\sim 140:1$) originates primarily from the second term of Eq.~\eqref{eq:BF}. 
The parameter space of a BBH system has 15 dimensions and it is not trivial to visualize to estimate the parameter-space volume.
However,  it is not difficult to obtain a factor of ${\sim}10^2$ difference in parameter space volume over a 15-dimensional posterior.

We explore the Bayes factor and likelihood with our simulations.
The simulated signals are generated with NRSur, so we would expect NRSur to have similar or higher maximum-likelihood with respect to the other models.  
This is necessarily true in the zero-noise case: the maximum likelihood is higher for NRSur (183.3) than XPHM (178.0) and XO4a (181.6). 
The Bayes factor supports NRSur over XPHM, but (perhaps surprisingly) supports XO4a over NRSur ($B_{\mathrm{NRSur, XO4a}}=-3.4$, or 30:1). 
The switch in preference between the Bayes factor and the likelihood is due to the prior term in Eq.~\eqref{eq:BF} and not due to which model fits the data better (NRSur, per the likelihood).
For simulations in Gaussian noise, the maximum likelihood is higher in NRSur than in XPHM in 18 cases over 20, and higher in NRSur than in XO4a in 12 cases over 20. 
The Bayes factor supports NRSur versus XPHM in 18 cases over 20, and support XO4a over NRSur in 16 cases over 20.

In conclusion, both for the GW231123 data and the NRSur simulation in zero-noise, the maximum likelihoods from NRSur and XO4a are very close, meaning that both models fit the data comparably well. 
The Bayes factor in favor of XO4a is then induced by a less tightly constrained likelihood over the prior range. 
These results further highlight that Bayes factors should not be used to resolve waveform systematics. 

\section{Single-detector inference for GW231123}
\label{sec:HvsL}

GW inference is typically performed using the data from all detectors that were taking data at the time of the event \cite{GWTC4method}. 
In particular cases, for example when a GW event overlaps with a glitch,
individual-detector inference might offer insights on the impact of the glitch as it affects only one detector, for example for GW200129 \cite{payne2022curious}. 
Concerning GW231123, searches for non-Gaussian excess power in the time-frequency window around the event determined that nearby glitches have no measurable effect on the analysis \cite{GW231123}.
However, given the extreme properties of the event, the presence of waveform systematics, and nearby glitches, single-detector parameter estimation has been performed: Ref.~\cite{GW231123} reported the spins measured in each detector, and highlighted greater support for spins component aligned with the orbital angular momentum in LIGO Hanford data.

In the previous sections, we showed that Gaussian noise impacts inference for signals like GW231123 by shifting the posterior distributions (Sec.~\ref{sec:impact}), and that such shifts can increase or reduce waveform systematics (Sec.~\ref{sec:wf_syst}). 
Since different detectors have different noise realizations, we investigate the expected degree of overlap between single-detector posteriors for GW231123-like signals in this section.
We show that the differences observed in GW231123 are consistent with Gaussian noise. 

\subsection{Comparison of LIGO Hanford-only and LIGO Livingston-only inference}
\label{subsec:H_L_GW231123}

\begin{figure}
        \centering 
        \includegraphics[width=0.48\textwidth]{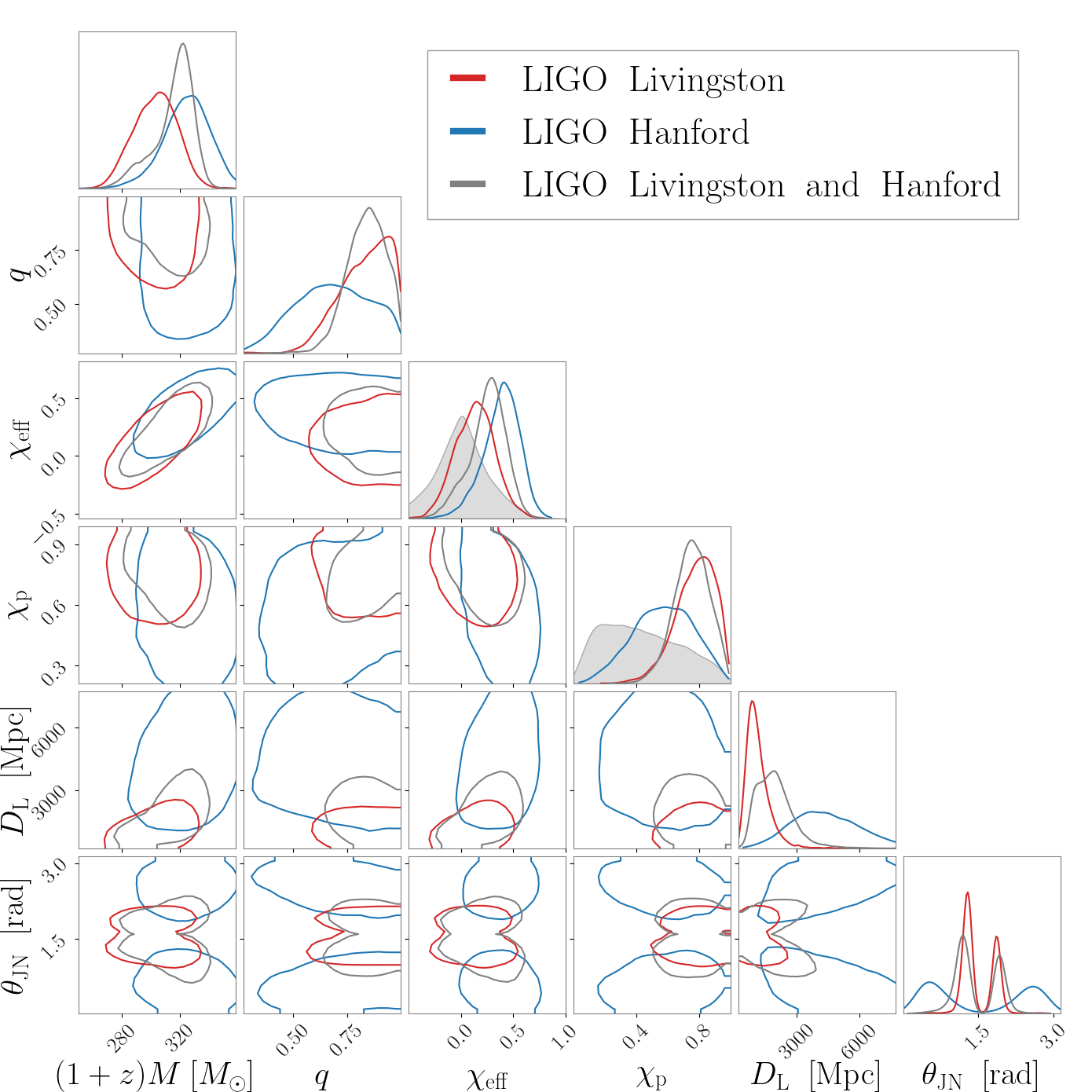} 
        \caption{GW231123 source inference with LIGO Hanford (blue), LIGO Livingston (red), and when using both detectors (gray) with the NRSur model. The contours and the vertical lines show the 90\% credible intervals. The gray shaded areas indicate the $\chi_{\rm{eff}}$ and $\chi_{\rm{p}}$ priors.}
        \label{fig:H_L_corner_plot} 
\end{figure}

First, we present in more detail the GW231123 source properties inferred using LIGO Hanford-only data and LIGO Livingston-only data~\cite{GW231123,zenodo_v2} in Fig. \ref{fig:H_L_corner_plot}.  
All results reported in this section employ the NRSur model. 
LIGO Hanford-only data and LIGO Livingston-only data are referred in the following with superscripts $^{\rm{H}}$ and $^{\rm{L}}$ respectively.  
The detector-frame total masses are $M^{\rm{H}}=327^{+26}_{-21}\,M_\odot$ and $M^{\rm{L}}=303^{+24}_{-22}\,M_\odot$, consistent at the 60\% level.
LIGO Hanford data favor a higher value of $\chi_{\rm{eff}}$ 
and intermediate value of precession ($\chi_{\rm{eff}}^{\rm{H}}=0.43^{+0.32}_{-0.22}$, $\chi_{\rm{p}}^{\rm{H}}=0.60^{+0.31}_{-0.28}$), while the LIGO Livingston data favor high precession ($\chi_{\rm{eff}}^{\rm{L}}=0.14^{+0.31}_{-0.26}$, $\chi_{\rm{p}}^{\rm{L}}=0.80^{+0.24}_{-0.15}$). 
LIGO Hanford localizes the event at $D_{\rm{L}}^{\rm{H}}=4452^{+2585}_{-3140}$\,Mpc favoring face-on (or face-off) inclination, 
while LIGO Livingston places the event much closer ($D_{\rm{L}}^{\rm{L}}=998^{+529}_{-992}$\,Mpc) and with a better constrained inclination. 
The differences in localization reflect differences in the SNR of higher order modes: LIGO Livingston recovers the mode (3,3) at an SNR of ${\sim} 2.9$ and the mode (4,4) at SNR $\sim3.3$, while in LIGO Hanford the SNR in the higher order modes is lower (${\sim} 1.8$ in (3,3), and ${\sim} 1.1$ in (4,4)).
The total SNR is ${\sim}17.6$ in LIGO Livingston and ${\sim}13.6$ in LIGO Hanford.
The source properties measured with both detectors, each detector individually, and the JS divergences between the single-detector posteriors are reported in Table \ref{tab:231123_app}, in Appendix~\ref{app:single-detector}. 
Full-network results are mainly driven by LIGO Livingston as the signal has a larger SNR.

The differences shown in Fig. \ref{fig:H_L_corner_plot} look consistent with the Gaussian noise shifts discussed in Sec.~\ref{sec:impact}. 
Below, we quantify such statement.

\begin{figure*}
        \centering 
        \includegraphics[width=\textwidth]{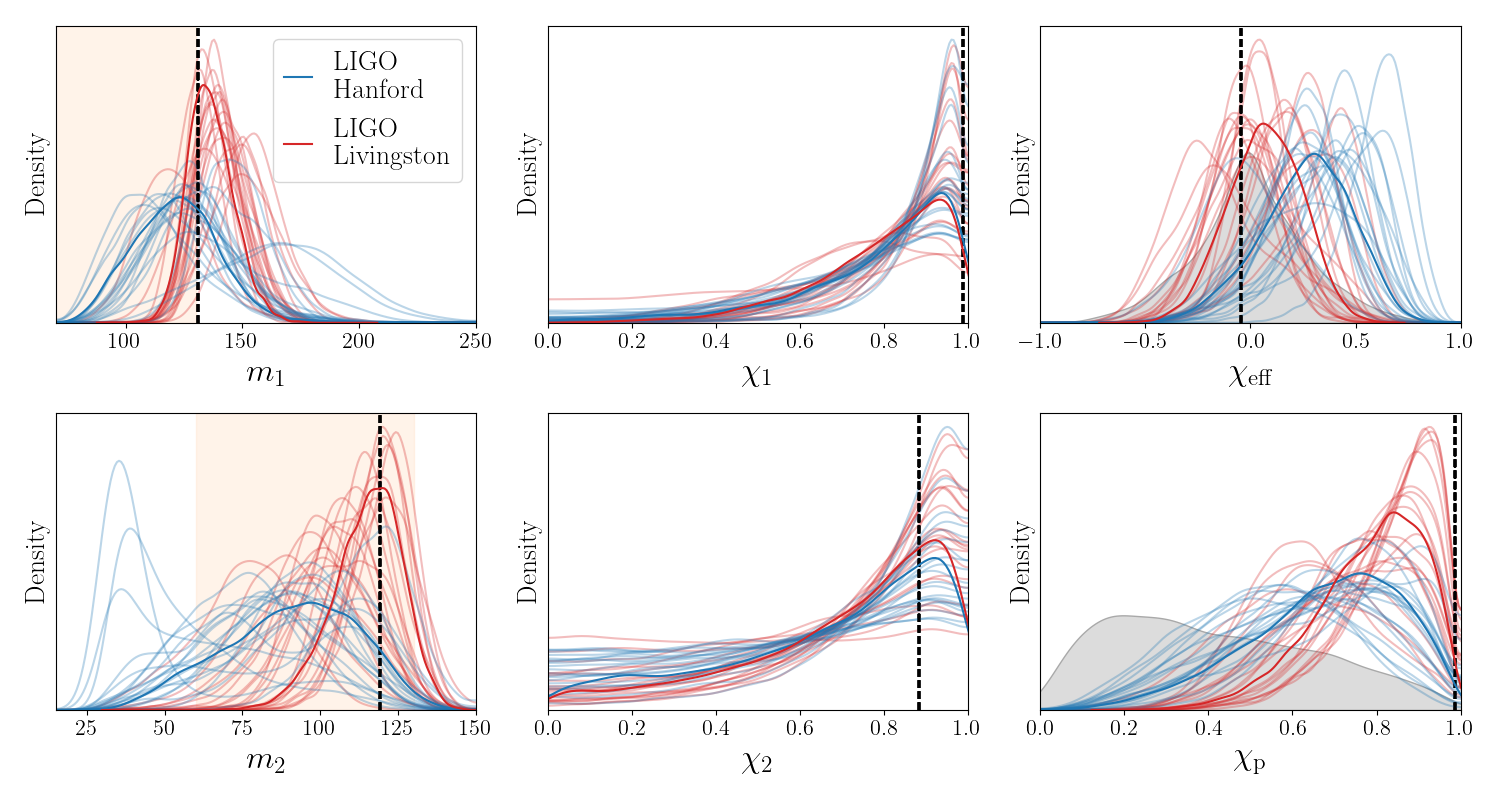} 
         
        \caption{Same as Fig. \ref{fig:Gaussian} when analyzing the simulated signal with each detector individually (LIGO Hanford in blue, LIGO Livingston in red). The darker lines indicate the posteriors from the zero-noise simulation. All analyses are carried out with the NRSur model.} 
        \label{fig:fig1_H_L} 
\end{figure*}

\begin{figure*}
        \centering 
        \includegraphics[width=0.32\textwidth]{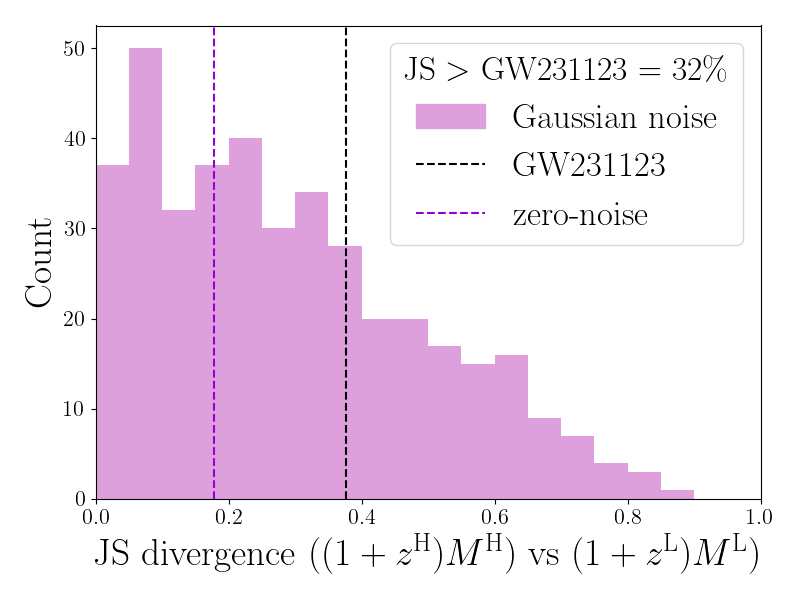} 
        \includegraphics[width=0.32\textwidth]{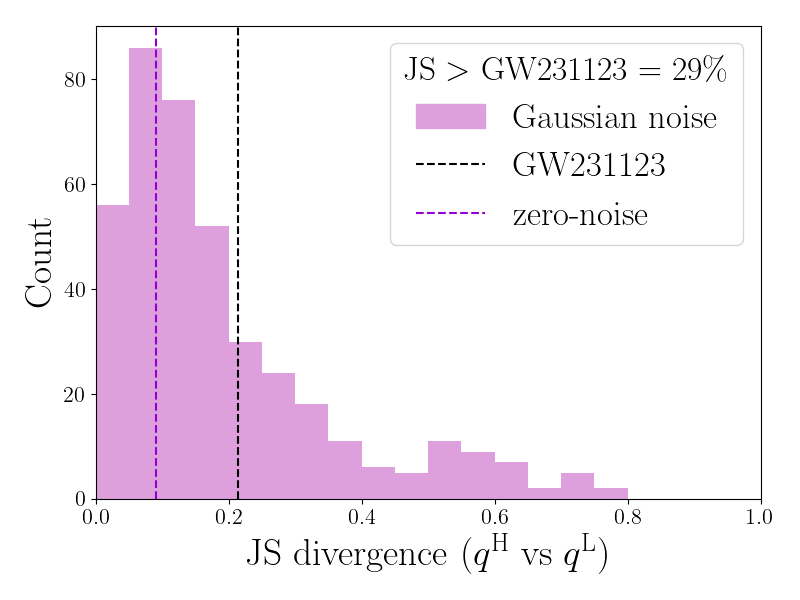} 
        \includegraphics[width=0.32\textwidth]{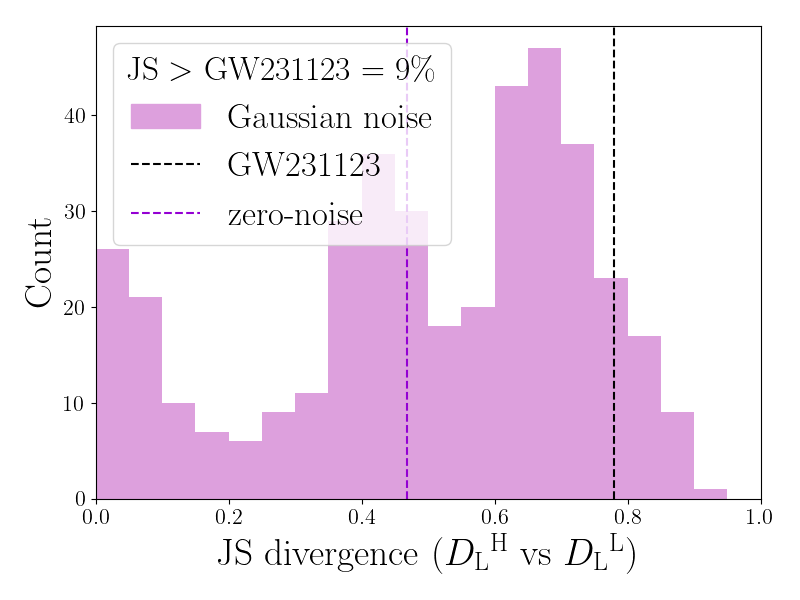}
        \includegraphics[width=0.32\textwidth]{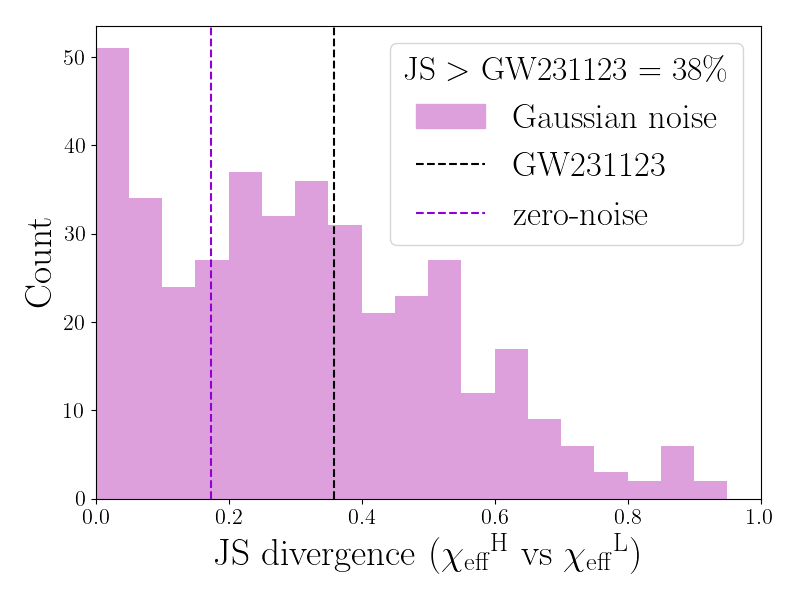} 
        \includegraphics[width=0.32\textwidth]{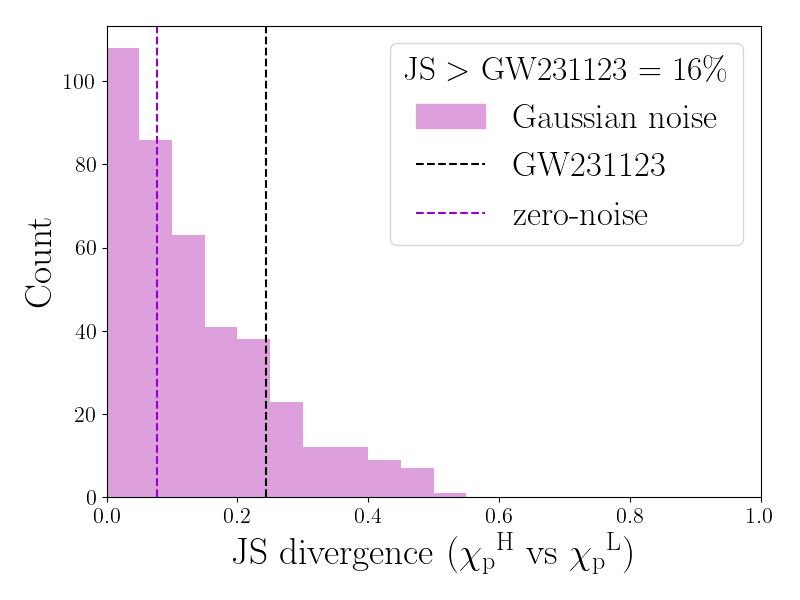} 
         
        \caption{JS divergence between posterior distributions of GW231123-like simulations in Gaussian noise recovered analyzing each detector independently. From top left: detector-frame total mass $M$, mass ratio $q$, luminosity distance $D_L$, effective inspiral spin $\chi_{\mathrm{eff}}$, precession spin $\chi_{\mathrm{p}}$. The waveform model used to generate the simulation and recover the signal is NRSur. The black dotted line indicates the JS for GW231123. The percentage of LIGO Livingston and LIGO Hanford pairs with a larger JS than what observed in GW231123 is reported in the legend. Differences in the single-detector inference in GW231123 are consistent with Gaussian noise expectations.
        } 
        \label{fig:JS} 
\end{figure*}


\subsection{Impact of Gaussian noise on single-detector inference}

Similarly to the analysis of Sec.~\ref{sec:impact}, here we explore the impact of Gaussian noise on single-detector inference and how it compares to the differences observed in GW231123, Fig. \ref{fig:H_L_corner_plot}.
We revisit the 20 Gaussian noise simulations from Sec.~\ref{sec:impact} and analyze each detector independently with NRSur.
Figure \ref{fig:fig1_H_L} shows the inferred masses and spins. 
When analyzing the signal with a single detector, the SNR is lower (optimal SNR$^{\mathrm{H}}\sim 13$, SNR$^{\mathrm{L}}\sim 17$), so the fluctuations of the posterior distribution are larger than when using both detectors in
Fig. \ref{fig:Gaussian}.  
The fluctuations are particularly evident in LIGO Hanford-only, for which the signal SNR is lower: posteriors for both component masses vary significantly, and $\chi_{\mathrm{eff}}$ and $\chi_{\mathrm{p}}$ fluctuate as well.

We quantify how often the single-detector posteriors in pair combinations have larger differences than what was observed in GW231123.
Since the Gaussian noise realizations in both detectors are independent, we calculate the JS divergence between single-detector posteriors for 400 2-detector combinations. 
The resulting JS divergences are reported in Figure \ref{fig:JS}: we find that in 32\% (29\%) of the pair combinations the posteriors of the detector-frame total mass (mass-ratio) have a larger JS than the one observed in GW231123. 
The effective and precessing spins measured in each detector individually are often (38\%, and 16\% respectively) more dissimilar than GW231123. 
In 40 pairs over 400, the distances are even less overlapping than in GW231123. 

These comparisons suggest that differences between single-detector inference larger than the ones observed in GW231123 can be often found when simulating high-mass high-spin signals in Gaussian noise. 
The differences between LIGO Hanford and LIGO Livingston parameter estimation on GW231123 do not suggest the presence of a data quality issue, such as transient noise, as they can be replicated considering only Gaussian noise.

\section{Parameter estimation prospects with LIGO proposed upgrades}
\label{sec:future}

The inference of the source properties of merger-dominated events such as GW231123 is challenging. 
Improvements in detector sensitivity, especially at low frequencies, will enable observations of GW events for more cycles and with higher SNR, improving the robustness of their interpretation.  
In this section, we explore inference prospects for high-mass and high-spin GW events like GW231123 with the expected sensitivity for LIGO detectors in the future.

\begin{figure}
        \centering 
        \includegraphics[width=0.43\textwidth]{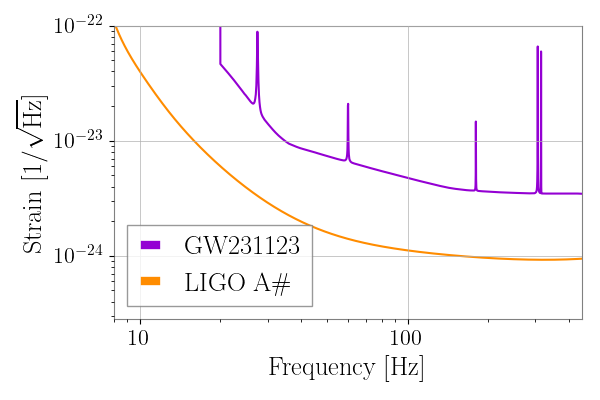}
        \includegraphics[width=0.45\textwidth]{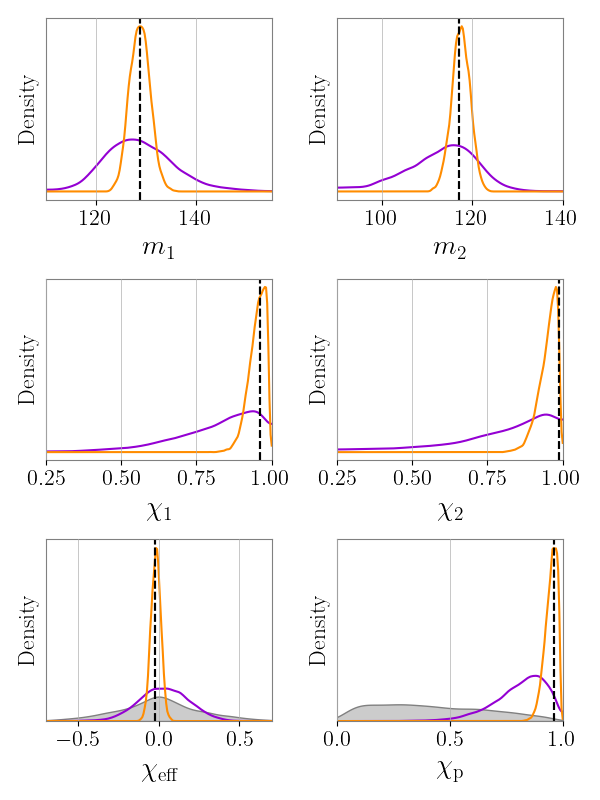}
        \caption{(\textit{Top panel}) LIGO A$\#$ PSD (orange) and median on-source PSD at the time of GW231123 in LIGO Livingston (violet). (\textit{Middle and bottom panels})  source-frame component masses, spin magnitudes, $\chi_{\mathrm{eff}}$ and $\chi_{\mathrm{p}}$ inferred by NRSur for a GW231123-like simulated signal in zero-noise with LIGO detectors with A$\#$ sensitivity (orange) and the sensitivity at the time of GW231123 (purple). The black dashed lines indicate the true values, and the gray shaded areas denote the $\chi_{\mathrm{eff}}$ and $\chi_{\mathrm{p}}$ prior distributions. With LIGO A\# sensitivity, we will be able to infer the properties of signal such as GW231123 with extraordinary accuracy.
        } 
        \label{fig:future} 
\end{figure}

LIGO A$\#$ (A-sharp) \cite{fritschel2022report} is a proposed upgrade to the Advanced LIGO detectors after the end of the fifth observing run (O5).
LIGO A$\#$ is envisioned to operate through mid-2030s, leading to the third generation of observatories, Cosmic Explorer \cite{Evans:2021gyd} in the US and Einstein Telescope \cite{ET:2019dnz} in Europe. 
The proposed upgrades include larger test masses, improved suspensions, higher laser power, increased squeezing efficiency at high frequencies, and better optical coatings.  
These improvements will significantly increase the sensitivity and boost detection rates \cite{Evans:2021gyd, ET:2019dnz}. 
The top panel of Fig. \ref{fig:future} compares the PSD at the time of GW231123 and the expected PSD for A$\#$: the latter is a factor of ${\sim}4$ lower at 50\,Hz.

To investigate LIGO A$\#$'s capabilities, we simulate a signal based again on the NRSur maximum-likelihood waveform in zero-noise, analyze it assuming two LIGO A$\#$ PSD detectors~\cite{a_sharp_psd} and NRSur.  
The optimal network SNR is $\sim102$, instead of ${\sim}20.7$ for GW231123. 
The minimum analysis frequency is conservatively kept at 20\,Hz, but this may be lowered with LIGO A$\#$, further increasing the SNR.  
Results are shown in Fig. \ref{fig:future} compared to the simulation using GW231123's O4a PSD: as expected given the higher SNR, the masses and spins posteriors are much narrower in LIGO A$\#$. 
The component masses ($m_1 = 128_{-9}^{+12}\,M_{\odot}$,  $m_2 = 114_{-15}^{+10}\,M_{\odot}$ with O4a sensitivity) remain consistent with A$\#$ ($m_1 = 128_{-3}^{+3}\,M_{\odot}$, $m_2 = 117_{-3}^{+3}\,M_{\odot}$) but the uncertainty decreases from ${\sim}8\%$ to ${\sim}2\%$ at the 90\% level. 
The spin magnitudes ($\chi_1=0.85_{-0.32}^{+0.13}$ and $\chi_2=0.85_{-0.46}^{+0.13}$ in O4a) increase getting more consistent with the true value ($\chi_1=0.96_{-0.06}^{+0.03}$ and $\chi_2=0.96_{-0.07}^{+0.03}$ with A$\#$).
The uncertainty decreases from ${\sim}37\%$ to ${\sim}6\%$ for the primary spin magnitude, and from ${\sim}54\%$ to ${\sim} 8\%$ for the secondary spin. 
The effective and precession spins are accurately inferred as well ($\chi_{\mathrm{eff}}=-0.02_{-0.05}^{+0.05}$ and $\chi_{\mathrm{p}}=0.95_{-0.05}^{+0.03}$ with A$\#$). 

Thus, the upgrades planned for LIGO A$\#$ will enable the precise and accurate measurement of the source properties of high-mass and high-spins GW events such as GW231123.

\section{Conclusions}
\label{sec:conclusions}

GW231123 is an exceptional GW event whose inferred masses and spins, under the BBH merger interpretation, challenge our current understanding of how such systems form. 
Source properties inference for such merger-dominated signals is challenging. 
Via a series of simulations, in this study we have revisited GW231123 inference and concluded that the properties inferred with the NRSur model are robust.

Firstly, significant differences in the inferred parameters between signal models have been reported \cite{GW231123}. 
Comparisons with numerical simulations have highlighted that NRSur is the most accurate model for such high-mass and high-spin systems and that there are simulations for which all the models exhibit systematics \cite{GW231123}.
In this work, we show that the waveform systematics observed in GW231123 can be replicated in the absence of noise for a signal described by the NRSur maximum-likelihood waveform.
This is in contrast to Ref.~\cite{GW231123} that reported inability to reproduce the systematics.
We attribute this difference to the fact that Ref.~\cite{GW231123} relied on existing numerical simulations, rather than waveforms that match the expected morphology of the GW231123 signal.

We then investigated whether the most relevant astrophysical conclusions from GW231123, namely the high masses and high spin magnitudes, are robust against Gaussian noise. 
To do so, we simulated the NRSur maximum-likelihood waveform into different Gaussian noise realizations and investigated how the posterior distributions fluctuate.  
We found that the impact of Gaussian noise on the measurements of mass and spin magnitudes with NRSur is minor: the key features of GW231123 (component masses within or above the mass gap, high spin magnitudes, and high precession) are consistently recovered. 
When using XO4a and XPHM models instead, the fluctuations in the secondary mass and in the spins are more pronounced, likely due to the combination of noise fluctuations and waveform systematics.
In the process, we showed that Gaussian noise can either enhance or hide  waveform systematics.
Discrepancies even larger than those observed in GW231123 can be produced when analyzing the NRSur maximum-likelihood waveform in Gaussian noise.

On the Bayes factor front, Ref.~\cite{GW231123} reported a ${\sim} 140:1$ preference XO4a over NRSur, which may naively suggest that XO4a is more trustworthy. 
However, we argued that the maximum-likelihood of NRSur and XO4a are in fact similar, meaning that these two models fit the data equally well. 
The difference in the Bayes factor does not arise because XO4a fits the data better, but because its posterior is less constrained.
Similar results can be reproduced in simulations where the signal is described by NRSur, suggesting that Bayes factors should not be used to resolve waveform systematics.

Finally, we tackled the issue of potential low data quality impacting each detector separately.
We demonstrated that the mild differences in the source properties of GW231123 inferred using LIGO Hanford and LIGO Livingston data independently need not indicate data-quality issues: larger differences can be often observed in simulations in Gaussian noise. Similarly to waveform systematics, a particular Gaussian noise realization can amplify or diminish such differences.

In the next years, planned upgrades to the LIGO detectors will further improve their sensitivity, allowing us to characterize systems like GW231123 more accurately. 
For example, an event like GW231123 will have an SNR of about 100 in the proposed LIGO A\# upgrade. 
This increase in sensitivity will reduce measurement uncertainties on the component masses from ${\sim}8\%$ to ${\sim}2\%$ (90\% credible level), and on the primary spin magnitude from ${\sim}37\%$ to ${\sim}6\%$.
GW231123 shows that such heavy, highly-spinning BBHs exist and future detectors will be able to extract their properties even more accurately.

\begin{acknowledgments}
We thank Harrison Siegel for helpful discussions and comments on the manuscript.
We thank Jacob Lange for helpful comments during internal LIGO review.

SB was supported by NSF Grant PHY-2309200.
KK was supported by the LIGO Laboratory Summer Undergraduate Research Fellowship program (LIGO SURF), and the California Institute of Technology Student-Faculty Programs.
KC was supported by NSF Grants PHY-2409001.
The Flatiron Institute is a division of the Simons Foundation.

This material is based upon work supported by NSF’s LIGO Laboratory 
which is a major facility fully funded by the 
National Science Foundation.
LIGO was constructed by the California Institute of Technology 
and Massachusetts Institute of Technology with funding from 
the National Science Foundation, 
and operates under cooperative agreement PHY-2309200. 
The authors are grateful
for computational resources provided by the LIGO Laboratory and supported by National Science Foundation
Grants PHY-0757058 and PHY-0823459.

\end{acknowledgments}

\appendix

\section{Waveform systematics using the XO4a maximum-likelihood simulation}\label{app:wf_syst}

\begin{figure*}
        \centering 
        \includegraphics[width=0.49\textwidth]{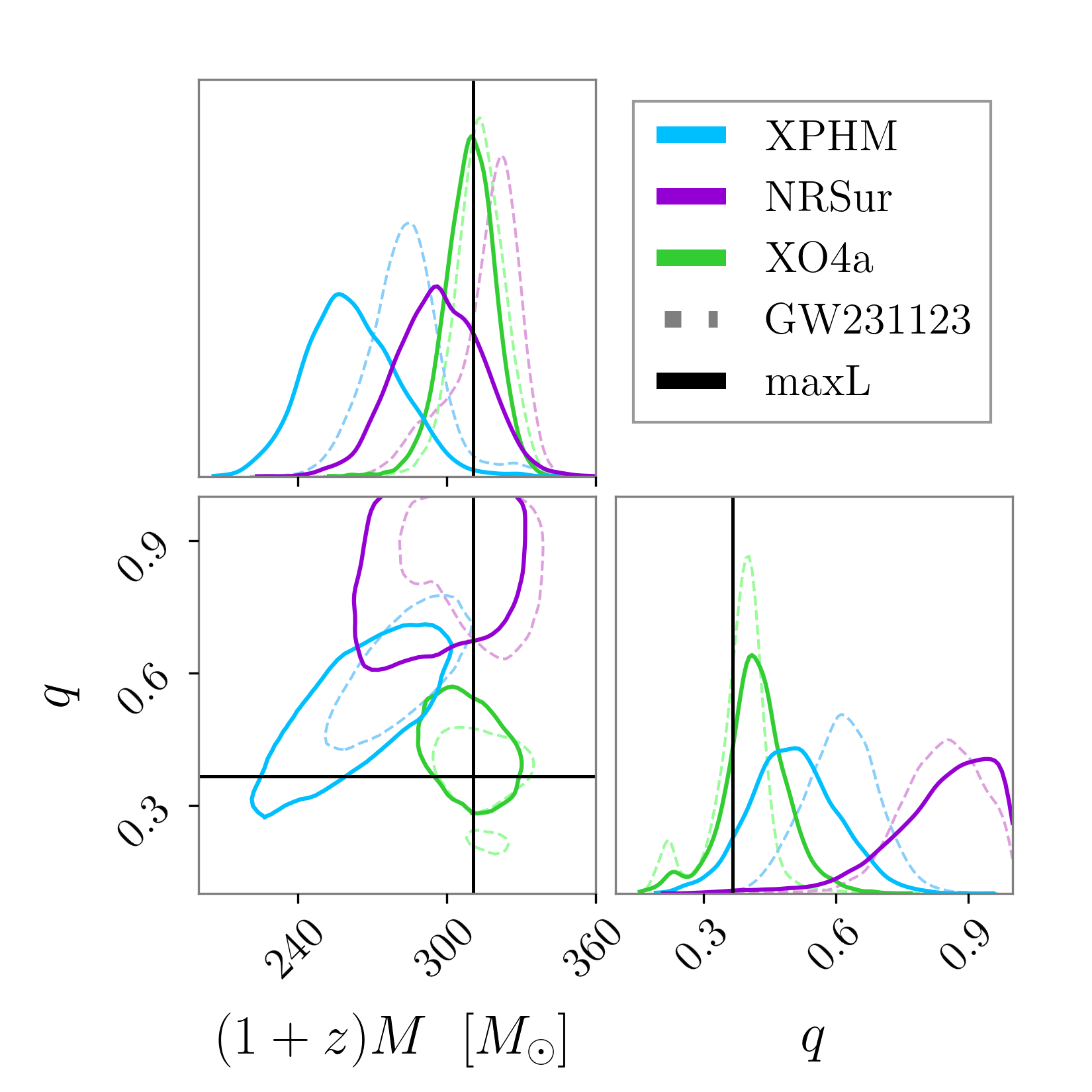}
        \includegraphics[width=0.49\textwidth]{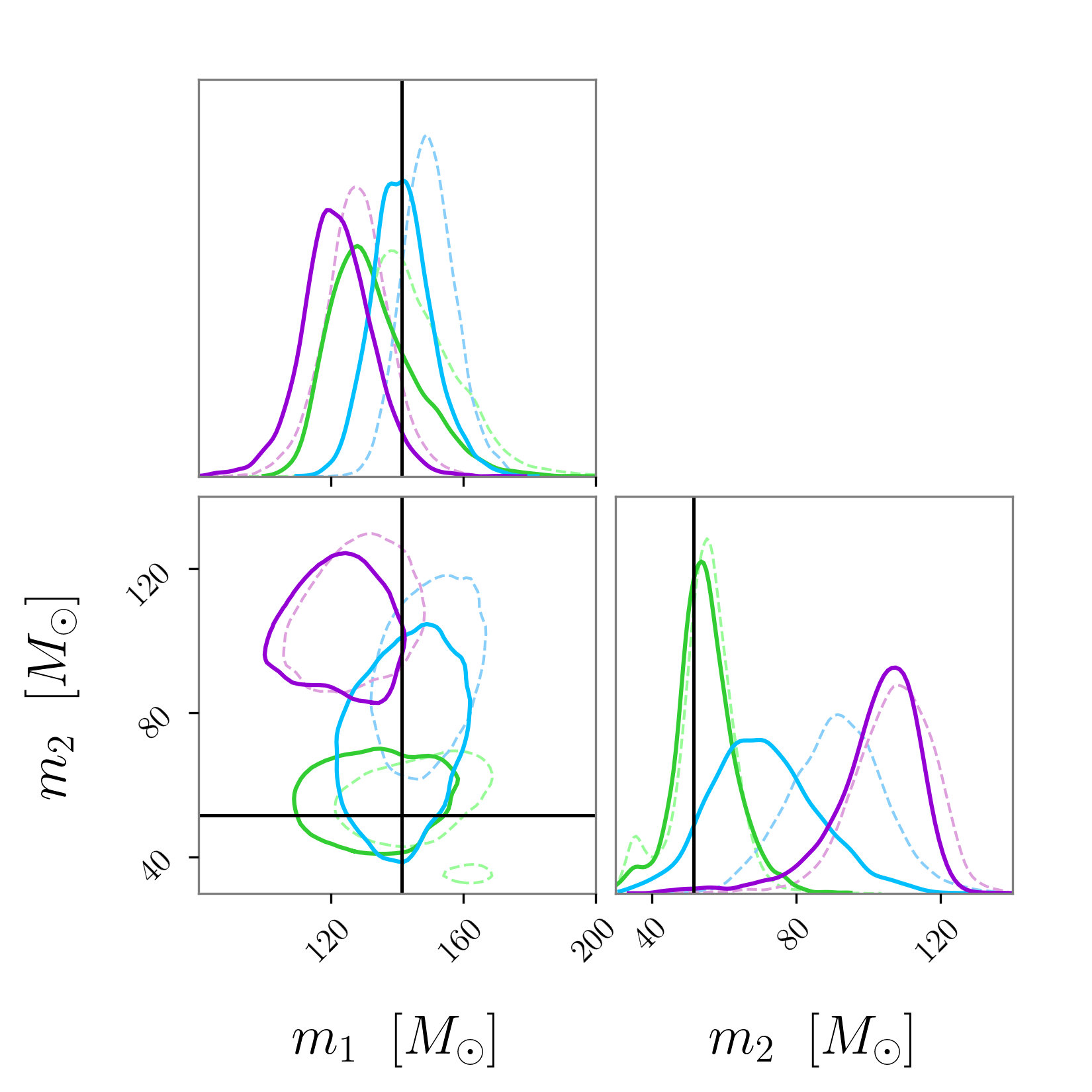}
        \includegraphics[width=0.32\textwidth]{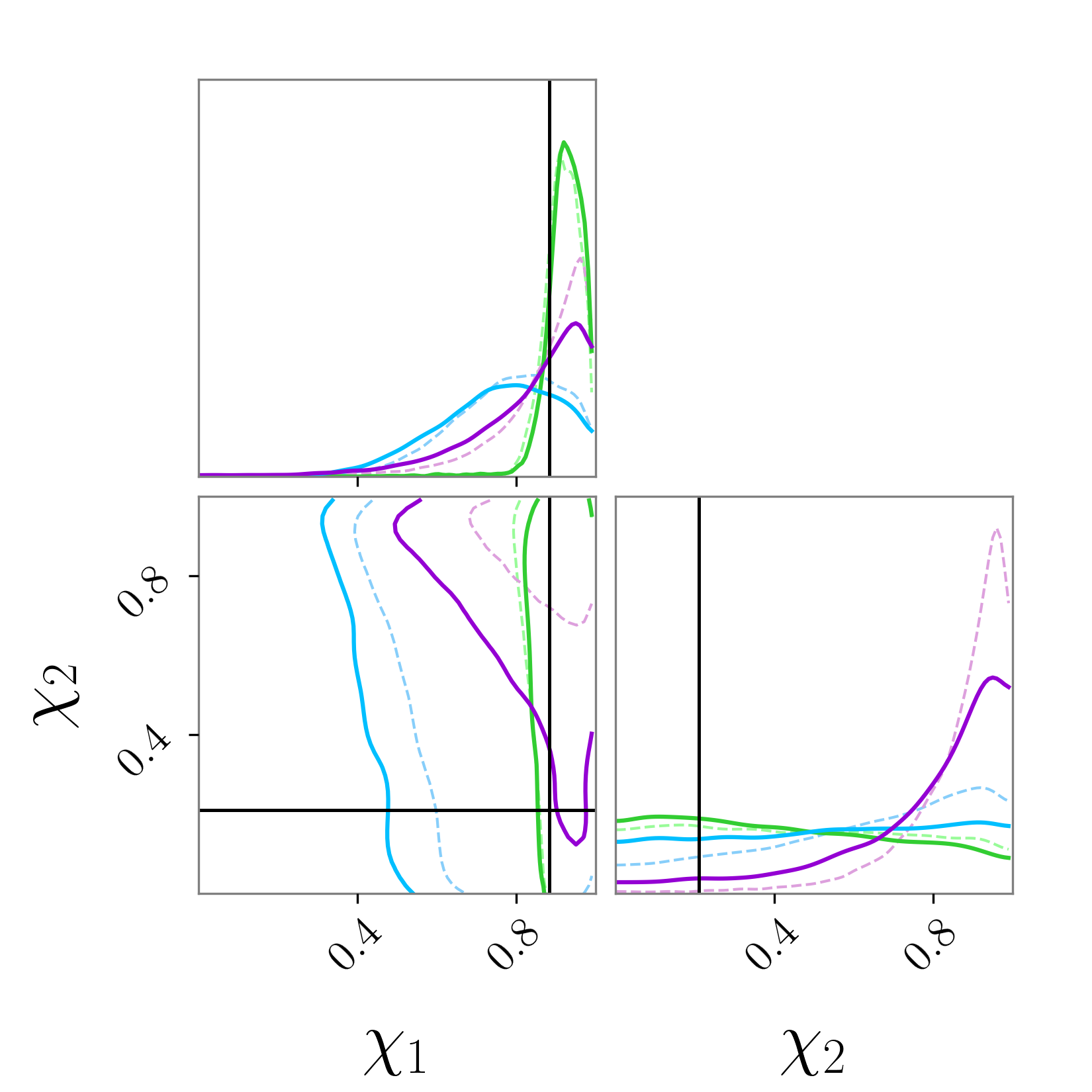}
        \includegraphics[width=0.32\textwidth]{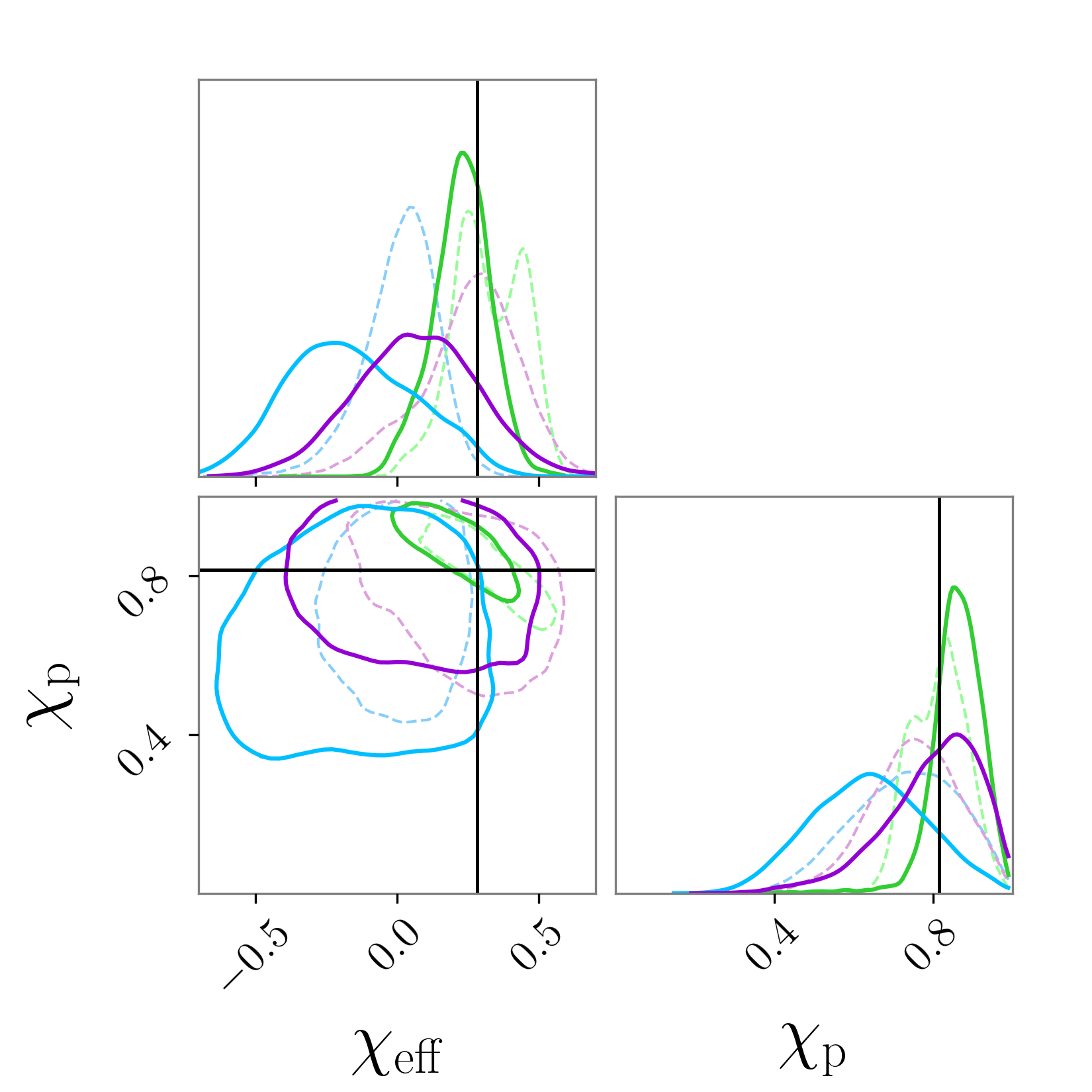}    \includegraphics[width=0.32\textwidth]{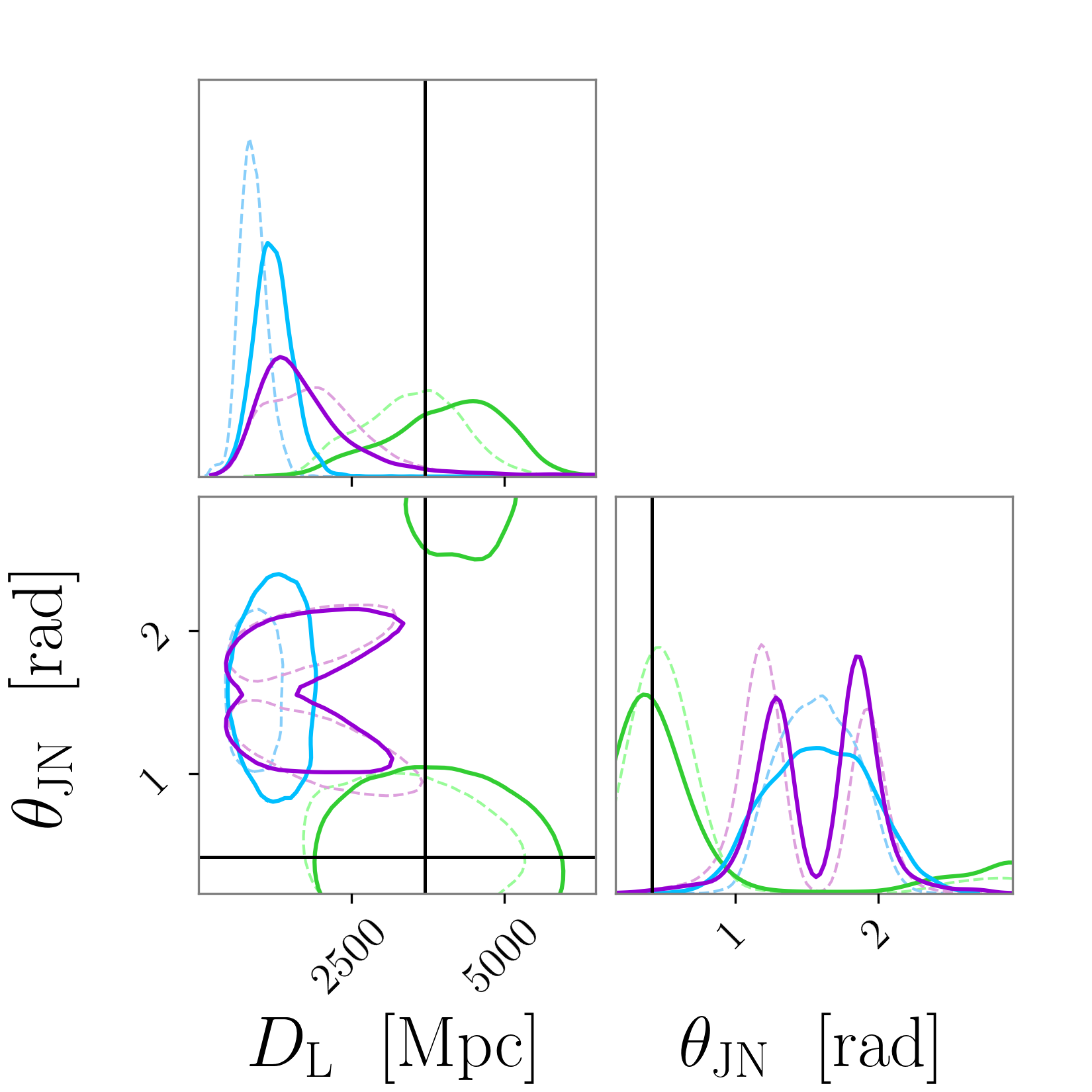}
        \caption{Similar to Fig. \ref{fig:wf_syst} but simulating the signal using the XO4a maximum-likelihood waveform instead of NRSur. The waveform systematics observed in GW231123 are reproduced in this case as well.} 
        \label{fig:wf_syst_XO4a} 
\end{figure*}

In Fig.~\ref{fig:maxL_3models}, we showed that the NRSur and XO4a maximum-likelihood waveforms are similar, even though the source properties inferred by the two models disagree substantially. 
In Sec.~\ref{sec:wf_syst} we showed that when we simulate a GW231123-like signal using the NRSur maximum-likelihood waveform, we can reproduce the systematics observed in GW231123.
Here, we repeat this analysis after simulating a GW231123-like signal using the XO4a maximum-likelihood waveform, in place of NRSur. 
Fig.~\ref{fig:wf_syst_XO4a} shows results in the same style as Fig.~\ref{fig:wf_syst}.

As expected, now, the XO4a inference is accurate for the masses, spins, and localization. 
XPHM and NRSur are biased in a way that is similar to the systematic differences observed in GW231123.
This is expected since the maximum-likelihood NRSur and XO4a waveforms are similar in Fig.~\ref{fig:maxL_3models}, even though the parameters that produce them are very different.
Overall, it is possible to reproduce the systematics observed in GW231123 in the absence of noise using both NRSur and XO4a maximum-likelihood waveforms. 

\section{LIGO Hanford and LIGO Livingston only analysis of GW231123}\label{app:single-detector}

Table \ref{tab:231123_app} reports the GW231123 source properties inferred with NRSur when using both LIGO detectors coherently, and when using each detector individually.

\renewcommand{\arraystretch}{1.4}
\begin{table*}
\begin{tabular}{l l l l l}
\hline
Parameter & LIGO Hanford & LIGO Livingston & Both\\
\hline
source-frame primary mass ($M_\odot$) & $114^{+34}_{-25}$ & $135^{+21}_{-16}$ & $128^{+16}_{-16}$\\
source-frame secondary mass ($M_\odot$) & $76^{+71}_{-25}$ & $114^{+43}_{-22}$ & $108^{+16}_{-20}$\\
source-frame total mass ($M_\odot$) & $192^{+47}_{-34}$ & $249^{+24}_{-24}$ & $237^{+23}_{-32}$\\
mass ratio $q$ & $0.69^{+0.27}_{-0.28}$ & $0.85^{+0.14}_{-0.23}$ & $0.85^{+0.15}_{-0.12}$ \\
primary spin magnitude $\chi_1$ & $0.85^{+0.13}_{-0.41}$ & $0.84^{+0.14}_{-0.35}$ & $0.90^{+0.10}_{-0.19}$ \\
secondary spin magnitude $\chi_2$ & $0.66^{+0.30}_{-0.58}$ & $0.84^{+0.14}_{-0.62}$ & $0.91^{+0.09}_{-0.22}$\\
eff. aligned spin parameter $\chi_{\rm{eff}}$ & $0.40^{+0.24}_{-0.33}$ & $0.14^{+0.30}_{-0.31}$ & $0.27^{+0.24}_{-0.35}$\\
eff. precessing spin parameter $\chi_{\rm{p}}$ & $0.58^{+0.29}_{-0.33}$ & $0.79^{+0.16}_{-0.25}$ & $0.76^{+0.19}_{-0.17}$ \\
luminosity distance $D_{\rm{L}}$ (Gpc) & $4.3^{+3.0}_{-2.4}$ & $1.0^{+1.1}_{-0.5}$ & $1.9^{+1.7}_{-1.0}$ \\
inclination angle $\theta_{\mathrm{JN}}$ (rad) & $1.0^{+1.8}_{-0.8}$ & $1.3^{+0.6}_{-0.2}$ & $1.3^{+0.8}_{-0.4}$ \\

optimal SNR & $12.5^{+1.6}_{-1.7}$ & $16.0^{+1.7}_{-1.7}$ &  $20.6^{+0.2}_{-0.3}$ \\

\end{tabular}
\caption{GW231123 source properties inferred by NRSur using only LIGO Hanford data (\textit{first column}), only LIGO Livingston data (\textit{second column}), and both detectors (\textit{third column}), as in Ref.~\cite{GW231123}). For each parameter, we report the median values and the 90\% credible interval.} \label{tab:231123_app}
\end{table*}

\bibliography{apssamp}

\begin{thebibliography}{67}%
\makeatletter
\providecommand \@ifxundefined [1]{%
 \@ifx{#1\undefined}
}%
\providecommand \@ifnum [1]{%
 \ifnum #1\expandafter \@firstoftwo
 \else \expandafter \@secondoftwo
 \fi
}%
\providecommand \@ifx [1]{%
 \ifx #1\expandafter \@firstoftwo
 \else \expandafter \@secondoftwo
 \fi
}%
\providecommand \natexlab [1]{#1}%
\providecommand \enquote  [1]{``#1''}%
\providecommand \bibnamefont  [1]{#1}%
\providecommand \bibfnamefont [1]{#1}%
\providecommand \citenamefont [1]{#1}%
\providecommand \href@noop [0]{\@secondoftwo}%
\providecommand \href [0]{\begingroup \@sanitize@url \@href}%
\providecommand \@href[1]{\@@startlink{#1}\@@href}%
\providecommand \@@href[1]{\endgroup#1\@@endlink}%
\providecommand \@sanitize@url [0]{\catcode `\\12\catcode `\$12\catcode `\&12\catcode `\#12\catcode `\^12\catcode `\_12\catcode `\%12\relax}%
\providecommand \@@startlink[1]{}%
\providecommand \@@endlink[0]{}%
\providecommand \url  [0]{\begingroup\@sanitize@url \@url }%
\providecommand \@url [1]{\endgroup\@href {#1}{\urlprefix }}%
\providecommand \urlprefix  [0]{URL }%
\providecommand \Eprint [0]{\href }%
\providecommand \doibase [0]{http://dx.doi.org/}%
\providecommand \selectlanguage [0]{\@gobble}%
\providecommand \bibinfo  [0]{\@secondoftwo}%
\providecommand \bibfield  [0]{\@secondoftwo}%
\providecommand \translation [1]{[#1]}%
\providecommand \BibitemOpen [0]{}%
\providecommand \bibitemStop [0]{}%
\providecommand \bibitemNoStop [0]{.\EOS\space}%
\providecommand \EOS [0]{\spacefactor3000\relax}%
\providecommand \BibitemShut  [1]{\csname bibitem#1\endcsname}%
\let\auto@bib@innerbib\@empty
\bibitem [{\citenamefont {Abac}\ \emph {et~al.}(2025{\natexlab{a}})\citenamefont {Abac} \emph {et~al.}}]{GW231123}%
  \BibitemOpen
  \bibfield  {author} {\bibinfo {author} {\bibfnamefont {A.~G.}\ \bibnamefont {Abac}} \emph {et~al.} (\bibinfo {collaboration} {LIGO Scientific, VIRGO, KAGRA}),\ }\href@noop {} {\bibfield  {journal} {\bibinfo  {journal} {Arxiv}\ } (\bibinfo {year} {2025}{\natexlab{a}})},\ \Eprint {http://arxiv.org/abs/2507.08219} {arXiv:2507.08219 [astro-ph.HE]} \BibitemShut {NoStop}%
\bibitem [{\citenamefont {Aasi}\ \emph {et~al.}(2015)\citenamefont {Aasi} \emph {et~al.}}]{LIGOScientific:2014pky}%
  \BibitemOpen
  \bibfield  {author} {\bibinfo {author} {\bibfnamefont {J.}~\bibnamefont {Aasi}} \emph {et~al.} (\bibinfo {collaboration} {LIGO Scientific}),\ }\href {\doibase 10.1088/0264-9381/32/7/074001} {\bibfield  {journal} {\bibinfo  {journal} {Class. Quant. Grav.}\ }\textbf {\bibinfo {volume} {32}},\ \bibinfo {pages} {074001} (\bibinfo {year} {2015})},\ \Eprint {http://arxiv.org/abs/1411.4547} {arXiv:1411.4547 [gr-qc]} \BibitemShut {NoStop}%
\bibitem [{\citenamefont {Abac}\ \emph {et~al.}(2025{\natexlab{b}})\citenamefont {Abac} \emph {et~al.}}]{gwtc4-intro}%
  \BibitemOpen
  \bibfield  {author} {\bibinfo {author} {\bibfnamefont {A.~G.}\ \bibnamefont {Abac}} \emph {et~al.} (\bibinfo {collaboration} {LIGO Scientific, VIRGO, KAGRA}),\ }\href@noop {} {\bibfield  {journal} {\bibinfo  {journal} {arXiv preprint arXiv:2508.18080}\ } (\bibinfo {year} {2025}{\natexlab{b}})},\ \Eprint {http://arxiv.org/abs/2508.18080} {arXiv:2508.18080 [gr-qc]} \BibitemShut {NoStop}%
\bibitem [{\citenamefont {Abac}\ \emph {et~al.}(2025{\natexlab{c}})\citenamefont {Abac}, \citenamefont {Abouelfettouh}, \citenamefont {Acernese}, \citenamefont {Ackley}, \citenamefont {Adhicary}, \citenamefont {Adhikari}, \citenamefont {Adhikari}, \citenamefont {Adhikari}, \citenamefont {Adkins}, \citenamefont {Afroz} \emph {et~al.}}]{GWTC4method}%
  \BibitemOpen
  \bibfield  {author} {\bibinfo {author} {\bibfnamefont {A.}~\bibnamefont {Abac}}, \bibinfo {author} {\bibfnamefont {I.}~\bibnamefont {Abouelfettouh}}, \bibinfo {author} {\bibfnamefont {F.}~\bibnamefont {Acernese}}, \bibinfo {author} {\bibfnamefont {K.}~\bibnamefont {Ackley}}, \bibinfo {author} {\bibfnamefont {S.}~\bibnamefont {Adhicary}}, \bibinfo {author} {\bibfnamefont {D.}~\bibnamefont {Adhikari}}, \bibinfo {author} {\bibfnamefont {N.}~\bibnamefont {Adhikari}}, \bibinfo {author} {\bibfnamefont {R.}~\bibnamefont {Adhikari}}, \bibinfo {author} {\bibfnamefont {V.}~\bibnamefont {Adkins}}, \bibinfo {author} {\bibfnamefont {S.}~\bibnamefont {Afroz}},  \emph {et~al.},\ }\href@noop {} {\bibfield  {journal} {\bibinfo  {journal} {arXiv preprint arXiv:2508.18081}\ } (\bibinfo {year} {2025}{\natexlab{c}})}\BibitemShut {NoStop}%
\bibitem [{\citenamefont {Abac}\ \emph {et~al.}(2025{\natexlab{d}})\citenamefont {Abac}, \citenamefont {Abouelfettouh}, \citenamefont {Acernese}, \citenamefont {Ackley}, \citenamefont {Adamcewicz}, \citenamefont {Adhicary}, \citenamefont {Adhikari}, \citenamefont {Adhikari}, \citenamefont {Adhikari}, \citenamefont {Adkins} \emph {et~al.}}]{gwtc4results}%
  \BibitemOpen
  \bibfield  {author} {\bibinfo {author} {\bibfnamefont {A.}~\bibnamefont {Abac}}, \bibinfo {author} {\bibfnamefont {I.}~\bibnamefont {Abouelfettouh}}, \bibinfo {author} {\bibfnamefont {F.}~\bibnamefont {Acernese}}, \bibinfo {author} {\bibfnamefont {K.}~\bibnamefont {Ackley}}, \bibinfo {author} {\bibfnamefont {C.}~\bibnamefont {Adamcewicz}}, \bibinfo {author} {\bibfnamefont {S.}~\bibnamefont {Adhicary}}, \bibinfo {author} {\bibfnamefont {D.}~\bibnamefont {Adhikari}}, \bibinfo {author} {\bibfnamefont {N.}~\bibnamefont {Adhikari}}, \bibinfo {author} {\bibfnamefont {R.}~\bibnamefont {Adhikari}}, \bibinfo {author} {\bibfnamefont {V.}~\bibnamefont {Adkins}},  \emph {et~al.},\ }\href@noop {} {\bibfield  {journal} {\bibinfo  {journal} {arXiv preprint arXiv:2508.18082}\ } (\bibinfo {year} {2025}{\natexlab{d}})}\BibitemShut {NoStop}%
\bibitem [{\citenamefont {{Farmer}}\ \emph {et~al.}(2019)\citenamefont {{Farmer}}, \citenamefont {{Renzo}}, \citenamefont {{de Mink}}, \citenamefont {{Marchant}},\ and\ \citenamefont {{Justham}}}]{2019ApJ...887...53F}%
  \BibitemOpen
  \bibfield  {author} {\bibinfo {author} {\bibfnamefont {R.}~\bibnamefont {{Farmer}}}, \bibinfo {author} {\bibfnamefont {M.}~\bibnamefont {{Renzo}}}, \bibinfo {author} {\bibfnamefont {S.~E.}\ \bibnamefont {{de Mink}}}, \bibinfo {author} {\bibfnamefont {P.}~\bibnamefont {{Marchant}}}, \ and\ \bibinfo {author} {\bibfnamefont {S.}~\bibnamefont {{Justham}}},\ }\href {\doibase 10.3847/1538-4357/ab518b} {\bibfield  {journal} {\bibinfo  {journal} {\apj}\ }\textbf {\bibinfo {volume} {887}},\ \bibinfo {eid} {53} (\bibinfo {year} {2019})},\ \Eprint {http://arxiv.org/abs/1910.12874} {arXiv:1910.12874 [astro-ph.SR]} \BibitemShut {NoStop}%
\bibitem [{\citenamefont {Woosley}\ and\ \citenamefont {Heger}(2021)}]{Woosley:2021xba}%
  \BibitemOpen
  \bibfield  {author} {\bibinfo {author} {\bibfnamefont {S.~E.}\ \bibnamefont {Woosley}}\ and\ \bibinfo {author} {\bibfnamefont {A.}~\bibnamefont {Heger}},\ }\href {\doibase 10.3847/2041-8213/abf2c4} {\bibfield  {journal} {\bibinfo  {journal} {Astrophys. J. Lett.}\ }\textbf {\bibinfo {volume} {912}},\ \bibinfo {pages} {L31} (\bibinfo {year} {2021})},\ \Eprint {http://arxiv.org/abs/2103.07933} {arXiv:2103.07933 [astro-ph.SR]} \BibitemShut {NoStop}%
\bibitem [{\citenamefont {Hendriks}\ \emph {et~al.}(2023)\citenamefont {Hendriks}, \citenamefont {van Son}, \citenamefont {Renzo}, \citenamefont {Izzard},\ and\ \citenamefont {Farmer}}]{Hendriks:2023yrw}%
  \BibitemOpen
  \bibfield  {author} {\bibinfo {author} {\bibfnamefont {D.~D.}\ \bibnamefont {Hendriks}}, \bibinfo {author} {\bibfnamefont {L.~A.~C.}\ \bibnamefont {van Son}}, \bibinfo {author} {\bibfnamefont {M.}~\bibnamefont {Renzo}}, \bibinfo {author} {\bibfnamefont {R.~G.}\ \bibnamefont {Izzard}}, \ and\ \bibinfo {author} {\bibfnamefont {R.}~\bibnamefont {Farmer}},\ }\href@noop {} {\bibfield  {journal} {\bibinfo  {journal} {Mon. Not. Roy. Astron. Soc.}\ }\textbf {\bibinfo {volume} {526}},\ \bibinfo {pages} {4130} (\bibinfo {year} {2023})},\ \Eprint {http://arxiv.org/abs/2309.09339} {arXiv:2309.09339 [astro-ph.HE]} \BibitemShut {NoStop}%
\bibitem [{\citenamefont {Belczynski}\ \emph {et~al.}(2020)\citenamefont {Belczynski} \emph {et~al.}}]{Belczynski:2017gds}%
  \BibitemOpen
  \bibfield  {author} {\bibinfo {author} {\bibfnamefont {K.}~\bibnamefont {Belczynski}} \emph {et~al.},\ }\href {\doibase 10.1051/0004-6361/201936528} {\bibfield  {journal} {\bibinfo  {journal} {Astron. Astrophys.}\ }\textbf {\bibinfo {volume} {636}},\ \bibinfo {pages} {A104} (\bibinfo {year} {2020})},\ \Eprint {http://arxiv.org/abs/1706.07053} {arXiv:1706.07053 [astro-ph.HE]} \BibitemShut {NoStop}%
\bibitem [{\citenamefont {Qin}\ \emph {et~al.}(2018)\citenamefont {Qin}, \citenamefont {Fragos}, \citenamefont {Meynet}, \citenamefont {Andrews}, \citenamefont {S\o{}rensen},\ and\ \citenamefont {Song}}]{Qin:2018vaa}%
  \BibitemOpen
  \bibfield  {author} {\bibinfo {author} {\bibfnamefont {Y.}~\bibnamefont {Qin}}, \bibinfo {author} {\bibfnamefont {T.}~\bibnamefont {Fragos}}, \bibinfo {author} {\bibfnamefont {G.}~\bibnamefont {Meynet}}, \bibinfo {author} {\bibfnamefont {J.}~\bibnamefont {Andrews}}, \bibinfo {author} {\bibfnamefont {M.}~\bibnamefont {S\o{}rensen}}, \ and\ \bibinfo {author} {\bibfnamefont {H.~F.}\ \bibnamefont {Song}},\ }\href {\doibase 10.1051/0004-6361/201832839} {\bibfield  {journal} {\bibinfo  {journal} {Astron. Astrophys.}\ }\textbf {\bibinfo {volume} {616}},\ \bibinfo {pages} {A28} (\bibinfo {year} {2018})},\ \Eprint {http://arxiv.org/abs/1802.05738} {arXiv:1802.05738 [astro-ph.SR]} \BibitemShut {NoStop}%
\bibitem [{\citenamefont {Fuller}\ and\ \citenamefont {Ma}(2019)}]{Fuller:2019sxi}%
  \BibitemOpen
  \bibfield  {author} {\bibinfo {author} {\bibfnamefont {J.}~\bibnamefont {Fuller}}\ and\ \bibinfo {author} {\bibfnamefont {L.}~\bibnamefont {Ma}},\ }\href {\doibase 10.3847/2041-8213/ab339b} {\bibfield  {journal} {\bibinfo  {journal} {Astrophys. J. Lett.}\ }\textbf {\bibinfo {volume} {881}},\ \bibinfo {pages} {L1} (\bibinfo {year} {2019})},\ \Eprint {http://arxiv.org/abs/1907.03714} {arXiv:1907.03714 [astro-ph.SR]} \BibitemShut {NoStop}%
\bibitem [{\citenamefont {van Son}\ \emph {et~al.}(2020)\citenamefont {van Son}, \citenamefont {de~Mink}, \citenamefont {Broekgaarden}, \citenamefont {Renzo}, \citenamefont {Justham}, \citenamefont {Laplace}, \citenamefont {Moran-Fraile}, \citenamefont {Hendriks},\ and\ \citenamefont {Farmer}}]{vanSon:2020zbk}%
  \BibitemOpen
  \bibfield  {author} {\bibinfo {author} {\bibfnamefont {L.~A.~C.}\ \bibnamefont {van Son}}, \bibinfo {author} {\bibfnamefont {S.~E.}\ \bibnamefont {de~Mink}}, \bibinfo {author} {\bibfnamefont {F.~S.}\ \bibnamefont {Broekgaarden}}, \bibinfo {author} {\bibfnamefont {M.}~\bibnamefont {Renzo}}, \bibinfo {author} {\bibfnamefont {S.}~\bibnamefont {Justham}}, \bibinfo {author} {\bibfnamefont {E.}~\bibnamefont {Laplace}}, \bibinfo {author} {\bibfnamefont {J.}~\bibnamefont {Moran-Fraile}}, \bibinfo {author} {\bibfnamefont {D.~D.}\ \bibnamefont {Hendriks}}, \ and\ \bibinfo {author} {\bibfnamefont {R.}~\bibnamefont {Farmer}},\ }\href {\doibase 10.3847/1538-4357/ab9809} {\bibfield  {journal} {\bibinfo  {journal} {Astrophys. J.}\ }\textbf {\bibinfo {volume} {897}},\ \bibinfo {pages} {100} (\bibinfo {year} {2020})},\ \Eprint {http://arxiv.org/abs/2004.05187} {arXiv:2004.05187 [astro-ph.HE]} \BibitemShut {NoStop}%
\bibitem [{\citenamefont {Gerosa}\ and\ \citenamefont {Fishbach}(2021)}]{Gerosa:2021mno}%
  \BibitemOpen
  \bibfield  {author} {\bibinfo {author} {\bibfnamefont {D.}~\bibnamefont {Gerosa}}\ and\ \bibinfo {author} {\bibfnamefont {M.}~\bibnamefont {Fishbach}},\ }\href {\doibase 10.1038/s41550-021-01398-w} {\bibfield  {journal} {\bibinfo  {journal} {Nature Astron.}\ }\textbf {\bibinfo {volume} {5}},\ \bibinfo {pages} {749} (\bibinfo {year} {2021})},\ \Eprint {http://arxiv.org/abs/2105.03439} {arXiv:2105.03439 [astro-ph.HE]} \BibitemShut {NoStop}%
\bibitem [{\citenamefont {Delfavero}\ \emph {et~al.}(2025)\citenamefont {Delfavero}, \citenamefont {Ray}, \citenamefont {Cook}, \citenamefont {Nathaniel}, \citenamefont {McKernan}, \citenamefont {Ford}, \citenamefont {Postiglione}, \citenamefont {McPike},\ and\ \citenamefont {O'Shaughnessy}}]{delfavero2025prospects}%
  \BibitemOpen
  \bibfield  {author} {\bibinfo {author} {\bibfnamefont {V.}~\bibnamefont {Delfavero}}, \bibinfo {author} {\bibfnamefont {S.}~\bibnamefont {Ray}}, \bibinfo {author} {\bibfnamefont {H.}~\bibnamefont {Cook}}, \bibinfo {author} {\bibfnamefont {K.}~\bibnamefont {Nathaniel}}, \bibinfo {author} {\bibfnamefont {B.}~\bibnamefont {McKernan}}, \bibinfo {author} {\bibfnamefont {K.}~\bibnamefont {Ford}}, \bibinfo {author} {\bibfnamefont {J.}~\bibnamefont {Postiglione}}, \bibinfo {author} {\bibfnamefont {E.}~\bibnamefont {McPike}}, \ and\ \bibinfo {author} {\bibfnamefont {R.}~\bibnamefont {O'Shaughnessy}},\ }\href@noop {} {\bibfield  {journal} {\bibinfo  {journal} {arXiv preprint arXiv:2508.13412}\ } (\bibinfo {year} {2025})}\BibitemShut {NoStop}%
\bibitem [{\citenamefont {Paiella}\ \emph {et~al.}(2025)\citenamefont {Paiella}, \citenamefont {Ugolini}, \citenamefont {Spera}, \citenamefont {Branchesi},\ and\ \citenamefont {Sedda}}]{paiella2025assembling}%
  \BibitemOpen
  \bibfield  {author} {\bibinfo {author} {\bibfnamefont {L.}~\bibnamefont {Paiella}}, \bibinfo {author} {\bibfnamefont {C.}~\bibnamefont {Ugolini}}, \bibinfo {author} {\bibfnamefont {M.}~\bibnamefont {Spera}}, \bibinfo {author} {\bibfnamefont {M.}~\bibnamefont {Branchesi}}, \ and\ \bibinfo {author} {\bibfnamefont {M.~A.}\ \bibnamefont {Sedda}},\ }\href@noop {} {\bibfield  {journal} {\bibinfo  {journal} {arXiv preprint arXiv:2509.10609}\ } (\bibinfo {year} {2025})}\BibitemShut {NoStop}%
\bibitem [{\citenamefont {Passenger}\ \emph {et~al.}(2025)\citenamefont {Passenger}, \citenamefont {Banagiri}, \citenamefont {Thrane}, \citenamefont {Lasky}, \citenamefont {Borchers}, \citenamefont {Fishbach},\ and\ \citenamefont {Ye}}]{passenger2025gw231123}%
  \BibitemOpen
  \bibfield  {author} {\bibinfo {author} {\bibfnamefont {L.}~\bibnamefont {Passenger}}, \bibinfo {author} {\bibfnamefont {S.}~\bibnamefont {Banagiri}}, \bibinfo {author} {\bibfnamefont {E.}~\bibnamefont {Thrane}}, \bibinfo {author} {\bibfnamefont {P.~D.}\ \bibnamefont {Lasky}}, \bibinfo {author} {\bibfnamefont {A.}~\bibnamefont {Borchers}}, \bibinfo {author} {\bibfnamefont {M.}~\bibnamefont {Fishbach}}, \ and\ \bibinfo {author} {\bibfnamefont {C.~S.}\ \bibnamefont {Ye}},\ }\href@noop {} {\bibfield  {journal} {\bibinfo  {journal} {arXiv preprint arXiv:2510.14363}\ } (\bibinfo {year} {2025})}\BibitemShut {NoStop}%
\bibitem [{\citenamefont {Gottlieb}\ \emph {et~al.}(2025)\citenamefont {Gottlieb}, \citenamefont {Metzger}, \citenamefont {Issa}, \citenamefont {Li}, \citenamefont {Renzo},\ and\ \citenamefont {Isi}}]{gottlieb2025spinning}%
  \BibitemOpen
  \bibfield  {author} {\bibinfo {author} {\bibfnamefont {O.}~\bibnamefont {Gottlieb}}, \bibinfo {author} {\bibfnamefont {B.~D.}\ \bibnamefont {Metzger}}, \bibinfo {author} {\bibfnamefont {D.}~\bibnamefont {Issa}}, \bibinfo {author} {\bibfnamefont {S.~E.}\ \bibnamefont {Li}}, \bibinfo {author} {\bibfnamefont {M.}~\bibnamefont {Renzo}}, \ and\ \bibinfo {author} {\bibfnamefont {M.}~\bibnamefont {Isi}},\ }\href@noop {} {\bibfield  {journal} {\bibinfo  {journal} {arXiv preprint arXiv:2508.15887}\ } (\bibinfo {year} {2025})}\BibitemShut {NoStop}%
\bibitem [{\citenamefont {Croon}\ \emph {et~al.}(2025)\citenamefont {Croon}, \citenamefont {Sakstein},\ and\ \citenamefont {Gerosa}}]{croon2025can}%
  \BibitemOpen
  \bibfield  {author} {\bibinfo {author} {\bibfnamefont {D.}~\bibnamefont {Croon}}, \bibinfo {author} {\bibfnamefont {J.}~\bibnamefont {Sakstein}}, \ and\ \bibinfo {author} {\bibfnamefont {D.}~\bibnamefont {Gerosa}},\ }\href@noop {} {\bibfield  {journal} {\bibinfo  {journal} {arXiv preprint arXiv:2508.10088}\ } (\bibinfo {year} {2025})}\BibitemShut {NoStop}%
\bibitem [{\citenamefont {Popa}\ and\ \citenamefont {de~Mink}(2025)}]{popa2025very}%
  \BibitemOpen
  \bibfield  {author} {\bibinfo {author} {\bibfnamefont {S.~A.}\ \bibnamefont {Popa}}\ and\ \bibinfo {author} {\bibfnamefont {S.~E.}\ \bibnamefont {de~Mink}},\ }\href@noop {} {\bibfield  {journal} {\bibinfo  {journal} {arXiv preprint arXiv:2509.00154}\ } (\bibinfo {year} {2025})}\BibitemShut {NoStop}%
\bibitem [{\citenamefont {De~Luca}\ \emph {et~al.}(2025)\citenamefont {De~Luca}, \citenamefont {Franciolini},\ and\ \citenamefont {Riotto}}]{de2025gw231123}%
  \BibitemOpen
  \bibfield  {author} {\bibinfo {author} {\bibfnamefont {V.}~\bibnamefont {De~Luca}}, \bibinfo {author} {\bibfnamefont {G.}~\bibnamefont {Franciolini}}, \ and\ \bibinfo {author} {\bibfnamefont {A.}~\bibnamefont {Riotto}},\ }\href@noop {} {\bibfield  {journal} {\bibinfo  {journal} {arXiv preprint arXiv:2508.09965}\ } (\bibinfo {year} {2025})}\BibitemShut {NoStop}%
\bibitem [{\citenamefont {Yuan}\ \emph {et~al.}(2025)\citenamefont {Yuan}, \citenamefont {Chen},\ and\ \citenamefont {Liu}}]{yuan2025gw231123}%
  \BibitemOpen
  \bibfield  {author} {\bibinfo {author} {\bibfnamefont {C.}~\bibnamefont {Yuan}}, \bibinfo {author} {\bibfnamefont {Z.-C.}\ \bibnamefont {Chen}}, \ and\ \bibinfo {author} {\bibfnamefont {L.}~\bibnamefont {Liu}},\ }\href@noop {} {\bibfield  {journal} {\bibinfo  {journal} {Physical Review D}\ }\textbf {\bibinfo {volume} {112}},\ \bibinfo {pages} {L081306} (\bibinfo {year} {2025})}\BibitemShut {NoStop}%
\bibitem [{\citenamefont {Liu}\ \emph {et~al.}(2025)\citenamefont {Liu}, \citenamefont {Wang}, \citenamefont {Tanikawa}, \citenamefont {Wu},\ and\ \citenamefont {Fujii}}]{liu2025formation}%
  \BibitemOpen
  \bibfield  {author} {\bibinfo {author} {\bibfnamefont {S.}~\bibnamefont {Liu}}, \bibinfo {author} {\bibfnamefont {L.}~\bibnamefont {Wang}}, \bibinfo {author} {\bibfnamefont {A.}~\bibnamefont {Tanikawa}}, \bibinfo {author} {\bibfnamefont {W.}~\bibnamefont {Wu}}, \ and\ \bibinfo {author} {\bibfnamefont {M.~S.}\ \bibnamefont {Fujii}},\ }\href@noop {} {\bibfield  {journal} {\bibinfo  {journal} {The Astrophysical Journal Letters}\ }\textbf {\bibinfo {volume} {993}},\ \bibinfo {pages} {L30} (\bibinfo {year} {2025})}\BibitemShut {NoStop}%
\bibitem [{\citenamefont {Tanikawa}\ \emph {et~al.}(2025)\citenamefont {Tanikawa}, \citenamefont {Liu}, \citenamefont {Wu}, \citenamefont {Fujii},\ and\ \citenamefont {Wang}}]{tanikawa2025gw231123}%
  \BibitemOpen
  \bibfield  {author} {\bibinfo {author} {\bibfnamefont {A.}~\bibnamefont {Tanikawa}}, \bibinfo {author} {\bibfnamefont {S.}~\bibnamefont {Liu}}, \bibinfo {author} {\bibfnamefont {W.}~\bibnamefont {Wu}}, \bibinfo {author} {\bibfnamefont {M.~S.}\ \bibnamefont {Fujii}}, \ and\ \bibinfo {author} {\bibfnamefont {L.}~\bibnamefont {Wang}},\ }\href@noop {} {\bibfield  {journal} {\bibinfo  {journal} {arXiv preprint arXiv:2508.01135}\ } (\bibinfo {year} {2025})}\BibitemShut {NoStop}%
\bibitem [{\citenamefont {Bartos}\ and\ \citenamefont {Haiman}(2025)}]{bartos2025accretion}%
  \BibitemOpen
  \bibfield  {author} {\bibinfo {author} {\bibfnamefont {I.}~\bibnamefont {Bartos}}\ and\ \bibinfo {author} {\bibfnamefont {Z.}~\bibnamefont {Haiman}},\ }\href@noop {} {\bibfield  {journal} {\bibinfo  {journal} {arXiv preprint arXiv:2508.08558}\ } (\bibinfo {year} {2025})}\BibitemShut {NoStop}%
\bibitem [{\citenamefont {K{\i}ro{\u{g}}lu}\ \emph {et~al.}(2025)\citenamefont {K{\i}ro{\u{g}}lu}, \citenamefont {Kremer},\ and\ \citenamefont {Rasio}}]{kirouglu2025beyond}%
  \BibitemOpen
  \bibfield  {author} {\bibinfo {author} {\bibfnamefont {F.}~\bibnamefont {K{\i}ro{\u{g}}lu}}, \bibinfo {author} {\bibfnamefont {K.}~\bibnamefont {Kremer}}, \ and\ \bibinfo {author} {\bibfnamefont {F.~A.}\ \bibnamefont {Rasio}},\ }\href@noop {} {\bibfield  {journal} {\bibinfo  {journal} {arXiv preprint arXiv:2509.05415}\ } (\bibinfo {year} {2025})}\BibitemShut {NoStop}%
\bibitem [{\citenamefont {Cuceu}\ \emph {et~al.}(2025)\citenamefont {Cuceu}, \citenamefont {Bizouard}, \citenamefont {Christensen},\ and\ \citenamefont {Sakellariadou}}]{Cuceu:2025fzi}%
  \BibitemOpen
  \bibfield  {author} {\bibinfo {author} {\bibfnamefont {I.}~\bibnamefont {Cuceu}}, \bibinfo {author} {\bibfnamefont {M.~A.}\ \bibnamefont {Bizouard}}, \bibinfo {author} {\bibfnamefont {N.}~\bibnamefont {Christensen}}, \ and\ \bibinfo {author} {\bibfnamefont {M.}~\bibnamefont {Sakellariadou}},\ }\href {\doibase 10.1103/zd8m-tzxd} {\bibfield  {journal} {\bibinfo  {journal} {Phys. Rev. D}\ } (\bibinfo {year} {2025}),\ 10.1103/zd8m-tzxd}\BibitemShut {NoStop}%
\bibitem [{\citenamefont {Fishbach}\ \emph {et~al.}(2020)\citenamefont {Fishbach}, \citenamefont {Farr},\ and\ \citenamefont {Holz}}]{Fishbach:2019ckx}%
  \BibitemOpen
  \bibfield  {author} {\bibinfo {author} {\bibfnamefont {M.}~\bibnamefont {Fishbach}}, \bibinfo {author} {\bibfnamefont {W.~M.}\ \bibnamefont {Farr}}, \ and\ \bibinfo {author} {\bibfnamefont {D.~E.}\ \bibnamefont {Holz}},\ }\href {\doibase 10.3847/2041-8213/ab77c9} {\bibfield  {journal} {\bibinfo  {journal} {Astrophys. J. Lett.}\ }\textbf {\bibinfo {volume} {891}},\ \bibinfo {pages} {L31} (\bibinfo {year} {2020})},\ \Eprint {http://arxiv.org/abs/1911.05882} {arXiv:1911.05882 [astro-ph.HE]} \BibitemShut {NoStop}%
\bibitem [{\citenamefont {Mandel}(2025)}]{Mandel:2025qnh}%
  \BibitemOpen
  \bibfield  {author} {\bibinfo {author} {\bibfnamefont {I.}~\bibnamefont {Mandel}},\ }\href@noop {} {\bibfield  {journal} {\bibinfo  {journal} {ArXiv}\ } (\bibinfo {year} {2025})},\ \Eprint {http://arxiv.org/abs/2509.05885} {arXiv:2509.05885 [astro-ph.HE]} \BibitemShut {NoStop}%
\bibitem [{\citenamefont {Miller}\ \emph {et~al.}(2024)\citenamefont {Miller}, \citenamefont {Isi}, \citenamefont {Chatziioannou}, \citenamefont {Varma},\ and\ \citenamefont {Mandel}}]{Miller:2023ncs}%
  \BibitemOpen
  \bibfield  {author} {\bibinfo {author} {\bibfnamefont {S.~J.}\ \bibnamefont {Miller}}, \bibinfo {author} {\bibfnamefont {M.}~\bibnamefont {Isi}}, \bibinfo {author} {\bibfnamefont {K.}~\bibnamefont {Chatziioannou}}, \bibinfo {author} {\bibfnamefont {V.}~\bibnamefont {Varma}}, \ and\ \bibinfo {author} {\bibfnamefont {I.}~\bibnamefont {Mandel}},\ }\href {\doibase 10.1103/PhysRevD.109.024024} {\bibfield  {journal} {\bibinfo  {journal} {Phys. Rev. D}\ }\textbf {\bibinfo {volume} {109}},\ \bibinfo {pages} {024024} (\bibinfo {year} {2024})},\ \Eprint {http://arxiv.org/abs/2310.01544} {arXiv:2310.01544 [astro-ph.HE]} \BibitemShut {NoStop}%
\bibitem [{\citenamefont {Miller}\ \emph {et~al.}(2025)\citenamefont {Miller}, \citenamefont {Isi}, \citenamefont {Chatziioannou}, \citenamefont {Varma},\ and\ \citenamefont {Hourihane}}]{Miller:2025eak}%
  \BibitemOpen
  \bibfield  {author} {\bibinfo {author} {\bibfnamefont {S.~J.}\ \bibnamefont {Miller}}, \bibinfo {author} {\bibfnamefont {M.}~\bibnamefont {Isi}}, \bibinfo {author} {\bibfnamefont {K.}~\bibnamefont {Chatziioannou}}, \bibinfo {author} {\bibfnamefont {V.}~\bibnamefont {Varma}}, \ and\ \bibinfo {author} {\bibfnamefont {S.}~\bibnamefont {Hourihane}},\ }\href@noop {} {\bibfield  {journal} {\bibinfo  {journal} {arXiv preprint arXiv:2505.14573}\ } (\bibinfo {year} {2025})},\ \Eprint {http://arxiv.org/abs/2505.14573} {arXiv:2505.14573 [gr-qc]} \BibitemShut {NoStop}%
\bibitem [{\citenamefont {Varma}\ \emph {et~al.}(2019)\citenamefont {Varma}, \citenamefont {Field}, \citenamefont {Scheel}, \citenamefont {Blackman}, \citenamefont {Gerosa}, \citenamefont {Stein}, \citenamefont {Kidder},\ and\ \citenamefont {Pfeiffer}}]{Varma:2019csw}%
  \BibitemOpen
  \bibfield  {author} {\bibinfo {author} {\bibfnamefont {V.}~\bibnamefont {Varma}}, \bibinfo {author} {\bibfnamefont {S.~E.}\ \bibnamefont {Field}}, \bibinfo {author} {\bibfnamefont {M.~A.}\ \bibnamefont {Scheel}}, \bibinfo {author} {\bibfnamefont {J.}~\bibnamefont {Blackman}}, \bibinfo {author} {\bibfnamefont {D.}~\bibnamefont {Gerosa}}, \bibinfo {author} {\bibfnamefont {L.~C.}\ \bibnamefont {Stein}}, \bibinfo {author} {\bibfnamefont {L.~E.}\ \bibnamefont {Kidder}}, \ and\ \bibinfo {author} {\bibfnamefont {H.~P.}\ \bibnamefont {Pfeiffer}},\ }\href {\doibase 10.1103/PhysRevResearch.1.033015} {\bibfield  {journal} {\bibinfo  {journal} {Phys. Rev. Research.}\ }\textbf {\bibinfo {volume} {1}},\ \bibinfo {pages} {033015} (\bibinfo {year} {2019})},\ \Eprint {http://arxiv.org/abs/1905.09300} {arXiv:1905.09300 [gr-qc]} \BibitemShut {NoStop}%
\bibitem [{\citenamefont {Colleoni}\ \emph {et~al.}(2025)\citenamefont {Colleoni}, \citenamefont {Vidal}, \citenamefont {Garc{\'\i}a-Quir{\'o}s}, \citenamefont {Ak{\c{c}}ay},\ and\ \citenamefont {Bera}}]{Colleoni:2024knd}%
  \BibitemOpen
  \bibfield  {author} {\bibinfo {author} {\bibfnamefont {M.}~\bibnamefont {Colleoni}}, \bibinfo {author} {\bibfnamefont {F.~A.~R.}\ \bibnamefont {Vidal}}, \bibinfo {author} {\bibfnamefont {C.}~\bibnamefont {Garc{\'\i}a-Quir{\'o}s}}, \bibinfo {author} {\bibfnamefont {S.}~\bibnamefont {Ak{\c{c}}ay}}, \ and\ \bibinfo {author} {\bibfnamefont {S.}~\bibnamefont {Bera}},\ }\href {\doibase 10.1103/PhysRevD.111.104019} {\bibfield  {journal} {\bibinfo  {journal} {Phys. Rev. D}\ }\textbf {\bibinfo {volume} {111}},\ \bibinfo {pages} {104019} (\bibinfo {year} {2025})},\ \Eprint {http://arxiv.org/abs/2412.16721} {arXiv:2412.16721 [gr-qc]} \BibitemShut {NoStop}%
\bibitem [{\citenamefont {Thompson}\ \emph {et~al.}(2024)\citenamefont {Thompson}, \citenamefont {Hamilton}, \citenamefont {London}, \citenamefont {Ghosh}, \citenamefont {Kolitsidou}, \citenamefont {Hoy},\ and\ \citenamefont {Hannam}}]{Thompson:2023ase}%
  \BibitemOpen
  \bibfield  {author} {\bibinfo {author} {\bibfnamefont {J.~E.}\ \bibnamefont {Thompson}}, \bibinfo {author} {\bibfnamefont {E.}~\bibnamefont {Hamilton}}, \bibinfo {author} {\bibfnamefont {L.}~\bibnamefont {London}}, \bibinfo {author} {\bibfnamefont {S.}~\bibnamefont {Ghosh}}, \bibinfo {author} {\bibfnamefont {P.}~\bibnamefont {Kolitsidou}}, \bibinfo {author} {\bibfnamefont {C.}~\bibnamefont {Hoy}}, \ and\ \bibinfo {author} {\bibfnamefont {M.}~\bibnamefont {Hannam}},\ }\href {\doibase 10.1103/PhysRevD.109.063012} {\bibfield  {journal} {\bibinfo  {journal} {Phys. Rev. D}\ }\textbf {\bibinfo {volume} {109}},\ \bibinfo {pages} {063012} (\bibinfo {year} {2024})},\ \Eprint {http://arxiv.org/abs/2312.10025} {arXiv:2312.10025 [gr-qc]} \BibitemShut {NoStop}%
\bibitem [{\citenamefont {Estell\'es}\ \emph {et~al.}(2022)\citenamefont {Estell\'es}, \citenamefont {Colleoni}, \citenamefont {Garc\'\i{}a-Quir\'os}, \citenamefont {Husa}, \citenamefont {Keitel}, \citenamefont {Mateu-Lucena}, \citenamefont {Planas},\ and\ \citenamefont {Ramos-Buades}}]{Estelles:2021gvs}%
  \BibitemOpen
  \bibfield  {author} {\bibinfo {author} {\bibfnamefont {H.}~\bibnamefont {Estell\'es}}, \bibinfo {author} {\bibfnamefont {M.}~\bibnamefont {Colleoni}}, \bibinfo {author} {\bibfnamefont {C.}~\bibnamefont {Garc\'\i{}a-Quir\'os}}, \bibinfo {author} {\bibfnamefont {S.}~\bibnamefont {Husa}}, \bibinfo {author} {\bibfnamefont {D.}~\bibnamefont {Keitel}}, \bibinfo {author} {\bibfnamefont {M.}~\bibnamefont {Mateu-Lucena}}, \bibinfo {author} {\bibfnamefont {M.~d.~L.}\ \bibnamefont {Planas}}, \ and\ \bibinfo {author} {\bibfnamefont {A.}~\bibnamefont {Ramos-Buades}},\ }\href {\doibase 10.1103/PhysRevD.105.084040} {\bibfield  {journal} {\bibinfo  {journal} {Phys. Rev. D}\ }\textbf {\bibinfo {volume} {105}},\ \bibinfo {pages} {084040} (\bibinfo {year} {2022})},\ \Eprint {http://arxiv.org/abs/2105.05872} {arXiv:2105.05872 [gr-qc]} \BibitemShut {NoStop}%
\bibitem [{\citenamefont {Ramos-Buades}\ \emph {et~al.}(2023)\citenamefont {Ramos-Buades}, \citenamefont {Buonanno}, \citenamefont {Estell\'es}, \citenamefont {Khalil}, \citenamefont {Mihaylov}, \citenamefont {Ossokine}, \citenamefont {Pompili},\ and\ \citenamefont {Shiferaw}}]{Ramos-Buades:2023ehm}%
  \BibitemOpen
  \bibfield  {author} {\bibinfo {author} {\bibfnamefont {A.}~\bibnamefont {Ramos-Buades}}, \bibinfo {author} {\bibfnamefont {A.}~\bibnamefont {Buonanno}}, \bibinfo {author} {\bibfnamefont {H.}~\bibnamefont {Estell\'es}}, \bibinfo {author} {\bibfnamefont {M.}~\bibnamefont {Khalil}}, \bibinfo {author} {\bibfnamefont {D.~P.}\ \bibnamefont {Mihaylov}}, \bibinfo {author} {\bibfnamefont {S.}~\bibnamefont {Ossokine}}, \bibinfo {author} {\bibfnamefont {L.}~\bibnamefont {Pompili}}, \ and\ \bibinfo {author} {\bibfnamefont {M.}~\bibnamefont {Shiferaw}},\ }\href {\doibase 10.1103/PhysRevD.108.124037} {\bibfield  {journal} {\bibinfo  {journal} {Phys. Rev. D}\ }\textbf {\bibinfo {volume} {108}},\ \bibinfo {pages} {124037} (\bibinfo {year} {2023})},\ \Eprint {http://arxiv.org/abs/2303.18046} {arXiv:2303.18046 [gr-qc]} \BibitemShut {NoStop}%
\bibitem [{\citenamefont {Ray}\ \emph {et~al.}(2025)\citenamefont {Ray}, \citenamefont {Banagiri}, \citenamefont {Thrane},\ and\ \citenamefont {Lasky}}]{Ray:2025rtt}%
  \BibitemOpen
  \bibfield  {author} {\bibinfo {author} {\bibfnamefont {A.}~\bibnamefont {Ray}}, \bibinfo {author} {\bibfnamefont {S.}~\bibnamefont {Banagiri}}, \bibinfo {author} {\bibfnamefont {E.}~\bibnamefont {Thrane}}, \ and\ \bibinfo {author} {\bibfnamefont {P.~D.}\ \bibnamefont {Lasky}},\ }\href@noop {} {\bibfield  {journal} {\bibinfo  {journal} {ArXiv}\ } (\bibinfo {year} {2025})},\ \Eprint {http://arxiv.org/abs/2510.07228} {arXiv:2510.07228 [gr-qc]} \BibitemShut {NoStop}%
\bibitem [{\citenamefont {Jan}\ \emph {et~al.}(2025)\citenamefont {Jan}, \citenamefont {Nicolella}, \citenamefont {Shoemaker},\ and\ \citenamefont {O'Shaughnessy}}]{Jan:2025zcm}%
  \BibitemOpen
  \bibfield  {author} {\bibinfo {author} {\bibfnamefont {A.}~\bibnamefont {Jan}}, \bibinfo {author} {\bibfnamefont {S.}~\bibnamefont {Nicolella}}, \bibinfo {author} {\bibfnamefont {D.}~\bibnamefont {Shoemaker}}, \ and\ \bibinfo {author} {\bibfnamefont {R.}~\bibnamefont {O'Shaughnessy}},\ }\href@noop {} {\bibfield  {journal} {\bibinfo  {journal} {ArXiv}\ } (\bibinfo {year} {2025})},\ \Eprint {http://arxiv.org/abs/2512.20060} {arXiv:2512.20060 [gr-qc]} \BibitemShut {NoStop}%
\bibitem [{\citenamefont {Nagar}\ \emph {et~al.}(2024)\citenamefont {Nagar}, \citenamefont {Gamba}, \citenamefont {Rettegno}, \citenamefont {Fantini},\ and\ \citenamefont {Bernuzzi}}]{Nagar:2024dzj}%
  \BibitemOpen
  \bibfield  {author} {\bibinfo {author} {\bibfnamefont {A.}~\bibnamefont {Nagar}}, \bibinfo {author} {\bibfnamefont {R.}~\bibnamefont {Gamba}}, \bibinfo {author} {\bibfnamefont {P.}~\bibnamefont {Rettegno}}, \bibinfo {author} {\bibfnamefont {V.}~\bibnamefont {Fantini}}, \ and\ \bibinfo {author} {\bibfnamefont {S.}~\bibnamefont {Bernuzzi}},\ }\href {\doibase 10.1103/PhysRevD.110.084001} {\bibfield  {journal} {\bibinfo  {journal} {Phys. Rev. D}\ }\textbf {\bibinfo {volume} {110}},\ \bibinfo {pages} {084001} (\bibinfo {year} {2024})},\ \Eprint {http://arxiv.org/abs/2404.05288} {arXiv:2404.05288 [gr-qc]} \BibitemShut {NoStop}%
\bibitem [{\citenamefont {Siegel}\ \emph {et~al.}(2025)\citenamefont {Siegel}, \citenamefont {Khusid}, \citenamefont {Isi},\ and\ \citenamefont {Farr}}]{siegel2025gw231123}%
  \BibitemOpen
  \bibfield  {author} {\bibinfo {author} {\bibfnamefont {H.}~\bibnamefont {Siegel}}, \bibinfo {author} {\bibfnamefont {N.~M.}\ \bibnamefont {Khusid}}, \bibinfo {author} {\bibfnamefont {M.}~\bibnamefont {Isi}}, \ and\ \bibinfo {author} {\bibfnamefont {W.~M.}\ \bibnamefont {Farr}},\ }\href@noop {} {\bibfield  {journal} {\bibinfo  {journal} {arXiv preprint arXiv:2511.02691}\ } (\bibinfo {year} {2025})}\BibitemShut {NoStop}%
\bibitem [{\citenamefont {Hu}\ \emph {et~al.}(2025)\citenamefont {Hu}, \citenamefont {Narola}, \citenamefont {Heynen}, \citenamefont {Wright}, \citenamefont {Veitch}, \citenamefont {Janquart},\ and\ \citenamefont {Van Den~Broeck}}]{Hu:2025lhv}%
  \BibitemOpen
  \bibfield  {author} {\bibinfo {author} {\bibfnamefont {Q.}~\bibnamefont {Hu}}, \bibinfo {author} {\bibfnamefont {H.}~\bibnamefont {Narola}}, \bibinfo {author} {\bibfnamefont {J.}~\bibnamefont {Heynen}}, \bibinfo {author} {\bibfnamefont {M.}~\bibnamefont {Wright}}, \bibinfo {author} {\bibfnamefont {J.}~\bibnamefont {Veitch}}, \bibinfo {author} {\bibfnamefont {J.}~\bibnamefont {Janquart}}, \ and\ \bibinfo {author} {\bibfnamefont {C.}~\bibnamefont {Van Den~Broeck}},\ }\href@noop {} {\bibfield  {journal} {\bibinfo  {journal} {arXiv}\ } (\bibinfo {year} {2025})},\ \Eprint {http://arxiv.org/abs/2512.17550} {arXiv:2512.17550 [gr-qc]} \BibitemShut {NoStop}%
\bibitem [{\citenamefont {Goyal}\ \emph {et~al.}(2025)\citenamefont {Goyal}, \citenamefont {Villarrubia-Rojo},\ and\ \citenamefont {Zumalacarregui}}]{Goyal:2025eqo}%
  \BibitemOpen
  \bibfield  {author} {\bibinfo {author} {\bibfnamefont {S.}~\bibnamefont {Goyal}}, \bibinfo {author} {\bibfnamefont {H.}~\bibnamefont {Villarrubia-Rojo}}, \ and\ \bibinfo {author} {\bibfnamefont {M.}~\bibnamefont {Zumalacarregui}},\ }\href@noop {} {\bibfield  {journal} {\bibinfo  {journal} {arXiv}\ } (\bibinfo {year} {2025})},\ \Eprint {http://arxiv.org/abs/2512.17631} {arXiv:2512.17631 [astro-ph.GA]} \BibitemShut {NoStop}%
\bibitem [{\citenamefont {Cutler}\ and\ \citenamefont {Flanagan}(1994)}]{cutler1994gravitational}%
  \BibitemOpen
  \bibfield  {author} {\bibinfo {author} {\bibfnamefont {C.}~\bibnamefont {Cutler}}\ and\ \bibinfo {author} {\bibfnamefont {E.~E.}\ \bibnamefont {Flanagan}},\ }\href@noop {} {\bibfield  {journal} {\bibinfo  {journal} {Physical Review D}\ }\textbf {\bibinfo {volume} {49}},\ \bibinfo {pages} {2658} (\bibinfo {year} {1994})}\BibitemShut {NoStop}%
\bibitem [{\citenamefont {Cutler}\ and\ \citenamefont {Vallisneri}(2007)}]{Cutler:2007mi}%
  \BibitemOpen
  \bibfield  {author} {\bibinfo {author} {\bibfnamefont {C.}~\bibnamefont {Cutler}}\ and\ \bibinfo {author} {\bibfnamefont {M.}~\bibnamefont {Vallisneri}},\ }\href {\doibase 10.1103/PhysRevD.76.104018} {\bibfield  {journal} {\bibinfo  {journal} {Phys. Rev. D}\ }\textbf {\bibinfo {volume} {76}},\ \bibinfo {pages} {104018} (\bibinfo {year} {2007})},\ \Eprint {http://arxiv.org/abs/0707.2982} {arXiv:0707.2982 [gr-qc]} \BibitemShut {NoStop}%
\bibitem [{\citenamefont {Vallisneri}(2008)}]{Vallisneri:2007ev}%
  \BibitemOpen
  \bibfield  {author} {\bibinfo {author} {\bibfnamefont {M.}~\bibnamefont {Vallisneri}},\ }\href {\doibase 10.1103/PhysRevD.77.042001} {\bibfield  {journal} {\bibinfo  {journal} {Phys. Rev. D}\ }\textbf {\bibinfo {volume} {77}},\ \bibinfo {pages} {042001} (\bibinfo {year} {2008})},\ \Eprint {http://arxiv.org/abs/gr-qc/0703086} {arXiv:gr-qc/0703086} \BibitemShut {NoStop}%
\bibitem [{\citenamefont {Udall}\ \emph {et~al.}(2025)\citenamefont {Udall}, \citenamefont {Bini}, \citenamefont {Chatziioannou}, \citenamefont {Davis}, \citenamefont {Hourihane}, \citenamefont {Lecoeuche}, \citenamefont {McIver},\ and\ \citenamefont {Miller}}]{udall2025inferring}%
  \BibitemOpen
  \bibfield  {author} {\bibinfo {author} {\bibfnamefont {R.}~\bibnamefont {Udall}}, \bibinfo {author} {\bibfnamefont {S.}~\bibnamefont {Bini}}, \bibinfo {author} {\bibfnamefont {K.}~\bibnamefont {Chatziioannou}}, \bibinfo {author} {\bibfnamefont {D.}~\bibnamefont {Davis}}, \bibinfo {author} {\bibfnamefont {S.}~\bibnamefont {Hourihane}}, \bibinfo {author} {\bibfnamefont {Y.}~\bibnamefont {Lecoeuche}}, \bibinfo {author} {\bibfnamefont {J.}~\bibnamefont {McIver}}, \ and\ \bibinfo {author} {\bibfnamefont {S.}~\bibnamefont {Miller}},\ }\href@noop {} {\bibfield  {journal} {\bibinfo  {journal} {arXiv preprint arXiv:2510.05029}\ } (\bibinfo {year} {2025})}\BibitemShut {NoStop}%
\bibitem [{\citenamefont {Branchesi}\ \emph {et~al.}(2023)\citenamefont {Branchesi}, \citenamefont {Maggiore}, \citenamefont {Alonso}, \citenamefont {Badger}, \citenamefont {Banerjee}, \citenamefont {Beirnaert}, \citenamefont {Belgacem}, \citenamefont {Bhagwat}, \citenamefont {Boileau}, \citenamefont {Borhanian} \emph {et~al.}}]{branchesi2023science}%
  \BibitemOpen
  \bibfield  {author} {\bibinfo {author} {\bibfnamefont {M.}~\bibnamefont {Branchesi}}, \bibinfo {author} {\bibfnamefont {M.}~\bibnamefont {Maggiore}}, \bibinfo {author} {\bibfnamefont {D.}~\bibnamefont {Alonso}}, \bibinfo {author} {\bibfnamefont {C.}~\bibnamefont {Badger}}, \bibinfo {author} {\bibfnamefont {B.}~\bibnamefont {Banerjee}}, \bibinfo {author} {\bibfnamefont {F.}~\bibnamefont {Beirnaert}}, \bibinfo {author} {\bibfnamefont {E.}~\bibnamefont {Belgacem}}, \bibinfo {author} {\bibfnamefont {S.}~\bibnamefont {Bhagwat}}, \bibinfo {author} {\bibfnamefont {G.}~\bibnamefont {Boileau}}, \bibinfo {author} {\bibfnamefont {S.}~\bibnamefont {Borhanian}},  \emph {et~al.},\ }\href@noop {} {\bibfield  {journal} {\bibinfo  {journal} {Journal of Cosmology and Astroparticle Physics}\ }\textbf {\bibinfo {volume} {2023}},\ \bibinfo {pages} {068} (\bibinfo {year} {2023})}\BibitemShut {NoStop}%
\bibitem [{\citenamefont {Lindblom}\ \emph {et~al.}(2008)\citenamefont {Lindblom}, \citenamefont {Owen},\ and\ \citenamefont {Brown}}]{Lindblom:2008cm}%
  \BibitemOpen
  \bibfield  {author} {\bibinfo {author} {\bibfnamefont {L.}~\bibnamefont {Lindblom}}, \bibinfo {author} {\bibfnamefont {B.~J.}\ \bibnamefont {Owen}}, \ and\ \bibinfo {author} {\bibfnamefont {D.~A.}\ \bibnamefont {Brown}},\ }\href {\doibase 10.1103/PhysRevD.78.124020} {\bibfield  {journal} {\bibinfo  {journal} {Phys. Rev. D}\ }\textbf {\bibinfo {volume} {78}},\ \bibinfo {pages} {124020} (\bibinfo {year} {2008})},\ \Eprint {http://arxiv.org/abs/0809.3844} {arXiv:0809.3844 [gr-qc]} \BibitemShut {NoStop}%
\bibitem [{\citenamefont {Miller}(2005)}]{miller2005accuracy}%
  \BibitemOpen
  \bibfield  {author} {\bibinfo {author} {\bibfnamefont {M.}~\bibnamefont {Miller}},\ }\href@noop {} {\bibfield  {journal} {\bibinfo  {journal} {Physical Review D—Particles, Fields, Gravitation, and Cosmology}\ }\textbf {\bibinfo {volume} {71}},\ \bibinfo {pages} {104016} (\bibinfo {year} {2005})}\BibitemShut {NoStop}%
\bibitem [{\citenamefont {{LIGO Scientific, Virgo, and KAGRA Collaboration}}(2025)}]{zenodo_v2}%
  \BibitemOpen
  \bibfield  {author} {\bibinfo {author} {\bibnamefont {{LIGO Scientific, Virgo, and KAGRA Collaboration}}},\ }\href {\doibase 10.5281/zenodo.16004263} {\enquote {\bibinfo {title} {Gw231123: a binary black hole merger with total mass 190-265\,$m_\odot$ --- data release},}\ } (\bibinfo {year} {2025})\BibitemShut {NoStop}%
\bibitem [{\citenamefont {Littenberg}\ and\ \citenamefont {Cornish}(2015)}]{Littenberg:2014oda}%
  \BibitemOpen
  \bibfield  {author} {\bibinfo {author} {\bibfnamefont {T.~B.}\ \bibnamefont {Littenberg}}\ and\ \bibinfo {author} {\bibfnamefont {N.~J.}\ \bibnamefont {Cornish}},\ }\href {\doibase 10.1103/PhysRevD.91.084034} {\bibfield  {journal} {\bibinfo  {journal} {Phys. Rev. D}\ }\textbf {\bibinfo {volume} {91}},\ \bibinfo {pages} {084034} (\bibinfo {year} {2015})},\ \Eprint {http://arxiv.org/abs/1410.3852} {arXiv:1410.3852 [gr-qc]} \BibitemShut {NoStop}%
\bibitem [{\citenamefont {Pankow}\ \emph {et~al.}(2018)\citenamefont {Pankow} \emph {et~al.}}]{Pankow:2018qpo}%
  \BibitemOpen
  \bibfield  {author} {\bibinfo {author} {\bibfnamefont {C.}~\bibnamefont {Pankow}} \emph {et~al.},\ }\href {\doibase 10.1103/PhysRevD.98.084016} {\bibfield  {journal} {\bibinfo  {journal} {Phys. Rev. D}\ }\textbf {\bibinfo {volume} {98}},\ \bibinfo {pages} {084016} (\bibinfo {year} {2018})},\ \Eprint {http://arxiv.org/abs/1808.03619} {arXiv:1808.03619 [gr-qc]} \BibitemShut {NoStop}%
\bibitem [{\citenamefont {Nissanke}\ \emph {et~al.}(2010)\citenamefont {Nissanke}, \citenamefont {Holz}, \citenamefont {Hughes}, \citenamefont {Dalal},\ and\ \citenamefont {Sievers}}]{Nissanke:2009kt}%
  \BibitemOpen
  \bibfield  {author} {\bibinfo {author} {\bibfnamefont {S.}~\bibnamefont {Nissanke}}, \bibinfo {author} {\bibfnamefont {D.~E.}\ \bibnamefont {Holz}}, \bibinfo {author} {\bibfnamefont {S.~A.}\ \bibnamefont {Hughes}}, \bibinfo {author} {\bibfnamefont {N.}~\bibnamefont {Dalal}}, \ and\ \bibinfo {author} {\bibfnamefont {J.~L.}\ \bibnamefont {Sievers}},\ }\href {\doibase 10.1088/0004-637X/725/1/496} {\bibfield  {journal} {\bibinfo  {journal} {Astrophys. J.}\ }\textbf {\bibinfo {volume} {725}},\ \bibinfo {pages} {496} (\bibinfo {year} {2010})},\ \Eprint {http://arxiv.org/abs/0904.1017} {arXiv:0904.1017 [astro-ph.CO]} \BibitemShut {NoStop}%
\bibitem [{\citenamefont {Ashton}\ \emph {et~al.}(2019)\citenamefont {Ashton} \emph {et~al.}}]{Ashton:2018jfp}%
  \BibitemOpen
  \bibfield  {author} {\bibinfo {author} {\bibfnamefont {G.}~\bibnamefont {Ashton}} \emph {et~al.},\ }\href {\doibase 10.3847/1538-4365/ab06fc} {\bibfield  {journal} {\bibinfo  {journal} {Astrophys. J. Suppl.}\ }\textbf {\bibinfo {volume} {241}},\ \bibinfo {pages} {27} (\bibinfo {year} {2019})},\ \Eprint {http://arxiv.org/abs/1811.02042} {arXiv:1811.02042 [astro-ph.IM]} \BibitemShut {NoStop}%
\bibitem [{\citenamefont {Romero-Shaw}\ \emph {et~al.}(2020)\citenamefont {Romero-Shaw} \emph {et~al.}}]{Romero-Shaw:2020owr}%
  \BibitemOpen
  \bibfield  {author} {\bibinfo {author} {\bibfnamefont {I.~M.}\ \bibnamefont {Romero-Shaw}} \emph {et~al.},\ }\href {\doibase 10.1093/mnras/staa2850} {\bibfield  {journal} {\bibinfo  {journal} {Mon. Not. Roy. Astron. Soc.}\ }\textbf {\bibinfo {volume} {499}},\ \bibinfo {pages} {3295} (\bibinfo {year} {2020})},\ \Eprint {http://arxiv.org/abs/2006.00714} {arXiv:2006.00714 [astro-ph.IM]} \BibitemShut {NoStop}%
\bibitem [{\citenamefont {Talbot}\ \emph {et~al.}(2025)\citenamefont {Talbot} \emph {et~al.}}]{Talbot:2025vth}%
  \BibitemOpen
  \bibfield  {author} {\bibinfo {author} {\bibfnamefont {C.}~\bibnamefont {Talbot}} \emph {et~al.},\ }\href@noop {} {\bibfield  {journal} {\bibinfo  {journal} {arXiv preprint arXiv:2508.11091}\ } (\bibinfo {year} {2025})}\BibitemShut {NoStop}%
\bibitem [{\citenamefont {Lin}(1991)}]{Lin:1991zzm}%
  \BibitemOpen
  \bibfield  {author} {\bibinfo {author} {\bibfnamefont {J.}~\bibnamefont {Lin}},\ }\href {\doibase 10.1109/18.61115} {\bibfield  {journal} {\bibinfo  {journal} {IEEE Trans. Info. Theor.}\ }\textbf {\bibinfo {volume} {37}},\ \bibinfo {pages} {145} (\bibinfo {year} {1991})}\BibitemShut {NoStop}%
\bibitem [{\citenamefont {Biscoveanu}\ \emph {et~al.}(2021)\citenamefont {Biscoveanu}, \citenamefont {Isi}, \citenamefont {Varma},\ and\ \citenamefont {Vitale}}]{Biscoveanu:2021nvg}%
  \BibitemOpen
  \bibfield  {author} {\bibinfo {author} {\bibfnamefont {S.}~\bibnamefont {Biscoveanu}}, \bibinfo {author} {\bibfnamefont {M.}~\bibnamefont {Isi}}, \bibinfo {author} {\bibfnamefont {V.}~\bibnamefont {Varma}}, \ and\ \bibinfo {author} {\bibfnamefont {S.}~\bibnamefont {Vitale}},\ }\href {\doibase 10.1103/PhysRevD.104.103018} {\bibfield  {journal} {\bibinfo  {journal} {Phys. Rev. D}\ }\textbf {\bibinfo {volume} {104}},\ \bibinfo {pages} {103018} (\bibinfo {year} {2021})},\ \Eprint {http://arxiv.org/abs/2106.06492} {arXiv:2106.06492 [gr-qc]} \BibitemShut {NoStop}%
\bibitem [{\citenamefont {Xu}\ and\ \citenamefont {Hamilton}(2023)}]{Xu:2022zza}%
  \BibitemOpen
  \bibfield  {author} {\bibinfo {author} {\bibfnamefont {Y.}~\bibnamefont {Xu}}\ and\ \bibinfo {author} {\bibfnamefont {E.}~\bibnamefont {Hamilton}},\ }\href {\doibase 10.1103/PhysRevD.107.103049} {\bibfield  {journal} {\bibinfo  {journal} {Phys. Rev. D}\ }\textbf {\bibinfo {volume} {107}},\ \bibinfo {pages} {103049} (\bibinfo {year} {2023})},\ \Eprint {http://arxiv.org/abs/2211.09561} {arXiv:2211.09561 [gr-qc]} \BibitemShut {NoStop}%
\bibitem [{\citenamefont {Abbott}\ \emph {et~al.}(2020)\citenamefont {Abbott} \emph {et~al.}}]{LIGOScientific:2020iuh}%
  \BibitemOpen
  \bibfield  {author} {\bibinfo {author} {\bibfnamefont {R.}~\bibnamefont {Abbott}} \emph {et~al.} (\bibinfo {collaboration} {LIGO Scientific, Virgo}),\ }\href {\doibase 10.1103/PhysRevLett.125.101102} {\bibfield  {journal} {\bibinfo  {journal} {Phys. Rev. Lett.}\ }\textbf {\bibinfo {volume} {125}},\ \bibinfo {pages} {101102} (\bibinfo {year} {2020})},\ \Eprint {http://arxiv.org/abs/2009.01075} {arXiv:2009.01075 [gr-qc]} \BibitemShut {NoStop}%
\bibitem [{\citenamefont {Romano}\ and\ \citenamefont {Cornish}(2017)}]{Romano:2016dpx}%
  \BibitemOpen
  \bibfield  {author} {\bibinfo {author} {\bibfnamefont {J.~D.}\ \bibnamefont {Romano}}\ and\ \bibinfo {author} {\bibfnamefont {N.~J.}\ \bibnamefont {Cornish}},\ }\href {\doibase 10.1007/s41114-017-0004-1} {\bibfield  {journal} {\bibinfo  {journal} {Living Rev. Rel.}\ }\textbf {\bibinfo {volume} {20}},\ \bibinfo {pages} {2} (\bibinfo {year} {2017})},\ \Eprint {http://arxiv.org/abs/1608.06889} {arXiv:1608.06889 [gr-qc]} \BibitemShut {NoStop}%
\bibitem [{\citenamefont {Hoy}(2022)}]{Hoy:2022tst}%
  \BibitemOpen
  \bibfield  {author} {\bibinfo {author} {\bibfnamefont {C.}~\bibnamefont {Hoy}},\ }\href {\doibase 10.1103/PhysRevD.106.083003} {\bibfield  {journal} {\bibinfo  {journal} {Phys. Rev. D}\ }\textbf {\bibinfo {volume} {106}},\ \bibinfo {pages} {083003} (\bibinfo {year} {2022})},\ \Eprint {http://arxiv.org/abs/2208.00106} {arXiv:2208.00106 [gr-qc]} \BibitemShut {NoStop}%
\bibitem [{\citenamefont {Hoy}\ \emph {et~al.}(2025)\citenamefont {Hoy}, \citenamefont {Ak{\c{c}}ay}, \citenamefont {Mac~Uilliam},\ and\ \citenamefont {Thompson}}]{hoy2025incorporation}%
  \BibitemOpen
  \bibfield  {author} {\bibinfo {author} {\bibfnamefont {C.}~\bibnamefont {Hoy}}, \bibinfo {author} {\bibfnamefont {S.}~\bibnamefont {Ak{\c{c}}ay}}, \bibinfo {author} {\bibfnamefont {J.}~\bibnamefont {Mac~Uilliam}}, \ and\ \bibinfo {author} {\bibfnamefont {J.~E.}\ \bibnamefont {Thompson}},\ }\href@noop {} {\bibfield  {journal} {\bibinfo  {journal} {Nature Astronomy}\ ,\ \bibinfo {pages} {1}} (\bibinfo {year} {2025})}\BibitemShut {NoStop}%
\bibitem [{\citenamefont {Payne}\ \emph {et~al.}(2022)\citenamefont {Payne}, \citenamefont {Hourihane}, \citenamefont {Golomb}, \citenamefont {Udall}, \citenamefont {Davis},\ and\ \citenamefont {Chatziioannou}}]{payne2022curious}%
  \BibitemOpen
  \bibfield  {author} {\bibinfo {author} {\bibfnamefont {E.}~\bibnamefont {Payne}}, \bibinfo {author} {\bibfnamefont {S.}~\bibnamefont {Hourihane}}, \bibinfo {author} {\bibfnamefont {J.}~\bibnamefont {Golomb}}, \bibinfo {author} {\bibfnamefont {R.}~\bibnamefont {Udall}}, \bibinfo {author} {\bibfnamefont {D.}~\bibnamefont {Davis}}, \ and\ \bibinfo {author} {\bibfnamefont {K.}~\bibnamefont {Chatziioannou}},\ }\href@noop {} {\bibfield  {journal} {\bibinfo  {journal} {Physical Review D}\ }\textbf {\bibinfo {volume} {106}},\ \bibinfo {pages} {104017} (\bibinfo {year} {2022})}\BibitemShut {NoStop}%
\bibitem [{\citenamefont {Fritschel}\ \emph {et~al.}(2022)\citenamefont {Fritschel}, \citenamefont {Kuns}, \citenamefont {Driggers}, \citenamefont {Effler}, \citenamefont {Lantz}, \citenamefont {Ottaway}, \citenamefont {Ballmer}, \citenamefont {Dooley}, \citenamefont {Adhikari}, \citenamefont {Evans} \emph {et~al.}}]{fritschel2022report}%
  \BibitemOpen
  \bibfield  {author} {\bibinfo {author} {\bibfnamefont {P.}~\bibnamefont {Fritschel}}, \bibinfo {author} {\bibfnamefont {K.}~\bibnamefont {Kuns}}, \bibinfo {author} {\bibfnamefont {J.}~\bibnamefont {Driggers}}, \bibinfo {author} {\bibfnamefont {A.}~\bibnamefont {Effler}}, \bibinfo {author} {\bibfnamefont {B.}~\bibnamefont {Lantz}}, \bibinfo {author} {\bibfnamefont {D.}~\bibnamefont {Ottaway}}, \bibinfo {author} {\bibfnamefont {S.}~\bibnamefont {Ballmer}}, \bibinfo {author} {\bibfnamefont {K.}~\bibnamefont {Dooley}}, \bibinfo {author} {\bibfnamefont {R.}~\bibnamefont {Adhikari}}, \bibinfo {author} {\bibfnamefont {M.}~\bibnamefont {Evans}},  \emph {et~al.},\ }\href@noop {} {\bibfield  {journal} {\bibinfo  {journal} {LIGO Document}\ }\textbf {\bibinfo {volume} {2200287}},\ \bibinfo {pages} {6} (\bibinfo {year} {2022})}\BibitemShut {NoStop}%
\bibitem [{\citenamefont {Evans}\ \emph {et~al.}(2021)\citenamefont {Evans} \emph {et~al.}}]{Evans:2021gyd}%
  \BibitemOpen
  \bibfield  {author} {\bibinfo {author} {\bibfnamefont {M.}~\bibnamefont {Evans}} \emph {et~al.},\ }\href@noop {} {\bibfield  {journal} {\bibinfo  {journal} {ArXiv}\ } (\bibinfo {year} {2021})},\ \Eprint {http://arxiv.org/abs/2109.09882} {arXiv:2109.09882 [astro-ph.IM]} \BibitemShut {NoStop}%
\bibitem [{\citenamefont {Maggiore}\ \emph {et~al.}(2020)\citenamefont {Maggiore} \emph {et~al.}}]{ET:2019dnz}%
  \BibitemOpen
  \bibfield  {author} {\bibinfo {author} {\bibfnamefont {M.}~\bibnamefont {Maggiore}} \emph {et~al.} (\bibinfo {collaboration} {ET}),\ }\href {\doibase 10.1088/1475-7516/2020/03/050} {\bibfield  {journal} {\bibinfo  {journal} {JCAP}\ }\textbf {\bibinfo {volume} {03}},\ \bibinfo {pages} {050} (\bibinfo {year} {2020})},\ \Eprint {http://arxiv.org/abs/1912.02622} {arXiv:1912.02622 [astro-ph.CO]} \BibitemShut {NoStop}%
\bibitem [{\citenamefont {Kuns}\ and\ \citenamefont {Fritschel}(2024)}]{a_sharp_psd}%
  \BibitemOpen
  \bibfield  {author} {\bibinfo {author} {\bibfnamefont {K.}~\bibnamefont {Kuns}}\ and\ \bibinfo {author} {\bibfnamefont {P.}~\bibnamefont {Fritschel}},\ }\href@noop {} {\bibfield  {journal} {\bibinfo  {journal} {LIGO Document}\ }\textbf {\bibinfo {volume} {T2300041-v1}} (\bibinfo {year} {2024})}\BibitemShut {NoStop}%
\end{thebibliography}%

\end{document}